%% file: main.tex
\documentclass[twocolumn, pra, aps, superscriptaddress, 10pt]{revtex4-2}
\usepackage[pdftex]{graphicx}
\usepackage{bm}
\usepackage{listings}
\usepackage{amsmath}
\usepackage{xcolor}
\usepackage[hidelinks]{hyperref}
\usepackage{dcolumn}
\usepackage[utf8]{inputenc}
\usepackage[T1]{fontenc}
\usepackage{etoolbox}

\makeatletter
\def\@email#1#2{%
 \endgroup
 \patchcmd{\titleblock@produce}
  {\frontmatter@RRAPformat}
  {\frontmatter@RRAPformat{\produce@RRAP{*#1\href{mailto:#2}{#2}}}\frontmatter@RRAPformat}
  {}{}
}%
\makeatother
\begin{document}

\title{GPAW: An open Python package for electronic-structure calculations}

\author{Jens Jørgen Mortensen}
\affiliation{Corresponding author: jjmo@dtu.dk}
\affiliation{CAMd, Department of Physics, Technical University of Denmark, 2800 Kgs. Lyngby, Denmark}
\author{Ask Hjorth Larsen}
\affiliation{CAMd, Department of Physics, Technical University of Denmark, 2800 Kgs. Lyngby, Denmark}
\author{Mikael Kuisma}
\affiliation{CAMd, Department of Physics, Technical University of Denmark, 2800 Kgs. Lyngby, Denmark}
\author{Aleksei V. Ivanov}
\affiliation{Riverlane Ltd, St Andrews House, 59 St Andrews Street, Cambridge, CB2 3BZ, United Kingdom}
\author{Alireza Taghizadeh}
\affiliation{CAMd, Department of Physics, Technical University of Denmark, 2800 Kgs. Lyngby, Denmark}
\author{Andrew Peterson}
\affiliation{School of Engineering, Brown University, Providence, Rhode Island, USA}
\author{Anubhab Haldar}
\affiliation{Department of Electrical and Computer Engineering, Boston University, United States of America.}
\author{Asmus Ougaard Dohn}
\affiliation{Department of Physics, Technical University of Denmark, 2800 Lyngby, Denmark; Science Institute and Faculty of Physical Sciences, VR-III, University of Iceland, Reykjavík 107, Iceland}
\author{Christian Schäfer}
\affiliation{Department of Physics, Chalmers University of Technology, SE-412 96 Gothenburg, Sweden}
\author{Elvar Örn Jónsson}
\affiliation{Science Institute and Faculty of Physical Sciences, University of Iceland, VR-III, 107 Reykjavík, Iceland}
\author{Eric D. Hermes}
\affiliation{Quantum-Si, 29 Business Park Drive, Branford, CT 06405, USA}
\author{Fredrik Andreas Nilsson}
\affiliation{CAMd, Department of Physics, Technical University of Denmark, 2800 Kgs. Lyngby, Denmark}
\author{Georg Kastlunger}
\affiliation{CatTheory, Department of Physics, Technical University of Denmark, 2800 Kgs. Lyngby, Denmark}
\author{Gianluca Levi}
\affiliation{Science Institute and Faculty of Physical Sciences, University of Iceland, VR-III, 107 Reykjavík, Iceland}
\author{Hannes Jónsson}
\affiliation{Science Institute and Faculty of Physical Sciences, University of Iceland, VR-III, 107 Reykjavík, Iceland}
\author{Hannu Häkkinen}
\affiliation{Departments of Physics and Chemistry, Nanoscience Center, University of Jyväskylä, FI-40014 Jyväskylä, Finland}
\author{Jakub Fojt}
\affiliation{Department of Physics, Chalmers University of Technology, SE-412 96 Gothenburg, Sweden}
\author{Jiban Kangsabanik}
\affiliation{CAMd, Department of Physics, Technical University of Denmark, 2800 Kgs. Lyngby, Denmark}
\author{Joachim Sødequist}
\affiliation{CAMd, Department of Physics, Technical University of Denmark, 2800 Kgs. Lyngby, Denmark}
\author{Jouko Lehtomäki}
\affiliation{Department of Applied Physics, Aalto University, P.O. Box 11100, 00076 Aalto, Finland}
\author{Julian Heske}
\affiliation{CAMd, Department of Physics, Technical University of Denmark, 2800 Kgs. Lyngby, Denmark}
\author{Jussi Enkovaara}
\affiliation{CSC -- IT Center for Science Ltd., P.O. Box 405, FI-02101 Espoo, Finland}
\author{Kirsten Trøstrup Winther}
\affiliation{SUNCAT Center for Interface Science and Catalysis, SLAC National Accelerator Laboratory, Menlo Park, California 94025, United States}
\author{Marcin Dulak}
\affiliation{CAMd, Department of Physics, Technical University of Denmark, 2800 Kgs. Lyngby, Denmark}
\author{Marko M. Melander}
\affiliation{Department of Chemistry, Nanoscience Center, University of Jyväskylä, FI-40014 Jyväskylä, Finland}
\author{Martin Ovesen}
\affiliation{CAMd, Department of Physics, Technical University of Denmark, 2800 Kgs. Lyngby, Denmark}
\author{Martti Louhivuori}
\affiliation{CSC -- IT Center for Science Ltd., P.O. Box 405, FI-02101 Espoo, Finland}
\author{Michael Walter}
\affiliation{University of Freiburg, FIT Freiburg Centre for Interactive Materials and Bioinspired Technologies, Georges-Köhler-Allee 105, 79110 Freiburg, Germany}
\author{Morten Gjerding}
\affiliation{CAMd, Department of Physics, Technical University of Denmark, 2800 Kgs. Lyngby, Denmark}
\author{Olga Lopez-Acevedo}
\affiliation{Biophysics of Tropical Diseases, Max Planck Tandem Group, University of Antioquia UdeA, 050010 Medellin, Colombia}
\author{Paul Erhart}
\affiliation{Department of Physics, Chalmers University of Technology, SE-412 96 Gothenburg, Sweden}
\author{Robert Warmbier}
\affiliation{School of Physics and Mandelstam Institute for Theoretical Physics, University of the Witwatersrand, 1 Jan Smuts Avenue, 2001, Johannesburg, South Africa}
\author{Rolf Würdemann}
\affiliation{Freiburger Materialforschungszentrum, Universität Freiburg, Stefan-Meier-Straße 21, D-79104 Freiburg, Germany}
\author{Sami Kaappa}
\affiliation{Computational Physics Laboratory, Tampere University, P.O. Box 692, FI-33014 Tampere, Finland}
\author{Simone Latini}
\affiliation{Nanomade, Department of Physics, Technical University of Denmark, 2800 Kgs. Lyngby, Denmark}
\author{Tara Maria Boland}
\affiliation{CAMd, Department of Physics, Technical University of Denmark, 2800 Kgs. Lyngby, Denmark}
\author{Thomas Bligaard}
\affiliation{Department of Energy Conversion and Storage, Technical University of Denmark, DK-2800 Lyngby, Denmark}
\author{Thorbjørn Skovhus}
\affiliation{CAMd, Department of Physics, Technical University of Denmark, 2800 Kgs. Lyngby, Denmark}
\author{Toma Susi}
\affiliation{University of Vienna, Faculty of Physics, Boltzmanngasse 5, 1090 Vienna, Austria}
\author{Tristan Maxson}
\affiliation{Department of Chemical and Biological Engineering, The University of Alabama, Tuscaloosa, AL 35487, USA}
\author{Tuomas Rossi}
\affiliation{CSC -- IT Center for Science Ltd., P.O. Box 405, FI-02101 Espoo, Finland}
\author{Xi Chen}
\affiliation{School of Physical Science and Technology, Lanzhou University, Lanzhou, Gansu 730000, China}
\author{Yorick Leonard A. Schmerwitz}
\affiliation{Science Institute and Faculty of Physical Sciences, University of Iceland, VR-III, 107 Reykjavík, Iceland}
\author{Jakob Schiøtz}
\affiliation{CAMd, Department of Physics, Technical University of Denmark, 2800 Kgs. Lyngby, Denmark}
\author{Thomas Olsen}
\affiliation{CAMd, Department of Physics, Technical University of Denmark, 2800 Kgs. Lyngby, Denmark}
\author{Karsten Wedel Jacobsen}
\affiliation{CAMd, Department of Physics, Technical University of Denmark, 2800 Kgs. Lyngby, Denmark}
\author{Kristian Sommer Thygesen}
\affiliation{CAMd, Department of Physics, Technical University of Denmark, 2800 Kgs. Lyngby, Denmark}

\date{\today}

\begin{abstract}
  \input{abstract}
\end{abstract}

\maketitle

\tableofcontents

\section{Introduction}
\input{intro/intro}
\section{Why GPAW?}
  \subsection{User's perspective}
  \input{overview-user/overview-user}

  \subsection{Developer's perspective}
  \input{overview-dev/overview-dev}
\section{Ground-state DFT}
  \label{sec:gs-dft}
  \subsection{Projector augmented-wave method}
  \input{paw/paw}
  \subsection{Numerical implementation}
    \subsubsection{Wave-function representations}
    \input{wavefunctions/wavefunctions}
    \subsubsection{Solving the Kohn--Sham equation}
    \input{ks-equation/ks-equation}
    \subsubsection{Updating wavefunctions in dynamics}
    \input{reuse-wfs/reuse-wfs}
    \subsubsection{Direct minimization}
    \input{directmin/directmin}
    \subsubsection{Convergence criteria}
    \input{conv-crit/conv-crit}
    \subsubsection{PAW data sets and pseudopotentials}
    \input{potentials/potentials}
    \subsubsection{Parallelization}
    \input{parallelization/parallelization}
    \subsubsection{GPU implementation}
    \input{gpu/gpu}
  \subsection{XC-functionals}
  \input{xc/xc}
    \subsubsection{Libxc and Libvdwxc}
    \input{libxc/libxc}
    \subsubsection{GLLB-sc}
    \input{xc/gllbsc}
    \subsubsection{Hubbard U}
    \input{hubbard-u/hubbard-u}
    \subsubsection{Hybrids}
    \input{hybrids/hybrids}
    \subsubsection{SIC}
    \input{sic/sic}
    \subsubsection{BEEF}
    \input{beef/beef}
\section{Ion dynamics}
\input{ion-dynamics/ion-dynamics}

\section{Magnetism and spin}
\label{sec:mag}
Many important technological applications utilize magnetic order or
manipulation of spin in materials.  GPAW has a wide range of
functionalities that faciliate the analysis of magnetic
properties. This includes calculations with non-collinear spin,
inclusion of external magnetic fields, spin--orbit coupling and spin
spiral calculations within the generalized Bloch theorem. The
implementaton of these features is described below while additional
methods for calculating magnetic excitations are described in section
\ref{sec:mag_response}.

  \subsection{Spin--orbit coupling}
  \input{spin-orbit/spin-orbit}

\input{mag-anisotropy/mag-anisotropy}
  \subsection{Self-consistent non-collinear magnetism}
  \input{non-col-scf/non-col-scf}
  \subsection{Orbital magnetization}
  \input{mag-orbmag/orbmag}
  \subsection{Constant B-field}
  \input{b-field/b-field}
  \subsection{Spin spirals}
  \input{spin-spirals/spin-spirals}
\section{Response functions and excitations}
Linear response functions are the bread and butter of condensed matter physics. Their
applications include the description of dielectric screening, optical and electron energy
loss spectra, many-body excitations, and ground state correlation energies. In this
section we describe the methods available in GPAW for calculating electronic response
functions as well as GW quasiparticle band structures and optical excitations from the
Bethe-Salpeter Equation (BSE). In addition to electronic response function, GPAW can also
calculate the transverse magnetic susceptibility, which holds information about the
magnetic excitations, e.g. magnons, and can be used to derive parameters for classical
Heisenberg spin models and to estimate magnetic transition temperatures. Finally, we
present methods to calculate Raman spectra of solids and quadratic optical response
tensors for describing second harmonics generation and the Pockels electro-optical effect.
  \subsection{Linear-response TDDFT}
  \input{linear-response-TDDFT/linear-response-TDDFT}
  \subsection{Dielectric function}
  \input{linear-response-TDDFT/dielectric-function}
  \subsection{Adiabatic-connection fluctuation-dissipation theorem}
  \input{acfdt/acfdt}
  \subsection{Magnetic response}
  \input{mag-excitations/mag-excitations}
    \subsubsection{Liechtenstein MFT}
    \input{mag-force-theorem/mag-force-theorem}
  \subsection{GW approximation}
  \input{gw/gw}
  \subsection{Bethe--Salpeter Equation (BSE)}
  \input{bse/bse}
  \subsection{Electron--phonon coupling}
  \input{electron-phonon/elph}
  \subsection{Raman spectrum}
  \input{raman/raman}
  \subsection{Quadratic optical response functions}
  \input{nlopt/nlopt}
\section{Real-time TDDFT}
  \input{rt-tddft/rt-tddft}
  \subsection{Circular dichroism for molecules}
  \input{rt-tddft/CD}
  \subsection{Radiation reaction potential}
  \input{rt-tddft/tddft-rr}
  \subsection{Ehrenfest dynamics}
  \input{rt-tddft/ehrenfest}
\section{Excited-state DFT methods}
  \subsection{Improved virtual orbitals}
  \input{ivos/ivos}
  \subsection{Variational excited-state calculations}
  \input{variational-excitedstate-calc/variational-excitedstate-calc}
    \subsubsection{Direct orbital optimization}
    \input{do-mom-gmf/do-mom-gmf}
    \subsubsection{Example applications of direct optimization}
    \input{do-mom-gmf/do-mom-gmf-applications}
\section{Other features}
  \label{sec:other}
  \subsection{Electric polarization}
    \input{berry/static-polarization}
  \subsection{Berry phases and band topology}
    \input{berry/topology}
  \subsection{Wannier functions}
    \input{wannier/wannier}
  \subsection{Point defect calculations with hybrid functionals}
  \input{point_defects/point_defects}
  \subsection{Point-group symmetry representations}
  \input{pointgroup-reps/pointgroup-reps}
  \subsection{Band-structure unfolding}
    \input{unfolding/unfolding}
  \subsection{The QEH model}
  \input{qeh/qeh}
  \subsection{Solvent models}
  \input{pcm/pcm}
  \subsection{Charged electrochemical interfaces}
  \input{elchem-charging/elchem-charging}
  \subsection{Constrained DFT}
    \input{cdft/cdft}
  \subsection{Orbital-free DFT}
  \input{ofdft/ofdft}
  \subsection{Zero-field splitting}
    \input{zfs/zfs}
  \subsection{Hyperfine coupling}
    \input{hyperfine/hyperfine}
\section{Outlook}
\input{outlook/outlook}
\section{Acknowledgements}
\input{acknowledgements}
\bibliographystyle{unsrt}
\bibliography{cleanedreferences.bib}

\end{document}

%% file: abstract.tex
We review the GPAW open-source Python package for electronic structure
calculations. GPAW is based on the projector-augmented wave method and
can solve the self-consistent density functional theory (DFT) equations
using three different wave-function representations, namely real-space
grids, plane waves, and numerical atomic orbitals. The three
representations are complementary and mutually independent and can be
connected by transformations via the real-space grid. This multi-basis
feature renders GPAW highly versatile and unique among similar codes. By
virtue of its modular structure, the GPAW code constitutes an ideal
platform for implementation of new features and methodologies. Moreover,
it is well integrated with the Atomic Simulation Environment (ASE)
providing a flexible and dynamic user interface. In addition to ground-state 
DFT calculations, GPAW supports many-body GW band structures,
optical excitations from the Bethe--Salpeter Equation (BSE), variational
calculations of excited states in molecules and solids via direct optimization,
and real-time propagation of the Kohn--Sham equations within time-dependent DFT.
A range of more advanced methods to describe magnetic excitations and
non-collinear magnetism in solids are also now available. In addition, GPAW
can calculate non-linear optical tensors of solids, charged crystal
point defects, and much more. Recently, support of GPU acceleration
has been achieved with minor modifications of the GPAW code thanks to
the CuPy library. We end the review with an outlook describing some future
plans for GPAW.

%% file: intro/intro.tex
The electronic-structure (ES) problem, i.e.~the solution to the
time-independent Schr\"{o}dinger equation for a collection of electrons and atomic
nuclei, forms the starting point for the quantum-mechanical treatment of
matter. Indeed, all chemical and physical properties of any substance
(solid, molecule, surface, etc.) can in principle be obtained from the
energies and wave functions that constitute the solution. A
pioneering step towards solving the many-body ES problem was the
formulation and formal proof of density functional theory (DFT) by
Hohenberg and Kohn in 1964~\cite{hohenberg1964inhomogeneous}, and a
practical scheme for its solution by Kohn and Sham in
1965~\cite{kohn1965self}. Today, most codes solving the ES problem from
first principles are based on DFT. Such codes are extremely powerful and
allow one to determine the atomic structure of solids and molecules
containing hundreds of atoms with a relative error below
1\%~\cite{perdew1996generalized,becke1993density,heyd2003hybrid}. Once the
atomic structure of the compound has been solved, its properties
(electronic, magnetic, optical, topological, etc.) can in principle be
determined. The evaluation of properties often involves theories beyond
the formal DFT framework to account for effects such as temperature and
lattice vibrations~\cite{torrie1977nonphysical,souvatzis2008entropy},
many-body interactions in excited states~\cite{hybertsen1986electron,golze2019gw}, or time
dependence~\cite{casida1998molecular,onida2002electronic}. As such,
first-principles atomistic calculations often involve two successive phases:
the solution of the ground-state ES problem (including ion dynamics) and the subsequent evaluation of
physical properties. This review is structured accordingly as Secs. \ref{sec:gs-dft}-\ref{sec:ion_dynamics} deal with the first phase while Secs. \ref{sec:mag}-\ref{sec:other} are devoted to the second.

In recent years, the scientific significance of ES codes has shifted from
a useful tool to describe and understand matter at the atomic scale to
an independent driver of the discovery and development of new
materials~\cite{norskov2009towards,
jain2013commentary,curtarolo2013high,haastrup2018computational,
castelli2012computational}.
This change in scope has been fueled by the exponential increase in
computer power accompanied by improved numerical
algorithms~\cite{marek2014elpa,liu2016cubic} as well as the use of workflow
management software for high-throughput
computations~\cite{jain2015fireworks,pizzi2016aiida,myqueue,gjerding2021atomic}
and the adoption of machine-learning techniques to leverage the rapidly
growing data generated by ES
codes~\cite{schmidt2019recent,bartok2010gaussian,rupp2012fast}. In parallel
with these capacity-extending developments, continuous progress in the
fundamental description of exchange--correlation effects has advanced the
predictive power of ES calculations to a level where they rival
experiments in terms of accuracy for many important
properties~\cite{teale2022dft,ren2012random,sun2015strongly,olsen2013random,
tkatchenko2009accurate,dion2004van}.

The GPAW code was originally intended as a Python-based multigrid solver
of the basic DFT equations within the projector-augmented wave (PAW)
formalism~\cite{gpaw1}. The name GPAW accordingly was an
abbreviation for ``grid-based projector-augmented waves''. Today, other
choices than regular grids for representations of the wave functions
exist in GPAW, but the name has stuck. During the years 2005--2010, GPAW
evolved to a full-blown DFT package~\cite{gpaw2} supporting most of the
functionality expected from a modern ES code, in
addition to a few more specialised features including real-time
propagation of wave functions~\cite{WalHakLeh08} and an alternative
basis of numerical atomic orbitals (referred to as the LCAO
basis)~\cite{larsen2009localized} to supplement the real-space grid. In
2011, a plane-wave (PW) basis set was also implemented. Today, the
possibility to use three different types of basis sets and even combining
them within a single run remains a unique feature of GPAW,
rendering the code very versatile.

The implementation of the PW basis set laid the groundwork for GPAW's
linear-response module, which today supports the calculation of linear
response functions~\cite{yan2011linear}, total energies from the adiabatic
connection fluctuation-dissipation
theorem~\cite{olsen2013random,olsen2012extending}, the GW self-energy
method for quasiparticle band structures~\cite{huser2013quasiparticle}, the
Bethe--Salpeter Equation (BSE) for optical
excitations~\cite{yan2012optical}, and more. The code also supports a wide range
of features related to the calculation of magnetism and spin--orbit
effects. Examples include spin-spiral calculations using the generalized Bloch
theorem~\cite{Sodequist2023a}, external magnetic fields, orbital
magnetization, magnetic anisotropy~\cite{Torelli2018}, adiabatic magnon
dispersions from the magnetic force theorem~\cite{Durhuus2023}, and dynamic
magnetic response from TDDFT~\cite{Skovhus2021}. For solids, the \textbf{k}-space
Berry phases can be computed directly from the Bloch orbitals and may be
used to obtain the spontaneous polarization~\cite{Kruse2023}, Born
effective charges, piezoelectric response tensors~\cite{Gjerding2021} and
various indices characterising the band topology~\cite{Olsen2018}.

In addition, GPAW can compute the localisation matrices forming the basis
for the construction of Wannier functions with e.g. the Atomic Simulation
Environment (ASE)~\cite{larsen2017atomic} or Wannier90~\cite{mostofi2008wannier90}. 
Electrostatic corrections to the formation
energies of charged point defects in insulators are implemented as are
calculations of the hyperfine coupling and zero-field splitting for
localised electron spins. GPAW also offers the possibility to perform time-independent,
variational calculations of localised electronic excitations, in e.g. molecules or at
crystal point defects, using direct orbital optimisation strategies implemented for all
three types of basis sets~\cite{Schmerwitz2023, Ivanov2021, Levi2020jctc}.
This provides an efficient and robust alternative to traditional ``$\Delta$SCF'' approaches.
GPAW can also be used to describe ultrafast electron dynamics within time-dependent density
functional theory (TDDFT) with wave functions represented either on a real
space grid~\cite{WalHakLeh08} or in the LCAO basis~\cite{KuiSakRos15}. The latter can
provide a significant speed-up due to the relatively small size of the
basis~\cite{RosLehSak15,RosKuiPus17,makkonen2021real}. The LCAO representation also forms the
basis for calculation of electron--phonon couplings as well as non-linear
optical spectra such as Raman scattering~\cite{taghizadeh2020library} (which can
alternatively be obtained in the PW mode as a finite difference of the
dielectric tensor), second-harmonics generation~\cite{taghizadeh2021two},
and shift currents~\cite{sauer2023shift} using higher-order perturbation
theory.



%% file: overview-user/overview-user.tex
There are dozens of electronic-structure codes available for the
interested user. The codes differ in their license (in particular,
whether open or proprietary), the underlying programming language
(e.g. Fortran, C, Python), their treatment of core electrons
(all-electron versus pseudopotentials), the employed representations
of the wave functions (plane waves, atom-centered orbitals, 
real-space grids), and the beyond-DFT features they support. Why 
should one choose GPAW?

In this section, we describe some of the features that make GPAW
interesting from the point of view of a common user who wants to
perform electronic-structure calculations. The next section
focuses on its possibilities for more advanced users, who perhaps
want to modify the code or implement completely new functionalities.

A first point to note is that GPAW is written almost exclusively in
Python and is directly integrated with the Atomic Simulation
Environment. This integration with ASE makes the setup, control,
and analysis of calculations easy and flexible. The programming
language is of course a key issue for developers, but also the
common user benefits from Python and the ASE/GPAW integration. A
typical stand-alone program only offers a fixed (though of course 
possibly large) set of tasks that it can perform, while Python scripting
allows for a more flexible use of the code. This could for example mean
combining several different GPAW calculations in new ways. Another
advantage is that ``inner parts'' of the code like the density or the
Kohn--Sham eigenvalues are directly accessible in a structured format
within Python for further analysis. It is even possible to ``open up the
main loop'' of GPAW and have access to, inspect, and also modify key
quantities during program execution (see Fig.~\ref{fig:code}).

As already mentioned in the introduction, GPAW distinguishes itself
from other available ES codes by supporting three different ways of
representing the wave functions. The most commonly used basis set is
plane waves (PW), which is appropriate for small or medium-size
systems, where high precision is required. Convergence is easily
and systematically controlled by tuning the cut-off energy.  A large
number of advanced features and ``beyond-DFT'' methods are available in
the PW mode. These include the calculation of hybrid functionals,
RPA total energies, linear-response TDDFT, and
many-body perturbation theory techniques like GW and the
Bethe--Salpeter equations. The new GPU implementation also uses
the PW mode.

The wave functions can alternatively be represented on real-space grids, which
was the original approach in GPAW. The implementation of this so-called
finite-difference (FD) mode relies on multi-grid solutions of the
Poisson and Kohn--Sham equations. The FD mode allows for more flexible
boundary conditions than the PW mode, which is restricted to periodic
supercells. The boundary conditions may for example be taken to
reflect the charge distribution in the unit cell. Calculations in the
FD mode can be systematically converged through lowering of the grid
spacing, but the approach to full convergence is slower than in
the PW mode. The FD mode is particularly well suited for large systems because the
wave-function representation allows for large-scale parallelization
through real-space decomposition. Furthermore, it is possible to
perform time-propagation TDDFT including Ehrenfest dynamics in this mode.

The third representation of the wave functions is a basis of numerical
atom-centered orbitals in the linear combination of atomic orbitals (LCAO)
mode. The size of the basis set can be varied through inclusion of
more angular momentum channels, additional orbitals within a channel,
or polarization functions. GPAW comes with a standard set of orbitals,
but a basis-set generator is included with the code so that
users may construct different basis sets depending on their needs and
requirements. The LCAO mode is generally less accurate than the PW and
FD modes, but it allows for the treatment of considerably larger
systems -- more than ten thousand atoms. It is also possible to study 
electron dynamics through a fast implementation of time-propagation DFT,
and Ehrenfest dynamics is under development.

As explained, the different modes have different virtues and
limitations, and it can therefore be an advantage to apply several
modes in a project. For larger systems, it is for example possible to
divide a structure optimization into two steps. First, an optimization
is performed with the fast LCAO basis leading to an approximately
correct structure. This is then followed by an optimization in either
the PW or FD mode, which now requires much fewer steps because of the
good initial configuration. Due to the ASE/Python interface this
combined calculation can easily be performed within a single script.

Since GPAW was originally created with the FD mode only, and the LCAO mode was
added next, some features have been implemented for only those modes. Examples
are real-time TDDFT (see section~\ref{sec:rttddft}) and electron--phonon coupling
(see section~\ref{sec:elph}). Conversely, some new features
only work for the PW mode, which was added after the real-space modes.  Examples are
RPA total energies (see section~\ref{sec:acfdt}) and calculation of the stress
tensor. To summarize, given the limitations just mentioned, users should
most of the time use PW or LCAO mode and the choice will depend on the
accuracy needed and the resources available.

%% file: overview-dev/overview-dev.tex
The GPAW source code is written in the Python and C languages and is hosted
on GitLab~\cite{gitlab} licensed under the GNU General Public
License v3.0. This ensures the transparency of all features and allows developers
to fully customise their experience and contribute new features to the community.

\begin{figure}
\fbox{\includegraphics[width=.65\linewidth]{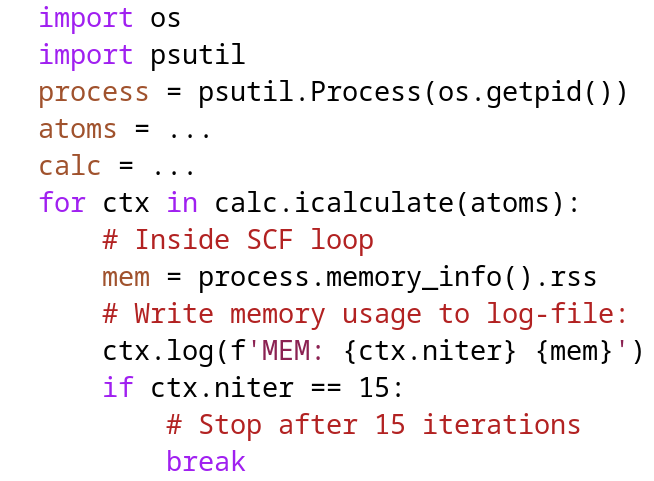}\hspace{0.325\linewidth}}
\caption{
The variable \texttt{calc} is the ground-state DFT calculator object and its
\texttt{icalculate} method yields a context object at every self-consistent field (SCF) step.
As seen, one can use this in a for-loop to
implement special logic for termination of the SCF iterations or for
diagnostics. In this example, the memory usage is written to the log-file
for the first 15 SCF iterations.
\label{fig:code}
}
\end{figure}

An advantage of having a Python code is that the Python script
you write to carry out your calculations will have access to everything inside a
GPAW calculation. An example showing the power and flexibility this affords is the
possibility to have user-code inserted inside the
self-consistent field (SCF) loop as demonstrated in Fig.~\ref{fig:code}.

At the time of this writing (July 2023), GPAW has two versions of the
ground-state DFT code in the main branch of the code.  There is the older
version that has grown organically since the birth of GPAW:  it has many
features, but also a lot of technical debt that makes it harder to
maintain and less ideal to build new features on top of.  The newer ground-state
code addresses these issues by having a better overall design.

The new design greatly improves the ease of implementation of
new features. The goal is to have the new code be feature-complete so that
it can pass the complete test suite and then delete
the old code once that is achieved.
At the moment, we recommend that all production calculations are done with the old code and that work on new features is
done on top of the new code even though certain features are not yet production-ready.
Three new features, not present in the old code base, have already been implemented based on the
new code: GPU implementation of PW mode calculations (see section~\ref{sec:gpu}), reuse of the wave functions after unit-cell changes during
cell optimization, and spin-spiral calculations (see section~\ref{sec:spirals}).

GPAW uses pytest~\cite{pytest} for
its test suite that currently consists of approximately
1600 unit and integration tests (see Table \ref{loc}).  A subset of
those tests runs as part of GitLab's continuous integration (CI) thereby
checking the correctness of every code change.  Unfortunately, the full test
suite is too time-consuming to run as part of CI, so we run that nightly
both in serial as well as in parallel using MPI.

Many of the code examples in GPAW's documentation, exercises and tutorials~\cite{tutorials} require resources (time and
number of CPUs) beyond what would make sense to run as part of the pytest
test suite. For that, we use MyQueue~\cite{myqueue} to submit those scripts as
jobs to a local supercomputer every weekend. At the moment this amounts to
approximately 5200 core-hours of calculations.

As can be seen from Table~\ref{loc}, the majority of the code is written
in Python, which is an interpreted language that is easy to read, write
and debug.

\begin{table}
  \caption{Number of files and number of lines of code in the
git repository of GPAW.  The Python source-code files are split into three parts:
the actual code, the test suite, and code examples in the documentation.
}
\begin{ruledtabular}
\begin{tabular}{lrr}
Type of file           & Files &  Lines \\
\hline
Python (the code)      &   513 & 146,604 \\
C                      &    80 &  19,719 \\
Python (test suite)    &   681 &  47,147 \\
Python (documentation) &   744 &  32,014 \\
\end{tabular}
\label{loc}
\end{ruledtabular}
\end{table}

Interpreter-executed code will not run as
efficiently as code that is compiled to native machine code. It is therefore
important to make sure that the places in the code where most of the time is
spent (hot spots) are in native machine code and not in the interpreter.  GPAW achieves
this by implementing the hot spots in C-code with Python wrappers that can be called
from the Python code.
Examples of such computationally intensive tasks are applying a
finite-difference stencil to a uniform grid, interpolating from one uniform
grid to another, or calculation of overlaps between projector functions and
wave functions. In addition, we have Python interfaces to the numerical
libraries FFTW~\cite{fftw}, ScaLAPACK~\cite{scalapack}, ELPA~\cite{marek2014elpa}, BLAS, Libxc~\cite{libxc,lehtola_recent_2018},
libvdwxc~\cite{libvdwxc}, and MPI. Finally, GPAW
makes heavy use of the NumPy~\cite{numpy} and
Scipy~\cite{2020SciPy-NMeth} Python
packages. NumPy provides us with the \texttt{numpy.ndarray} data type which
is an $N$-dimensional array that we use for storing wave functions, electron
densities, potentials, matrices like the overlap matrix or the LCAO wave
function coefficients, and much more.  The use of NumPy arrays allows us to use
the many sub-modules of SciPy to manipulate data.  This also gives us
an efficient memory layout allowing us to simply pass a pointer to
the memory whenever we need to call the C-code from the Python code. With
this strategy, we can get away with having most of the code written in a
relatively slow interpreted language and still have most of the time spent in
highly efficient C-code or optimized numerical libraries.

The advantage of the original FD-mode, where
there are no Fourier transforms of the wave functions, is that the
algorithms should parallelize well for large systems. In practice, it has
turned out that FD-mode has a number of disadvantages: 1) Due to integrals over the
unit cell being done as sums over grid-points, there will be a small periodic
energy variation as you translate atoms and the period of the variation will
be equal to the grid-spacing used (the so-called egg-box error); 2) The
system sizes that are typically most interesting for applications of DFT are
too small for the parallel scalability to be the decisive advantage; 3) The memory
used to store the wave functions on uniform grids in real-space is
significant. In contrast, the PW mode has practically no egg-box error, is very
efficient for the most typical system sizes and often uses a factor of 10 less
memory compared to an FD mode calculation of similar accuracy. The main
advantages for LCAO mode are low memory usage and high efficiency for large
systems; for small unit cells with many $\mathbf{k}$-points the PW mode is most
efficient.
{\color{black}
One disadvantage of LCAO-mode is egg-box errors: LCAO-mode uses the same
uniform grids as used in FD-mode for integration of matrix elements like
$\langle \Phi_\mu | \tilde v | \Phi_\nu \rangle$ and therefore has similar
egg-box energy variation.  A second disadvantage of LCAO-mode is that
as for any localized basis-set,
reaching the complete basis-set limit is more involved compared to PW and FD
modes. This can have severe consequences even for ground state calculations of difficult systems such as
Cr$_2$ for example\cite{wurdemann_density_2015}.
In PW or FD modes the complete basis set limit is easy to reach by simply
increasing the number of plane-waves or grid-points, respectively,
which leads to a smooth convergence\cite{gpaw1}.
At the
moment we only provide double-$\zeta$ polarized (DZP) basis sets and
going beyond DZP is left for users to do themselves.
}

%% file: paw/paw.tex
The diverging Coulomb potential causes rapid oscillations in electronic
wave functions near the nuclei, and special care is required to be able
to work with smooth wave functions.
The projector augmented-wave (PAW) method by Bl\"ochl~\cite{Blochl1994} is a widely
adopted generalization of pseudopotential methods, utilizing their strength
of smooth pseudo wave functions while retaining a mapping from all-electron
wave functions ($|\psi_{n} \rangle$) to pseudo wave functions
($|\tilde \psi_{n} \rangle$).

The crux is to define a linear transformation $\hat T$
from pseudo to all-electron space,
\begin{equation}
    \hat T |\tilde \psi \rangle = | \psi \rangle,
    \label{eq:pawtransform}
\end{equation}
where
\begin{equation}
    \hat T = 1 + \sum_a \sum_i \left(
    | \phi^a_i \rangle - |\tilde \phi^a_i \rangle
    \right) \langle \tilde p^a_{i} |.
\end{equation}
Here $\tilde p^a_i(\mathbf r)$, $\phi^a_i(\mathbf r)$ and $\tilde \phi^a_i(\mathbf r)$
are called projectors, partial waves and pseudo-partial waves, respectively.
The pseudo-partial waves and projectors are chosen to be biorthogonal,
$\int {\rm d}{\mathbf r} \,
\tilde \phi^a_i(\mathbf r) \tilde p^a_j(\mathbf r) = \delta_{ij}$,
allowing for an approximate closure relation
\begin{equation}
\sum_i \tilde \phi^a_i(\mathbf r) \tilde p^a_i(\mathbf r')  \approx \delta(\mathbf r-\mathbf r'),
\end{equation}
which is utilized heavily to obtain an efficient, but all-electron, picture.
In addition to biorthogonality, the pseudo and all-electron partial waves are chosen to be equal outside the PAW augmentation sphere
cutoff radius $\tilde \phi_{i}(\mathbf r) = \phi_i(\mathbf r)$, $r > r_c.$

The basic recipe for converting an operator $\hat O$ is
$\langle \psi_n| \hat O \psi_n \rangle =
 \langle \tilde \psi_n| \underbrace{\hat T^\dagger \hat O \hat T}_{\hat{\tilde O}} \tilde \psi_n \rangle$.
For example, the all-electron Kohn--Sham equations
\begin{equation}
   \hat H[n({\mathbf r})] |\psi_n \rangle
= \epsilon_{n}
| \psi_n \rangle,
\end{equation}
where $\hat H$ is the single-particle all-electron Kohn--Sham Hamiltonian operator,
can be transformed to their PAW counterparts:
\begin{equation}
\hat T^\dagger \hat H[n({\mathbf r})] \hat T | \tilde \psi_n \rangle
= \epsilon_{n}
\hat T^\dagger \hat T
| \tilde \psi_n \rangle.
\end{equation}
We have used Eq.~\ref{eq:pawtransform}, and also multiplied with
$\hat T^\dagger$ from left to make its dual space the pseudo one, i.e.
$\langle \tilde \psi| \in \mathcal{H}^*$ can act from left.
This results in a PAW Hamiltonian and in PAW overlap operators as follows:
\begin{align}
    \hat{\tilde H} &=
    \hat T^\dagger \hat H[n({\mathbf r})] \hat T \nonumber\\&=
    -\frac{1}{2}\nabla^2 + \tilde v_{\rm KS}(\mathbf r)
    + \sum_a \sum_{ii'} |\tilde p^a_i \rangle \Delta H^a_{ii'}
     \langle \tilde p^a_{i'} |,
    \label{dh}
    \\
    \hat{\tilde{S}}
    &=
    \hat T^\dagger \hat T =
    1
    + \sum_a \sum_{ii'} |\tilde p^a_i \rangle \Delta S^a_{ii'}
     \langle \tilde p^a_{i'} |.
\end{align}
Terms such as $\Delta H^a_{ii'}$ and $\Delta S^a_{ii'}$ represent so-called PAW-corrections.
In each part of the description, which handles a particular kind of operators, such as kinetic energy,
spin-operators, or the electrostatic potential, respective PAW-corrections must be calculated.
The most crucial ones are precalculated, such as overlap, kinetic energy, Coulomb,
and stored in the `setup'-file, which also stores the partial waves and projectors.
As an example, the overlap PAW corrections are precalculated to setup as follows:
\begin{equation}
    \Delta S^a_{ii} =
    \int {\rm d}{\mathbf r}\,
    \phi_i^a(\mathbf r)
    \phi_{i'}^a(\mathbf r)
  - \int {\rm d}{\mathbf r}
    \tilde \phi_i^a(\mathbf r)
    \tilde \phi_{i'}^a(\mathbf r).
    \label{eq:ds}
\end{equation}

We further define the atomic density matrices as
\begin{equation}
D^a_{\sigma ii'} = \sum_{n} f_{n}
    \langle
    \tilde \psi_{\sigma n} |
    \tilde p^{a}_i
    \rangle
    \langle \tilde p^{a}_{i'} | \tilde \psi_{\sigma n} \rangle .
    \label{eq:atomicdensitymatrix}
\end{equation}
The atomic density matrix contains all information required to construct PAW corrections to any local all-electron expectation value:
\begin{equation}
    \langle O \rangle = \sum_n \langle \tilde{\psi}_n | \tilde O | \tilde{\psi}_n \rangle
    + \sum_a \sum_{ii'} D^a_{ii'} \Delta O^a_{i'i}.
\end{equation}
The all-electron atomic density can be constructed as
\begin{equation}
    n^{a}_\sigma(\mathbf r) = \sum_{ii'} D^a_{\sigma ii'}
    \phi^{a*}_i(\mathbf r)
    \phi^a_{i'}(\mathbf r) +
    n_{\text{core}}^a,
    \label{allelectrondensity}
\end{equation}
and the corresponding equation for pseudo-densities holds with $n \rightarrow \tilde{n}$ and
$\phi \rightarrow \tilde{\phi}$.

Since the exchange--correlation (xc) potential is non-linear, the PAW corrections must be evaluated explicitly.
The xc PAW corrections are performed by constructing the atomic all-electron and pseudo-electron densities as given by Eq.~\ref{allelectrondensity}:
\begin{eqnarray}
    \Delta H^a_{{\rm xc},ii'} &=&
    \int {\rm d}{\mathbf r}
    \phi_i^a(\mathbf r)
    v_{\rm xc}[n^a](\mathbf r)
    \phi_{i'}^a(\mathbf r) \nonumber\\
  &&- \int {\rm d}{\mathbf r}
    \tilde \phi_i^a(\mathbf r)
    v_{\rm xc}[\tilde n^a](\mathbf r)
    \tilde \phi_{i'}^a(\mathbf r).
\end{eqnarray}
This integral is numerically evaluated in a Cartesian product of a
Lebedev angular grid and a non-uniform radial grid (denser mesh closer to
the nucleus) for each atom.

%% file: wavefunctions/wavefunctions.tex
\label{sec:wf-rep}
GPAW supports three representations for smooth wave functions.
The plane wave (PW)
\begin{equation}
	\tilde \psi_{{\mathbf k}n}(r) =
	\sum_{\mathbf G} C_{n{\mathbf G}} e^{i ({\mathbf k + \mathbf G}) \cdot \mathbf r}
\end{equation}
and linear combination of atomic orbitals (LCAO)
\begin{equation}
	\tilde \psi_{{\mathbf k}n}(r) =
	\sum_{\mu} C_{n{\mathbf k}\mu} \Phi_{\mathbf k \mu}(\mathbf r)
\end{equation}
representations rely on basis functions. The finite difference (FD) mode
relies on a representation of kinetic energy operator on a
uniform Cartesian grid.

\subsubsection{All-electron quantities}

The beauty of the PAW method is that you never \emph{need} to transform the pseudo
wave functions to all-electron wave functions, but you \emph{can} do it if you want
to. GPAW has tools for interpolating the pseudo wave functions to a fine
real-space grid and adding the PAW corrections. A fine real-space grid is
needed to properly represent the cusp and all the oscillations necessary for the
wave function to be orthogonal to all the frozen core states.

GPAW also has tools for calculating the all-electron electrostatic potential.
This is useful for transmission electron microscopy (TEM)
simulations~\cite{susi_efficient_2019}. Most TEM simulations have relied on the
so-called independent atom model (IAM), where the specimen potential is
described as a superposition of isolated atomic potentials. While this is
often sufficient, there is increasing interest in understanding the influence
of valence bonding~\cite{madsen_ab_2021}. This can be investigated by a TEM
simulation code such as \emph{ab}TEM~\cite{madsen_abtem_2021}, which can
directly use \emph{ab initio} scattering potentials from GPAW.

%% file: ks-equation/ks-equation.tex
The default method for solving the Kohn--Sham equation for PW and FD modes is
to do iterative diagonalization combined with density mixing; for LCAO mode
we do a full diagonalization of the Hamiltonian. Alternatively, one can do
direct minimization as described in the next section.

For PW and FD modes, we need an initial guess of the wave functions. For this,
we calculate the effective potential from a superposition of atomic densities
and diagnonalize an LCAO Hamiltonian in a small basis set consisting of all
the pseudo partial waves corresponding to bound atomic states.

Each step in the self-consistent field (SCF) loop consists of the following operations:
1) Diagonalization of the Hamiltonian in the subspace of the current wave
functions (skipped for LCAO).
2) One or more steps through the iterative
eigensolver (except for LCAO, where a full
diagonalization is performed).
3) Update of eigenvalues and occupation numbers.
4) Density mixing and symmetrization. See previous work~\cite{gpaw1,gpaw2} and
Ref.~\cite{Kresse1996} for details.

GPAW has two kinds of Poisson equation solvers: direct solvers based on
Fourier transforms or Fourier-sine transforms, and iterative multi-grid
solvers (Jacobi or Gauss-Seidel). The default is to use a direct solver
whereas the iterative solvers may be chosen for larger systems where they can
be more efficient.

For 0-, 1- and 2-dimensional systems, the default boundary conditions are to
have the potential go to zero on the cell boundaries. This becomes a
problem for systems involving large dipole moments. The potential due to the
dipole is long-ranged and, thus, the converged potential requires large
vacuum sizes. For molecules (0D systems), the boundary
conditions can be improved by adding multipole moment corrections to the
density so that the corresponding multipoles of the density vanish. The
potential of these corrections is added to the obtained potential. The
same trick is used to handle charged systems. For slabs (2D systems), a
dipole layer can be added to account for differences in the work functions on
the two sides of the slab.

Methods for calculating occupation numbers are the Fermi-Dirac,
Marzari-Vanderbilt~\cite{cold-smearing} and Methfessel-Paxton distributions as
well as the tetrahedron method and the improved tetrahedron
method~\cite{tetra}.

%% file: reuse-wfs/reuse-wfs.tex
Simulations commonly move the atoms without changing other parameters.
If an atom moves only slightly, we would expect most of the charge in its
immediate vicinity to move along with it. We use this to compute an improved
guess for the wave functions in the next self-consistency loop with FD or PW
mode where the eigensolver is iterative.

Near the atoms, the dual basis of pseudo-partial waves and projectors
is almost complete, i.e.
\begin{equation}
  \sum_{i} |\phi_i^a \rangle \langle \tilde p_i^a| = 1\quad \textrm{(near atom $a$)}.
\end{equation}
If an atom moves by $\Delta \mathbf R^a$,
the wave functions $\tilde \psi_n(\mathbf r)$ are updated by rigidly
moving the projection $\sum_{i} \tilde \phi_i^a(\mathbf r) \langle \tilde p_i^a | \tilde \psi_n \rangle$ along with it, i.e.,
\begin{eqnarray}
  \tilde \psi_n^{\mathrm{new}}(\mathbf r) =
  \tilde \psi_n(\mathbf r)
  &+& \sum_{ai} \tilde \phi_i^a(\mathbf r - \Delta \mathbf R^a) \langle \tilde p_i^a | \tilde \psi_n \rangle \nonumber\\
    &-& \sum_{ai} \tilde \phi_i^a(\mathbf r) \langle \tilde p_i^a | \tilde \psi_n \rangle.
\end{eqnarray}
As the partial waves on different atoms are not orthonormal, this
expression generally ``double-counts'' contributions, resulting in
wave functions that are to some extent unphysical. Nevertheless, we
have found that this simple method achieves a significant speedup ($\sim$15\% in realistic structure optimisations) compared to not updating the wave functions.

The method could be further improved by using the LCAO
basis set and the overlap matrix to prevent double-counting.

%% file: directmin/directmin.tex
\label{sec:directmin}
Direct orbital minimization~\cite{Lehtola2020, Voorhis2002, Payne1992, Head-Gordon1988} is a robust alternative to the conventional eigensolver and density mixing routines.
The orbitals can be expressed as a unitary transformation of a set of reference, or auxiliary, orbitals $\mathbf{\Psi}_0$:
\begin{equation}
 \mathbf{\Psi} = \mathbf{U}\mathbf{\Psi}_0
\end{equation}
In the direct minimization (DM) method implemented in GPAW~\cite{ivanov_directmin_2021, Ivanov2021}, the unitary matrix $\mathbf{U}$ is parametrized as an exponential transformation, i.e. $\mathbf{U} = e^{\mathbf{A}}$, where ${\mathbf{A}}$ is an anti-Hermitian matrix (${\mathbf{A}} = -{\mathbf{A}^\dagger}$). The energy can be considered as a functional of both $\mathbf{A}$ and $\mathbf{\Psi}_0$:
\begin{equation}
 E[\mathbf{\Psi}] = \mathcal{F}[\mathbf{A},\mathbf{\Psi}_0]
\end{equation}
Thus, in general, the optimal orbitals corresponding to the minimum of the energy functional can be found in a dual loop procedure. First, the energy is minimized with respect to the elements of $\mathbf{A}$, and second, the functional $\mathcal{L}[\mathbf{\Psi}_0] = \underset{\mathbf{A}} {\rm min\,}\mathcal{F}[\mathbf{A},\mathbf{\Psi}_0]$ is minimized with respect to $\mathbf{\Psi}_0$:
\begin{equation}
 \underset{\mathbf{\Psi}} {\rm min\,} E[\mathbf{\Psi}] = \underset{\mathbf{\Psi}_0} {\rm min\,}
 \underset{\mathbf{A}} {\rm min\,} \mathcal{F}[{\mathbf{A}},\mathbf{\Psi}_0]
\end{equation}
Since anti-Hermitian matrices form a linear space, the inner loop minimization can use well-established local minimization strategies such as efficient quasi-Newton methods with inexact line search, e.g. the limited-memory Broyden--Fletcher--Goldfarb--Shanno (L-BFGS) algorithm. The outer loop minimization follows the gradient $\partial \mathcal{L}/\partial \mathbf{\Psi}_0$ projected on the tangent space at $\mathbf{\Psi}_0$.

The GPAW formulation of DM is applicable with all representations of the
orbitals available in GPAW, as well as Kohn--Sham (unitary invariant) and
orbital density dependent (nonunitary invariant) energy functionals, and
can be used for both finite and extended systems. In LCAO calculations,
the reference orbitals are expressed as a linear combination of the
atomic basis functions $\mathbf{\Phi}$, $\mathbf{\Psi}_0=\mathbf{\Phi}\mathbf{C}_0$,
where the matrix of coefficients $\mathbf{C}_0$ is fixed.
Therefore, only a
minimization with respect to the elements of the matrix $\mathbf{A}$
is required~\cite{ivanov_directmin_2021}. For plane wave and real-space
grid representations, a minimization in the space tangent to the
reference orbitals is sufficient if the functional is unitary invariant.
Otherwise, if the functional is nonunitary invariant, such as when
self-interaction correction is used (see section~\ref{sec:sic}), an
inner loop minimization in the occupied--occupied block of the matrix
$\mathbf{A}$ is performed to make the energy stationary with respect to
unitary transfomation of the occupied orbitals~\cite{Ivanov2021}.

The DM method avoids diagonalization of the Hamiltonian at each step and
as a result it usually involves a smaller computational effort. The DM
method has also been shown to be more robust than conventional
eigensolvers and density mixing in calculations of molecules and
extended systems~\cite{ivanov_directmin_2021}. However, the current
implementation does not support finite-temperature distribution of
occupation numbers and thus can only be used for systems with a finite
band gap.

%% file: conv-crit/conv-crit.tex



The modular architecture of GPAW allows the user to have precise control over how the SCF loop decides that the electronic structure has converged to sufficient precision.
GPAW contains simple keywords for common convergence criteria, such as ``energy'', ``forces'', (electron) ``density'', and ``work function'', which are sufficient for the most common use cases.

Internally, all convergence criteria are instances of convergence classes, and for each step through the SCF loop, each convergence class is called.
When any convergence class is called, it is passed a context that contains the current state of the calculation, such as the wavefunctions and the Hamiltonian.
The criterion can thus pull relevant data from the calculation to decide if it is converged.
Because the convergence criterion itself is an object, it can store information, such as previous values of the energy for comparison to the new value.
When all convergence criteria report that they are converged, the calculation as a whole is considered to be converged and terminates.

This modular nature gives the user full control over how each convergence criterion operates.
For example, the user can easily ask the energy criterion to check the differences in the last four values of the energy, rather than the last three.
If a convergence criterion itself is expensive to calculate, it can make sense to not check it until the rest of the convergence criteria are met.
This can be accomplished by activating an internal ``calculate\_last'' flag within the convergence criterion.

Users can easily add their own custom convergence criteria to the SCF loop.
If a user would like to use a criterion not included by default with GPAW, it is straightforward to write a new criterion as a Python class, and pass this to the convergence dictionary of GPAW.
For example, if one wanted to be sure the bandgap of a semiconductor was converged, the criterion could check the bandgap at each iteration and compare it to stored values from previous iterations, and report that the calculation is converged when the peak-to-peak variance among the last $n$ iterations is below a given threshold.

%% file: potentials/potentials.tex
GPAW can read PAW datasets from (possibly compressed) XML-files following
the PAW-XML specification~\cite{paw-xml}. Dataset files for most of the
periodic table can be downloaded from the GPAW webpage or installed with the
\texttt{gpaw install-data} command-line tool. The datasets are available
for the LDA, PBE, revPBE, RPBE and GLLBSC functionals. The electronic
structure code Abinit~\cite{abinit} also reads the PAW-XML format
allowing GPAW and Abinit to share PAW dataset collections
such as the Jollet-Torrent-Holzwarth collection~\cite{JTH}.

Specialized datasets can be generated with the \texttt{gpaw dataset}
command-line tool. This allows one to tweak the properties of a dataset.
Some examples could be: 1) add more/less semi-core states; 2)
increase/decrease the augmentation sphere radius to make the pseudo wave
functions more/less smooth; 3) add/remove projector functions and
corresponding pseudo and all-electron partial waves; or 4) base the PAW
dataset on a different XC-functional. These changes will affect the
accuracy and cost of the calculations.

GPAW is also able to use norm-conserving pseudopotentials (NCPP) such as
HGH~\cite{hgh} and pseudopotentials in the UPF format, such as
SG15~\cite{sg15}. Non-local NCPPs can be considered an approximation to
PAW: in the PAW description, the non-local part of the Hamiltonian (the
term containing $\Delta H_{ij}^a$ in Eq.~\eqref{dh}) will adapt to the
environment, whereas for a NCPP, $\Delta H_{ij}^a$ will be diagonal and have
a fixed value taken from a reference atom. Because of the
norm-conservation, NCPPs will have $\Delta S_{ij}^a=0$.

%% file: parallelization/parallelization.tex
GPAW can parallelize over various degrees of freedoms depending on the
type of calculation, and implements multiple algorithms for achieving
good parallel performance and scalability. In calculations involving \textbf{k}-points,
parallelization over them is typically the most efficient, as
little communication is needed during the summation of wave functions to
calculate the density and any derived quantities. As the number of \textbf{k}-points
is often limited especially in large systems, parallelization is also
possible over real-space grids in the FD and LCAO modes, as well as over plane waves
in the PW mode. All modes also support parallelization over electronic
bands, which is particularly efficient for real-time TDDFT where the
time-propagation of each band can be carried out independently.
Additional parallelization possibilities exist depending on calculations,
such as over electron--hole pairs in linear-respose TDDFT calculations.

Parallelization is done mainly with MPI. In the FD and LCAO modes, it is
possible in addition to use also OpenMP within shared-memory nodes,
which can improve performance when the number of CPU cores per
node is large. Dense linear algebra, such as matrix diagonalization and
Cholesky decomposition, can be carried out with the parallel ScaLAPACK or ELPA
libraries. This applies to both the direct diagonalization in the
LCAO mode as well as to subspace diagonalizations in iterative Davidson and
RMM-DIIS methods in the FD and PW modes.

For ground-state calculations, GPAW will divide the cores into three
MPI-communicators: \textbf{k}-points, bands and domain. When parallelizing over \textbf{k}-points
and/or bands, all the cores of the \textbf{k}-point and/or band communicators will have
a copy of the density (possibly distributed over the domain communicator).
GPAW has the option to redistribute the density from the domain-communicator
to all cores so that operations such as evaluating the XC-energy and solving
the Poisson equation can be done more efficiently.

For each \textbf{k}-point, all the cores in the band and domain communicators will
cooperate on calculating matrix elements of the Hamiltonian and the overlap
operators. Dense-matrix linear-algebra operations on those matrices can be
done on a single core (most efficient for small systems) or with ScaLAPACK where the
matrices are distributed over either all or some of the cores from the pool of cores in
the band and domain communicators.

One drawback of Python is that in large parallel calculations its import mechanism
may incur a heavy load on the filesystem because all the parallel processes are trying to
read the same Python files. GPAW tries to alleviate this by a
special ‘‘broadcast imports'' mechanism: during initial module imports only a
single process loads the modules, afterwards MPI is used to broadcast the data to all
the processes.

Parallel scalability depends strongly on the calculation mode and the
system. The FD mode offers the best scalability for high core counts as only
nearest-neighbour communication is needed across domains over a domain
decomposition.
In PW mode, the limiting factor is all-to-all communication in parallelization
over plane waves. In LCAO mode, communications arise from multi-center
integrals of basis functions across domains. At best, GPAW scales to tens or
hundreds of nodes in supercomputers.

%% file: gpu/gpu.tex
\label{sec:gpu}
The GPU implementation of GPAW works both on NVIDIA and AMD GPUs, targeting
either CUDA or HIP backends, respectively. GPAW
uses a combination of manually written GPU kernels, external GPU libraries
(such as cuBLAS / hipBLAS), and CuPy~\cite{cupy}. CuPy offers an easy-to-use
Python interface for GPUs centered around a NumPy-like GPU array and makes
many hardware details completely transparent to the end-user.

In the manually written GPU kernels, both GPU backends (CUDA and HIP) are
targeted using a header-only porting approach~\cite{hop} in which generic
GPU identifiers are translated to vendor-specific identifiers at compile
time. For example, to allocate GPU memory, the identifier {\tt gpuMalloc}
is used in the code, that is then translated either to {\tt cudaMalloc} or
{\tt hipMalloc} depending on which GPU backend is targeted. This allows us
to avoid unnecessary code duplication and still target multiple hardware
platforms natively.

An earlier GPU implementation of GPAW~\cite{Hakala2013, Hakala2015} served as the
starting point for the recent work on a new GPU code based on the rewritten
ground-state code. The objects that store quantities like
$\tilde\psi_n(\mathbf{r})$,
$\tilde p_i^a(\mathbf{r})$,
$\langle\tilde p_i^a | \tilde\psi_n\rangle$,
$D_{ii'}^a$,
$\tilde n(\mathbf{r})$ and
$\tilde v(\mathbf{r})$, use Numpy arrays for the CPU code and CuPy arrays when running on a GPU. At the moment, GPUs can be used for total-energy calculations with LDA/GGA in the PW mode.

Parallelization to multiple GPUs is done using MPI. Each MPI rank is assigned
a single GPU and communication between the GPUs is handled by MPI. Support for
GPU-aware MPI makes it possible to do direct GPU-to-GPU communication without
unnecessary memory copies between the GPU device and the host CPU.

%% file: xc/xc.tex
Exchange--correlation (XC) functionals provide a mapping between the interacting and the non-interacting system of
electrons. In Kohn--Sham DFT, the density is built from a set of occupied non-interacting single-particle
orbitals $\psi_i$:  $n = \sum_i f_i \left | \psi_i \right |^2$, with $f_i$ denoting the occupation numbers, leading
to the same density as the interacting system. The total energy in DFT
is expressed as a sum of density
functionals for the different contributions:
\begin{eqnarray}
E_\mathrm{tot} [n] = T_\mathrm{S} [n] + V_\mathrm{ext} [n] + U_\mathrm{H} [n] + E_\mathrm{xc} [n],\label{eq:xc_intro_Etot}
\end{eqnarray}
where $T_\mathrm{S} [n]$ denotes the kinetic energy of the non-interacting system, $V_\mathrm{ext} [n]$
the energy of the density in the external potential, $U_\mathrm{H} [n]$ the classical Coulomb energy of the
density with itself, and $E_\mathrm{xc} [n]$ the so called exchange--correlation energy, which collects all
energy contributions missing in the prior terms and therefore provides a mapping between the interacting
and the non-interacting system of electrons. While the first three terms can be calculated exactly, even the
form of $E_\text{xc}$ is unknown and although proven to exist and be exact in principle, it
has to be approximated in practice.  A huge number of approaches belonging to several families
exists~\cite{hohenberg1964inhomogeneous,kohn1965self,perdew_jacobs_2001}. Several of these
approximations are available in GPAW, and an overview is given in the following.

%% file: libxc/libxc.tex
The \emph{libxc} library~\cite{lehtola_recent_2018} provides implementations of several
(semi-)local variants of the XC functional, given by the LDA, GGA, and MGGA families.
These are available in GPAW by a combination of their names
from \emph{libxc} as, e.g. ``GGA\_X\_PBE+GGA\_C\_PBE'' for PBE~\cite{perdew1996generalized}.
Additionally, GPAW provides its own implementation of several
(semi-)local functionals called by their short-names, e.g.
TPSS~\cite{tao_climbing_2003},
PBE~\cite{perdew1996generalized} and
LDA, the latter with
the correlation of Perdew and Wang~\cite{perdew_accurate_1992}. Several hybrids, see below,
are implemented in GPAW with the support of the \emph{libxc} library for their local parts.

For fully non-local van der Waals functionals, like the vdW-DF
functional~\cite{dion2004van}, GPAW uses the efficient fast Fourier-transform convolution
algorithm by Roman-Perez and Soler~\cite{perez2009} as implemented in the \emph{libvdwxc}
library~\cite{libvdwxc}.

%% file: xc/gllbsc.tex
GPAW has an implementation of the solid-state modification of
the Gritsenko--van Leeuwen--van Lenthe--Baerends exchange--correlation model potential (GLLB-sc)~\cite{GriLeeLen95, KuiOjaEnk10}.
This potential has been shown to improve
the description of the band gap~\cite{castelli2012computational}
and the $d$-electron states in noble metals~\cite{KuiSakRos15, YanJacThy11, YanJacThy12, RahTibRos20}.

%% file: hubbard-u/hubbard-u.tex
DFT+U calculations using a Hubbard-like correction can be performed in GPAW to improve the
treatment of Coulombic interactions of localized electrons. This correction is most
commonly applied to the valence orbitals of transition metals to assist with obtaining
experimental band gaps of oxides~\cite{Bashyal_2018}, which may otherwise be
underestimated~\cite{PhysRevLett.105.196403,https://doi.org/10.1002/qua.560280846}.
Also, formation energies and magnetic states are often improved due to a more correct
description of the electronic structure.
It may
also be applied to main-group elements such as N, O, P, or S, but this is less commonly
done~\cite{PhysRevB.101.165117}. 

In the formalism chosen in GPAW~\cite{PhysRevB.57.1505},
one uses a single $U^\text{eff}$ parameter rather than seperate $U$ and $J$ parameters for
on-site Coulombic and on-site exchange, respectively~\cite{PhysRevB.52.R5467}. The
correction influences the calculation by applying an energy penalty to the system:

\begin{equation}
U^\text{eff} = U - J,
\end{equation}
\begin{equation}
E_\text{DFT+U}=E_\text{DFT}+\sum_{a,i}\frac{U^\text{eff}_i}{2}\textup{Tr}\left(\rho^{a}_i-\rho^{a}_i\rho^{a}_i\right),
\end{equation}
where the sum runs over the atoms $a$ and orbitals $i$ for which the correction should be
applied. $E_{\text{DFT}}$ is calculated by a standard GPAW calculation and is corrected to
$E_{\text{DFT+U}}$ by penalizing the energy such that fully occupied or fully unoccupied orbitals
are stabilized. The magnitude of the correction depends on $U^\text{eff}_i$, and the atomic
orbital occupation matrix ($\rho^a_i$) controls which orbitals contribute to the correction
based on their occupation.
%
In principle,
any orbital on any element that is partially occupied can be corrected. 

GPAW supports
normalization of the occupation matrix, accounting for the truncated portion of the wave
function outside the augmentation sphere. To maintain consistency with other codes that do
not support normalization, this normalization can be disabled, but large disagreements are
expected when applied to $p$ orbitals. GPAW is one of the few codes that currently supports
multiple simultaneous corrections to orbitals on the same atom; this is useful when two
types of orbitals, such as $p$ and $d$ orbitals, are nearly degenerate but both are
partially occupied. 

There is no $U^\text{eff}$ that is strictly correct, but methods such
as RPA~\cite{aryasetiawan2011constrained} or linear
response~\cite{Cococcioni_2005,moore2022highthroughput} can allow for a value to be
calculated from first principles. More commonly, $U^\text{eff}$ is chosen semi-empirically
to fit experimental properties such as formation energy~\cite{PhysRevB.73.195107}, band
gap~\cite{Taib_2019}, or more recently by machine learning
predictions~\cite{yu2020machine}.

%% file: hybrids/hybrids.tex
\label{sec:range_separated}

Hybrid functionals, especially range-separated functionals (see
below), correct problems present in calculations utilizing
(semi-)local functionals such as the wrong asymptotic behavior of the effective
potential leading to an improper description of Rydberg
excitations~\cite{tawada_long-range-corrected_2004}, the improper
calculation of the total energy against fractional charges leading
to the charge delocalization error~\cite{cohen_insights_2008}, and the
wrong description of (long-range) charge transfer excitations due to
the locality of the exchange
hole~\cite{Dreuw2003,baerends_kohnsham_2013,kummel_charge-transfer_2017}.

The exchange--correlation energy $E_\mathrm{xc}$,
can be split into the contributions from exchange, $E_\mathrm{x}$, and
from correlation, $E_\mathrm{c}$~\cite{kohn1965self,baer_tuned_2010}.
Hybrid functionals combine the exchange
from the (semi-)local functionals with the exchange from Hartree--Fock theory (HF).
Global hybrids such as
PBE0~\cite{adamo_toward_1998} mix the exchange from DFT with the exchange from
HF by a global fixed amount.
Range-separated functionals (RSF)
add a separation
function $\omega_\mathrm{RSF}$~\cite{yanai_new_2004,akinaga_range-separation_2008}
to express the Coulomb kernel in the exchange integrals as
\begin{align}
  \frac{1}{r_{12}} &=
  \underbrace{\frac{1 - \left [ \alpha + \beta \left( 1 - \omega_\mathrm{RSF} \left( \gamma, r_{12} \right) \right) \right ]}{r_{12}}}_{\mathrm{SR}} \notag \\
    &+ \underbrace{\frac{\left [ \alpha + \beta \left( 1 - \omega_\mathrm{RSF} \left( \gamma, r_{12} \right) \right) \right ]}{r_{12}}}_{\mathrm{LR}},
	\label{eq:hybrids_rsfomega}
\end{align}
where $r_{12}=|\vec{r}_1-\vec{r}_2|$.
Here
$\alpha$ and $\beta$ are mixing parameters for global and range-separated mixing, respectively.
$\omega_\mathrm{RSF}$ is a soft function with values ranging from
one for $r_{12} = 0$ to zero for $r_{12} \to \infty$, where the
decay is controlled by the parameter $\gamma$.
Long-range-corrected RSFs such as LCY-PBE~\cite{seth_range-separated_2012}
use (semi-)local approximations for short-range (SR) interaction and
apply HF exchange for long-range (LR) interaction.
Short-range-corrected RSFs such as HSE~\cite{heyd2003hybrid}
reverse this approach.
The parameter
$\gamma$ is either fixed or can be varied to match
criteria of the ideal functional, e.g. that the energy
of the highest occupied molecular orbital matches the
ionization potential~\cite{livshits_well-tempered_2007,baer_tuned_2010}.

Details on the FD-mode implementation of long-range corrected RSF can be
found in Refs.~\cite{wurdemann_berechnung_2016,wurdemann_charge_2018}.
In general, the FD-mode implementation of hybrids is limited to molecules, and
forces have not been implemented.  The PW-mode implementation of hybrids
handles $\mathbf{k}$-points, exploits symmetries and can calculate forces.

%% file: sic/sic.tex
\label{sec:sic}
A fully self-consistent and variational implementation of the Perdew-Zunger self-interaction
correction~\cite{perdew_zunger_1981} (PZ-SIC) is available in GPAW.
It corrects the various problems with (semi-)local Kohn--Sham functionals
mentioned above in the context of hybrid functionals.
Atomic forces are available in the GPAW implementation with all three types of basis sets.
A corrected KS functional has the form
\begin{equation}
 E^\mathrm{SIC}[n_1, \dots, n_n] =
 E^\mathrm{KS}[n] - \alpha \sum_i^\mathrm{occ}\left(U_\mathrm{H}[n_i]
 + E_\mathrm{xc}[n_i, 0] \right)
\end{equation}
where the self-Coulomb and self-XC of each occupied orbital density
is subtracted from the Kohn--Sham energy functional. Due to the explicit dependence
on the orbital densities, the corrected energy functional is not invariant
under a unitary transformation among occupied orbitals
and thereby not a KS functional.
As a result, the minimization of
$E^\mathrm{SIC}$ requires special direct minimization techniques
(see section~\ref{sec:directmin}) and delivers a specific set of (typically localized) optimal orbitals.
The calculations should be carried out using complex orbitals~\cite{Kluepfel_2011,Lehtola_2014,Lehtola_2016}.
The full PZ-SIC has been shown to give an over-correction to binding
energy as well as band gaps and improved results for these properties
are obtained by scaling the SIC by $\alpha=1/2$~\cite{Jonsson_2011,Kluepfel_2012},
while the long-range form of the effective potential, necessary for Rydberg state calculations,
requires the full correction~\cite{Gudmundsdottir_2013}.

PZ-SIC has been shown to give accurate results in cases where commonly used KS functionals fail.
This includes, for example, the Mn dimer where the PBE functional gives qualitatively incorrect
results while the corrected functional gives close agreement with high-level quantum chemistry
results as well as experimental measurements~\cite{IvanovMnDimer_2021}.
Another example is the defect state of a substitutional Al atom in
$\alpha$-quartz~\cite{Gudmundsdottir_2015}.
Also, PZ-SIC has been shown to improve values of the excitation energy of
molecules obtained in variational calculations of excited states~\cite{Ivanov2021,Schmerwitz2022}
(see section~\ref{sec:variationalexcited}).

%% file: beef/beef.tex
\label{sec:BEEF}

A great strength of Kohn--Sham DFT and its extensions is that a reasonably
high accuracy of physical, material, and chemical properties can be
obtained at a relatively moderate computational cost. DFT is thus often
used to simulate materials, reactions, and properties, where there de facto
exists no ``better'' alternative. Even though more accurate
electronic structure methods in principle might exist for a given
application, the poor scaling of systematically more accurate methods often
makes these computationally infeasible at the given system size that is
being studied using DFT. One is therefore often in the situation that the
accuracy of a given DFT calculation of some materials or chemical property
cannot be verified against e.g. a more accurate solution of the
Schr\"odinger equation, even on the biggest available supercomputers.

On the other hand, the wealth of available XC functionals
naturally allows one to look at how sensitive a DFT result is on the choice
of functional, and often accuracy is therefore being judged primarily by
applying a small set of different XC functions, especially if
no accurate benchmark theoretical simulation or experimental measurement
is available. A challenge, however, is that the different available
functionals are often known to be particularly good at simulating certain
properties and poor at others. It is thus not at all clear how much one
should trust a given functional for a given simulated property. The
Bayesian Error Estimation (BEE) class of
functionals~\cite{mortensen_bayesian_2005} attempts to develop a practical
framework to establishing an error estimate from a selected set of
functional ``ingredients''.

Assume that the XC model $M(a)$ is a function of a set of parameters, $a$,
that can be varied freely. If a benchmark data set, $D$, of highly accurate
properties established from experiments or higher-accuracy electronic
structure simulations is available, we can attempt to identify the ensemble
of models given by some distribution function, $P(M(a))$, such that the
most likely model in the ensemble, $M(a_0)$, makes an accurate predictions
for the benchmark data set, while the spread of ensemble reproduces the
spread between the predictions of the most likely model and the benchmark
data.

Bayes' theorem provides a natural framework to search for the
ensemble distribution. If a joint distribution between $M(a)$ and $D$
exists, which we would assume, then Bayes' theorm gives
\begin{equation}
P(M(a)\vert D)\propto P(D\vert M(a))P_0(M(a)),
\end{equation}
where $P_0(M(a))$ is the prior expectation to the distribution of models
before looking at the data, $D$, and $P(D\vert M(a))$ is a likelihood of
seeing the data given the model. To achieve a useful ensemble, much care
has to be put into finding a large enough set of varied and accurate data
for different materials and chemical properties, and much care has to be
applied as to how the ensemble is regularized to avoid
overfitting~\cite{BEEF,wellendorffMBEEFAccurateSemilocal2014}.

In all the
BEE functionals, choices are made such that ultimately the XC functional is
linear in $a$, and such that the distribution of $a$ ends up following a
multidimensional Normal distribution given by a regularized covariance
matrix $\Gamma$:
\begin{equation}
P(a)\propto \exp(-(a-a_0)\Gamma (a-a_0)),
\end{equation}
where $\Gamma$ has been scaled in such a way that the ensemble reproduce
the observed standard deviation between $M(a_0)$ and $D$.

\begin{figure}[tb]
    \centering
    \includegraphics{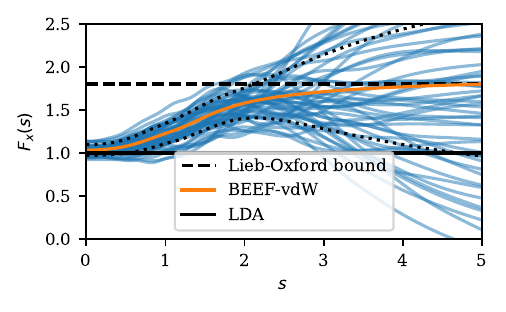}
    \caption{
Bayesian ensemble of XC functionals
around BEEF–vdW. Orange solid line is the BEEF–vdW
exchange enhancement factor, while the blue lines depict $F_\text{x}(s)$
for 50 samples of the randomly generated ensemble. Dotted black
lines mark the exchange model perturbations that yield DFT results
±1 standard deviation away from BEEF–vdW results.}
\label{fig:beef}
\end{figure}

Figure \ref{fig:beef} shows an ensemble of exchange enhancement factors
$F_x(s)$ from the BEEF-vdW functional, where $s=|\nabla n|/(2k_Fn)$ is the
reduced density gradient. The approach has given rise to several
functionals BEEF-vdW~\cite{BEEF},
mBEEF~\cite{wellendorffMBEEFAccurateSemilocal2014},
and mBEEF-vdW~\cite{mBEEF-vdW}, which all include
error estimation and are readily
available in GPAW. Through the ASE interface, one can for example utilize
the error ensembles to establish error estimates on Python-implemented
models that are using DFT simulations for their parametrization. An
example of this is the application to error estimates on adsorption scaling
relations, microkinetics, and materials selection~\cite{BEEFammonia}.

One
risk to the approach of establishing error estimates from a small selected
``ensemble''  of XC-functionals is clearly that if the simulated property
in question is poorly described by all functionals in the ensemble, then
the error estimate might be also become poor. This could for example be the
case if one tried to establish an error estimate for band gaps in oxides
or van der Waals bonding of adsorbates on surfaces based on an ensemble of
GGA XC-functionals, since no GGA functional may be accurate for
simulating either property.

%% file: ion-dynamics/ion-dynamics.tex
\label{sec:ion_dynamics}

GPAW can be employed as a `black-box' calculator, supplying energies and
forces to other programs such as e.g. ASE, which then perform optimization of
ground state geometries and reaction paths or carry out molecular dynamics.
In fact, this is a key design principle behind GPAW. Methodological
developments and general implementations that are not directly dependent on
fast access to detailed electronic structure information should preferably be
implemented externally to GPAW. This lead to maximal simplicity of the GPAW
code itself, while also allowing for the external code to be utilized and
tested with other electronic structure codes and atomistic potentials. A key
to the high efficiency of GPAW simulations involving ionic displacements is
the versatile implementation of constraints in ASE. Here many types of
constraints are readily accessible, from the simple removal of degrees of
freedom to more exotic constraints allowing for rigid molecule
dynamics~\cite{Ciccotti1982}, to harmonic restoring forces~\cite{Andy2014},
spacegroup preservation and combined ionic--unit cell
dynamics~\cite{Kaxiras1999}. Many algorithms are available for various 
structure optimization and molecular dynamics tasks.

\subsection{Structure relaxation}

Local structure optimization in GPAW is typically achieved through the
use of an optimizer from ASE. Here a larger range of standard optimizers are
available such as quasi-Newton algorithms including
BFGS~\cite{Broyden,Fletcher,Goldfarb,Shanno} and limited-memory
BFGS~\cite{LBFGS}, or Newtonian dynamics-based algorithms such as MDMin
and FIRE~\cite{FIRE}. That the optimizers have been implemented externally to
GPAW provides benefits in terms of simple procedures for restarting long
simulations and monitoring their convergence. Some optimizers from
SciPy~\cite{2020SciPy-NMeth} are also available through the open-source 
ASE package, which provides a simple way to interface any
optimizer in the SciPy format to GPAW. Preconditioning is implemented in an
accessible way~\cite{Packwood2016}, which often leads to significant
performance improvements. 

Of the classical optimization methods, the
quasi--Newton algorithms are often highly competitive. Here one builds up
information on the Hessian or inverse Hessian matrix from calculated forces,
ultimately leading to an accurate harmonic model of the potential energy
surface in the vicinity of the local minimum. Such algorithms can, however,
have problems both dealing with anharmonicity in the potential energy surface
and with any noise in the electronic structure simulations. It often makes sense 
instead to fit a Gaussian process to the calculated
energies and forces, and minimize within this model potential. This is
implemented as the so-called GPmin method in ASE and often converges on the
order of three times faster than the best quasi--Newton optimizers.~\cite{GPMin}

\subsection{Reaction paths and barriers}

Reliable calculations of energy barriers are of key importance for determining 
the rate of atomistic processes. In many quantum-chemical codes utilizing 
accurate atom-centered basis functions, this is
achieved using analytical second derivatives in the direct search for
first-order saddle points. This approach is less useful in the plane-wave based or
grid-based modes of GPAW. The Dimer method~\cite{DimerMethod} is
implemented in ASE and can be used with GPAW, but often one would like to
have an overview of the entire reaction path of an atomic-scale process to verify 
that the correct energy barrier for a process has been
determined, and to obtain a full overview of the atomistic mechanism. For
this purpose, the Nudged Elastic Band (NEB) method is typically employed through
the ASE interface to GPAW. Both the original method~\cite{NEB} and a range of
later improvements are
available~\cite{improvedtangent,CI,II,KolsbergNEB,LindgrenNEB,MakriPrecon}.
Special care has to be taken in selecting the optimizer for a NEB algorithm,
as this choice can have drastic influence on the convergence rate. For 
optimization of the reaction path, drastic performance improvements can be 
obtained by carrying out
the optimization in a surrogate machine-learning model fitted to the
potential energy surface~\cite{AndyNEB,OlipekkaNEB,JoseNEB}. GPAW has been
used to drive NEB calculations on both surface systems~\cite{ari2018,
stergaard2018, Arnarson2018, Wan2020} as well as in molecules~\cite{Wan2020,
Levi2020pccp}. 

\subsection{Global structure optimization}

GPAW is integrated with various tools for global optimization of structures, 
compositions, as well as materials
morphologies. Some quite generally applicable global optimization tools
available are Basin Hopping~\cite{BasinHopping}, Minima
Hopping~\cite{MinimaHopping}, and Genetic Algorithms~\cite{HammerGA}. Some of
the most powerful global optimization problems addressed using GPAW rely on
machine-learning accelerated global-optimization strategies. These strategies
have for example been applied to surfaces, clusters~\cite{Christiansen2022},
and crystal structures in general~\cite{Jennings2019}.
In other machine-learning accelerated global-optimization routines, GPAW 
was used to generate initial databases of surface and bulk systems, and for 
later model validation in a strategy
that uses Gaussian processes to generate surrogate potential-energy surfaces.
These were then explored with Bayesian-optimization techniques, achieving
speed-ups over more conventional methods of several orders of magnitude in
finding optimal structures of the systems under
investigation~\cite{gofee, Kaappa2021, Kaappa2021b}. The method has been augmented
through introducing extra (hyper)dimensions by interpolating between chemical
elements, which speeds up the efficiency of the global
search~\cite{larsen2022machinelearning}.
GPAW has been
integrated with a Covariance Matrix Adaptation Evolution Strategy (CMA-ES)
framework, providing energies and forces, generating training data, and
evaluating CMA-ES candidate structures~\cite{Wanzenbck2022}.

\subsection{Molecular dynamics and QM/MM}
\emph{Ab initio} molecular dynamics (MD) can be performed and analyzed through the ASE interface to GPAW. This include standard simulations such as constant-NVE, constant-NVT, and constant-NPT simulations. There is access to e.g. Langevin-, Andersen-, Nose-Hoover-, and Berendsen-dynamics. Due to the externalization of the dynamics, the development of novel algorithms and analysis tools becomes facile. Some examples of the use of GPAW span \emph{ab initio} MD studies
that have explored the liquid structure of water~\cite{Mgelhj2011} and the
water/Au(111) electrochemical interface~\cite{Hansen2016}.

Furthermore, GPAW is capable of working with external electrostatic potential terms from user-supplied point charge values and positions, enabling Quantum Mechanical (QM) / Molecular Mechanical (MM) simulations.
Since GPAW is from the outset designed around highly efficient grid operations, the computational overhead of
evaluating this potential as well as the resulting forces on the MM point charges from the QM density
is kept low~\cite{Dohn2017}. GPAW has been central in a range studies of ion dynamics in solution,
both in and out of equilibrium. By using the molecular-dynamics functionality of ASE, researchers
have performed QM/MM MD simulations of excited-state bond formation in photocatalyst model
systems~\cite{Dohn2014, vanDriel2016, Levi2018, Haldrup2019}, and electron transfer as well as coupled
solute-solvent conformational dynamics in photosensitizer systems~\cite{Dohn2016, Levi2020pccp}.
Work on polarizable embedding QM/MM within GPAW is ongoing~\cite{Jnsson2019, Dohn2019}.

%% file: spin-orbit/spin-orbit.tex
The spin--orbit coupling is typically completely dominated by regions close to the nuclei where the electrostatic potential becomes strong. Within the PAW augmentation sphere of atom $a$, the spin-$\sigma$ component of orbital $n$ is given by
\begin{equation}
\psi_{n\sigma}(\mathbf{r})=\sum_i\langle \tilde p^a_i|\tilde\psi_{n\sigma}\rangle\phi_i^a(\mathbf{r}).
\end{equation}
Assuming a spherically symmetric potential, the spin--orbit Hamiltonian for atom $a$ is then written as
\begin{equation}\label{eq:H_soc}
    \hat H_\mathrm{SOC}^a=\frac{\mu_\mathrm{B}}{\hbar m_\mathrm{e} ec^2}\frac{1}{r}\frac{\mathrm{d} V_r^a}{\mathrm{d}r}\hat{\mathbf{L}}\cdot\hat{\mathbf{S}},
\end{equation}
where $V_r^a$ is the radial electrostatic potential of atom $a$.
We evaluate $V_r^a$ as the spherical part of the XC- and Hartree-potential from the local expansion of the density given by Eq.~\ref{allelectrondensity}.
Since the partial waves
$\phi^a_i$ are eigenstates of the scalar-relativistic Hamiltonian, these are independent of
spin and it is straightforward to evaluate the action of $\hat{\mathbf{L}}$ on them.

\begin{figure}[tb]
    \centering
    \includegraphics{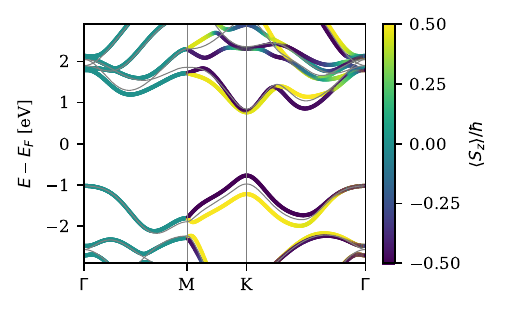}
    \caption{Band structure of WS$_2$ monolayer obtained from PBE with
    non-selfconsistent spin--orbit coupling. The colors indicate the
    expectation value of $S_z$ for each state. The grey lines show the band structure without spin--orbit coupling.}
    \label{fig:WS2_bands}
\end{figure}

The eigenenergies can be accurately calculated in a non-selfconsistent treatment of the spin--orbit coupling and may be obtained by diagonalizing the full Hamiltonian in a basis of scalar-relativistic orbitals~\cite{Olsen2016a}:
\begin{equation}\label{eq:nsc_soc}
H_{mn}=\varepsilon^0_m\delta_{mn}+\sum_{aij\sigma\sigma'}\langle\tilde\psi_{m\sigma}^0|\tilde p^a_i\rangle\langle\phi_i^a|\hat H^a_\mathrm{SOC,\sigma\sigma'}|\phi_j^a\rangle\langle \tilde p^a_j|\tilde\psi_{n\sigma'}^0\rangle.
\end{equation}
Here $\varepsilon^0_m$ and $\psi_{m}^0$ represent the scalar-relativistic eigenenergies and eigenstates, respectively. This constitutes a fast post-processing step for any scalar-relativistic calculation and only requires the projector overlaps $\langle \tilde p^a_j|\tilde\psi_{n\sigma'}^0\rangle$. It should be noted that this approach in principle requires convergence with respect to the number of scalar-relativistic states included in the basis, but the eigenvalues typically converge rather rapidly with respect to the basis. In Fig.~\ref{fig:WS2_bands}, we show the band structure of a WS$_2$ monolayer obtained from PBE with non-self-consistent spin--orbit coupling. The W atoms introduce strong spin--orbit coupling in this material and the valence band is split by 0.45~eV at the K point. The spin degeneracy is retained along the $\Gamma$-M line, which is left invariant by two non-commuting mirror symmetries.

%% file: mag-anisotropy/mag-anisotropy.tex
For magnetic materials the non-selfconsistent treatment of spin--orbit coupling is convenient for evaluating the magnetic anisotropy. The magnetic force theorem~\cite{Liechtenstein1987} implies that rotating the magnetic moments away from the ground state configuration yields a contribution to the energy, which is well approximated by the change in Kohn--Sham eigenvalues. The change in energy for a given orientation of the magnetization density can thus be obtained as
\begin{equation}\label{eq:mag_ani}
\Delta E^{\theta,\varphi}=\sum_nf_n^{\theta,\varphi}\varepsilon_n^{\theta,\varphi},
\end{equation}

where $\varepsilon_n^{\theta,\varphi}$ are the
eigenvalues obtained from diagonalizing
Eq.~(\ref{eq:nsc_soc}) with the spins rotated to a direction defined by the angles
$(\theta,\varphi)$ and $f_n^{\theta,\varphi}$ are the associated occupation numbers. This corresponds to rotating the xc-magnetic field (defined below), which will lead to different eigenvalues when spin--orbit coupling is included. In
two-dimensional magnets, an easy-axis anisotropy is decisive for magnetic
order~\cite{Mermin1966, Lado2017, Torelli2018} and Eq.~(\ref{eq:mag_ani}) is easily  applied
to high throughput computations of magnetic properties~\cite{Torelli2019b, Torelli2020}.

%% file: non-col-scf/non-col-scf.tex
The Kohn--Sham framework for treating non-collinear magnetism was developed by Barth and Hedin~\cite{Barth1972} and involves the spin density matrix
\begin{equation}
\rho_{\sigma\sigma'}(\mathbf{r}) = \sum_nf_n\psi^*_{n\sigma}(\mathbf{r})\psi_{n\sigma'}(\mathbf{r})
\end{equation}
as the basic variable. The electronic density and magnetization are then given by $n=\mathrm{Tr}[\rho]$ and $\mathbf{m}=\mathrm{Tr}[\boldsymbol\sigma\rho]$ respectively. The XC-potential acquires four components, which are the functional derivatives of the XC-energy with respect to the density matrix and may be represented as a $2\times2$ matrix acting on spinor Kohn--Sham states. These can be expressed in terms of the density and magnetization, which lead to the XC-part of the Kohn--Sham Hamiltonian
\begin{equation}
H_\mathrm{xc}=v_\mathrm{xc}+\mathbf{B}_\mathrm{xc}\cdot\boldsymbol{\sigma},
\end{equation}
where the scalar potential and XC-magnetic field are given by
\begin{equation}
v_\mathrm{xc}(\mathbf{r})=\frac{\delta E_\mathrm{xc}}{\delta n(\mathbf{r})}\qquad \mathbf{B}_\mathrm{xc}(\mathbf{r})=\frac{\delta E_\mathrm{xc}}{\delta \mathbf{m}(\mathbf{r})}.
\end{equation}
In GPAW, the self-consistent treatment of non-collinear spin is implemented within LDA where $\mathbf{B}_\mathrm{xc}$ is approximated as
\begin{equation}\label{eq:nc_lda}
\mathbf{B}_\mathrm{xc}^\mathrm{LDA}(\mathbf{r})=\frac{\delta E_\mathrm{xc}^\mathrm{LDA}}{\delta m(\mathbf{r})}\mathbf{\hat m}(\mathbf{r}).
\end{equation}
Here $m(\mathbf{r})=|\mathbf{m}(\mathbf{r})|$ is the magnitude and $\mathbf{\hat m}(\mathbf{r})$ is the direction of the magnetization. The generalization of Eq.~\ref{eq:nc_lda} to GGAs are plagued by formal as well as numerical problems~\cite{Scalmani2012, Kim2021} and the self-consistent solution of the non-collinear Kohn--Sham equations is presently restricted to LDA. It should be emphasized that the PAW formalism allows for a fully non-collinear treatment that does not rely on intra-atomic collinearity, which is often imposed in other electronic structure packages.

Spin--orbit coupling may be included by adding Eq.~\ref{eq:H_soc} to the Kohn--Sham Hamiltonian and this constitutes a fully self-consistent framework for spin--orbit calculations within LDA.

%% file: mag-orbmag/orbmag.tex
Current \emph{ab initio} methods for determining the orbital magnetization of a material involve either the modern theory, i.e. a
Berry-phase formula, or the atom-centered approximation (ACA), where contributions to the expectation value of the angular momentum operator
are restricted within atom-centered muffin-tin (MT) spheres with specified cutoff radii. The PAW formulation of the wave functions allows
for an approximation similar to the MT-ACA where only the PAW expansion of the wave functions is assumed to contribute significantly to the
expectation value of the angular momentum. The orbital magnetic moments can then be calculated through
\begin{equation}
    \mathbf{m}^a_\mathrm{orb}=-\frac{e}{2m_e}\sum_a\sum_{ii'}\sum_{\sigma}D^a_{\sigma ii'}
    \left<\phi^a_{i}\middle|\widehat\mathbf{L}\middle|\phi^a_{i'}\right>,
    \label{eq:orbmag}
\end{equation}
where $D^a_{\sigma ii'}$ are elements of the atomic density matrix as defined in Eq.~\eqref{eq:atomicdensitymatrix}, $\phi^a_i(\mathbf{r})$
are the \emph{bound} all-electron partial waves of atom $a$, and $\hat{\mathbf{L}}$ is the angular-momentum operator.\par
We note that this expression does not entail cutoff radii for the atomic contributions to the orbital magnetic moments; instead, the entire
all-electron partial waves are included although the all-electron atomic expansion is only formally exact inside the PAW spheres.
Additionally, this means that the unbounded, i.e. non-normalisable, all-electron partial waves must be excluded in Eq.~\eqref{eq:orbmag}\par
A prerequisite for nonzero orbital magnetization is broken time-reversal symmetry as represented by a complex Hamiltonian that is not
unitarily equivalent to a real counterpart. Practically this means
that a finite orbital magnetization requires magnetic order and either
inclusion of spin--orbit coupling or non-coplanar spin textures. In GPAW, the spin--orbit interaction can be included either
self-consistently in a non-collinear calculation or as a non-self-consistent post-processing step following a collinear calculation.
\begin{table}
\begin{ruledtabular}
\def\arraystretch{1.3}
\begin{tabular}{l|llll}
Crystal & \multicolumn{1}{c}{\textbf{bcc-Fe}} & \multicolumn{1}{c}{\textbf{fcc-Ni}}
        & \multicolumn{1}{c}{\textbf{fcc-Co}} & \multicolumn{1}{c}{\textbf{hcp-Co}} \\
Easy axis & \multicolumn{1}{c}{$[001]$} & \multicolumn{1}{c}{$[111]$}
          & \multicolumn{1}{c}{$[111]$} & \multicolumn{1}{c}{$[001]$} \\[2pt] \hline
PAW-ACA                           & $0.0611$ & $0.0546$ & $0.0845$ & $0.0886$ \\
MT-ACA~\cite{Ceresoli2010}        & $0.0433$ & $0.0511$ & $0.0634$ & $0.0868$ \\
Modern theory~\cite{Ceresoli2010} & $0.0658$ & $0.0519$ & $0.0756$ & $0.0957$ \\
Experiment~\cite{Meyer1961}       & $0.081$  & $0.053$  & $0.120$  & $0.133$
\end{tabular}
\caption{Calculated and measured values for the orbital magnetization in units of $\mu_\mathrm{B}$ per atom.}
\label{tab:orbmag}
\end{ruledtabular}
\end{table}
The orbital magnetization has been calculated using Eq.~\eqref{eq:orbmag} for the simple ferromagnets bcc-Fe, fcc-Ni, fcc-Co, and hcp-Co, where the
spin--orbit interaction is included non-self-consistently. The PAW-ACA results are displayed in Table \ref{tab:orbmag} along with MT-ACA
results, modern theory results, and measurements from experiment, demonstrating mainly that the PAW-ACA can be an improvement over the MT-ACA
and secondly that there is decent agreement between the PAW-ACA and the modern theory for these systems.

%% file: b-field/b-field.tex
\begin{figure}[tb]
    \centering
    \includegraphics{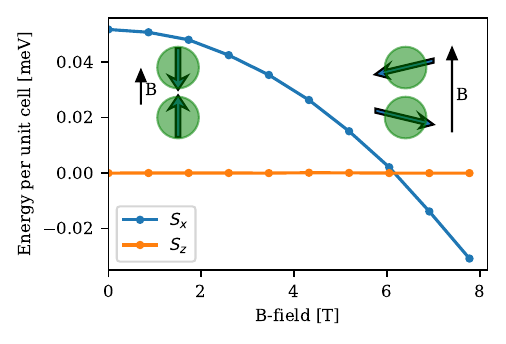}
    \caption{Spin-flop transition in Cr$_2$O$_3$. The alignment of the spins with respect to the magnetic field is sketched for small and large magnitudes of the field. The canting (small ferromagnetic component after the spin flop) is exaggerated for visualization. The actual canting is roughly $1^\circ$.}
    \label{fig:spin_flop}
\end{figure}

The coupling of the electronic spin magnetic moments to a constant external magnetic field $\mathbf{B}$ can be included by the addition of a Zeeman term in the Kohn--Sham Hamiltonian:
\begin{equation}
\Delta H_\mathrm{B}=\mathbf{B}\cdot\boldsymbol\sigma.
\end{equation}
As an example, we can consider the spin-flop transition in Cr$_2$O$_3$.
The ground state is anti-ferromagnetic with a weak anisotropy that prefers alignment of the
spins with the $z$-axis, and if a magnetic field is applied along the $z$-direction, it
becomes favorable to align the spins along a perpendicular direction (with a small
ferromagnetic component) at the critical field where the
Zeeman energy overcomes the
anisotropy. This is clearly a spin--orbit effect and one has to carry out the calculations
in a fully non-collinear framework with self-consistent spin--orbit coupling. In
Fig.~\ref{fig:spin_flop} we show the energy of the two alignments of spin as a function of
external magnetic field obtained with LDA+U ($U=2$~eV). The minimum energy configuration
changes from $S_z$ to $S_x$ alignment at 6 T, i.e., the spin flips at this critical field
value. This is in excellent agreement with the experimental value of
Ref.~\cite{PhysRev.130.183}. The critical field is, however, rather strongly dependent on
the chosen value of $U$ and increases by a factor of two in the absence of $U$. The magnetic
anisotropy (per unit cell) can be read off from the energy difference at $B=0$~T.

%% file: spin-spirals/spin-spirals.tex
\label{sec:spirals}
The ground-state magnetic structure of frustrated and/or chiral magnets is often non-collinear and may be incommensurate with the chemical unit cell. In the classical isotropic Heisenberg model,
the energy is always minimized by a
planar spin spiral~\cite{Kaplan1959}.
%
The energy of a general planar spin spiral can be evaluated efficiently within the chemical unit cell using the Generalized Bloch's Theorem (GBT)~\cite{Sandratskii1986}. 
In the GBT implementation of GPAW~\cite{Sodequist2023a}, there is no restriction on interatomic collinearity and thus it encodes a rotation of the all-electron magnetization density by an angle
$\varphi = \mathbf{q} \cdot \mathbf{R}_i$ upon translations of a lattice vector $\mathbf{R}_i$.
The Kohn--Sham equations can then be solved self-consistently for a fixed wave vector $\mathbf{q}$,
\begin{align}
    \hat H_\mathbf{q}(\mathbf{k}) \left|u_{\mathbf{q},n\mathbf{k}}\right\rangle
    =\varepsilon_{\mathbf{q},n\mathbf{k}} \left|u_{\mathbf{q},n\mathbf{k}}\right\rangle,
    \label{eq:gbt ks equations}
\end{align}
using only the periodic part of the generalized Bloch orbitals,
$|\psi_{\mathbf{q},n\mathbf{k}}\rangle = U_\mathbf{q}^\dag(\mathbf{r})e^{i\mathbf{k}\cdot\mathbf{r}} |u_{\mathbf{q},n\mathbf{k}}\rangle$,
where $U_\mathbf{q}$ denotes the generator of spin rotations about the $z$-axis:
\begin{align}\label{eq:U}
    U_\mathbf{q}(\mathbf{r})=
\begin{pmatrix}
e^{i\mathbf{q}\cdot\mathbf{r}/2} & 0\\
0 & e^{-i\mathbf{q}\cdot\mathbf{r}/2}
\end{pmatrix}.
\end{align}
The generalized Bloch Hamiltonian without spin--orbit coupling is given by
\begin{align}
    \hat H_\mathbf{q}(\mathbf{k})&=e^{-i\mathbf{k}\cdot\mathbf{r}}
U_\mathbf{q}(\mathbf{r})\hat H U_\mathbf{q}^\dag(\mathbf{r})e^{i\mathbf{k}\cdot\mathbf{r}},
\end{align}
and once Eq.~\eqref{eq:gbt ks equations} has been solved self-consistently for a given $\mathbf{q}$,
the corresponding spin-spiral energy is evaluated as usual,
$E_\mathrm{SS}(\mathbf{q}) = E_\mathbf{q}^\mathrm{DFT}[n,\mathbf{m}]$.

\begin{figure}[tb]
    \centering
    \includegraphics{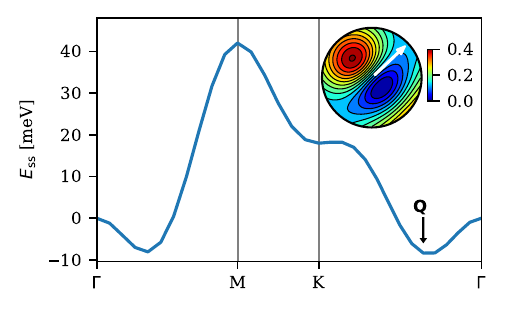}
    \caption{
    LSDA spin-spiral spectrum of the NiBr$_2$ monolayer (structure taken from the C2DB~\cite{haastrup2018computational}).
    The ground state has an incommensurate wave vector $\mathbf{Q} \simeq [0.1, 0.1, 0]$.
    The magnetic moment displays only weak longitudinal fluctuations
    and the band gap remains finite for all wave vectors $\mathbf{q}$,
    indicating that a Heisenberg model description of the material would be appropriate.
    The inset shows the spin--orbit correction to the spin-spiral energies (in meV)
    as a function of the normal vector $\mathbf{n}$ of the planar spin spiral.
    $\mathbf{n}$ is depicted in terms of its stereographic projection in the upper hemisphere above the monolayer plane.
    The spiral plane is found to be orthogonal to $\mathbf{Q}$
    and tilted slightly with respect to the out-of-plane direction.
    }
    \label{fig:NiBr2}
\end{figure}

Using the GPAW functionality to compute the spin-spiral energy as a function of $\mathbf{q}$,
one can then search for the ground-state wave vector $\mathbf{Q}$ that minimizes the energy.
In Fig.~\ref{fig:NiBr2}, we show that the monolayer NiBr$_2$ has an incommensurate spin-spiral ground state,
and that the local magnetic moment on the Ni atom depends only weakly on the wave vector $\mathbf{q}$.

Furthermore, the orientation of the ground state spin spiral can be obtained by including
spin--orbit coupling non-self-consistently in the projected spin--orbit approximation~\cite{sandratskii2017insight} and search for the orientation that minimizes the energy. The ordering vector $\mathbf{Q}$ and the normal vector $\mathbf{n}$ of the spiral plane thus constitute a complete specification of the magnetic ground state within the class of single-$\mathbf{q}$ states.
The normal vector is of particular interest
since spin spirals may lead to spontaneous breaking of the crystal symmetry,
and the normal vector largely determines the direction of magnetically induced spontaneous polarization~\cite{Sodequist2023a}.
In Fig.~\ref{fig:NiBr2}, we show that $\mathbf{n}$ is perpendicular to the wavevector $\mathbf{Q}$
in the monolayer NiBr$_2$ ground state, corresponding to a cycloidal spin spiral.


%% file: linear-response-TDDFT/linear-response-TDDFT.tex
\label{sec:LR-TDDFT}
To linear order, the change in electron density induced by a time-dependent external
(scalar) potential $V_{\mathrm{ext}}(\mathbf{r},t)$ is governed by the electronic
susceptibility, $\chi$:
\begin{equation}
  \delta n(\mathbf{r}, t) = \int_{-\infty}^\infty dt' \int d\mathbf{r}'\, \chi(\mathbf{r}, \mathbf{r}', t-t')  V_{\mathrm{ext}}(\mathbf{r}',t').
\end{equation}
The susceptibility is itself given by the Kubo formula~\cite{Kubo1957}
\begin{equation}
  \chi(\mathbf{r}, \mathbf{r}', t-t') = - \frac{i}{\hbar} \theta(t-t') \langle\left[\hat{n}_0(\mathbf{r},t), \hat{n}_0(\mathbf{r}',t')\right]\rangle_0,
\end{equation}
where the expectation value is taken with respect to the ground state at zero temperature and the
density operators are cast in the interaction picture.

In the noninteracting Kohn--Sham system, the susceptibility can be evaluated explicitly from
the Kohn--Sham orbitals, eigenvalues and occupations:
\begin{align}
  \chi_0(\mathbf{r}, \mathbf{r}', \omega) = &\lim_{\eta\rightarrow 0^+} 2\, \sum_{n\mathbf{k}} \sum_{m\mathbf{k}'}
                                            \left(f_{n\mathbf{k}} - f_{m\mathbf{k}'}\right)
                                            \nonumber \\
                                            &\times
                                            \frac{\psi^*_{n\mathbf{k}}(\mathbf{r})\psi_{m\mathbf{k}'}(\mathbf{r})
                                                  \psi^*_{m\mathbf{k}'}(\mathbf{r}')\psi_{n\mathbf{k}}(\mathbf{r}')}
                                                  {\hbar\omega - (\epsilon_{m\mathbf{k}'} - \epsilon_{n\mathbf{k}}) + i\hbar\eta}.
  \label{eq:chi0 real-space}
\end{align}
Based on the Kohn--Sham susceptibility, $\chi_0$, the many-body susceptibility can be calculated via a Dyson-like equation~\cite{Gross1985}:
\begin{align}
  \chi(\mathbf{r}, \mathbf{r}', \omega) = \chi_0(\mathbf{r}, \mathbf{r}', \omega)
                                          + &\iint d\mathbf{r}_1 d\mathbf{r}_2\, \chi_0(\mathbf{r}, \mathbf{r}_1, \omega)
                                          \nonumber \\
                                          &\times
                                          K_{\mathrm{Hxc}}(\mathbf{r}_1, \mathbf{r}_2, \omega) \chi(\mathbf{r}_2, \mathbf{r}', \omega).
  \label{eq:dyson elec. chi real space}
\end{align}
Here, electronic interactions are accounted for via the Hartree-XC kernel,
which is defined in terms of the effective potentials of time-dependent DFT (TDDFT)~\cite{Runge1984}:
\begin{equation}
  K_{\mathrm{Hxc}}(\mathbf{r}, \mathbf{r}', t - t') =v_c (\mathbf{r}-\mathbf{r}')+ f_{\mathrm{xc}}(\mathbf{r}, \mathbf{r}', t - t')
  \label{eq: Hxc kernel}
\end{equation}
where $v_c$ is the Coulomb interaction and $f_\mathrm{xc}$ is the XC-kernel defined as 
\begin{equation}
  f_{\mathrm{xc}}(\mathbf{r}, \mathbf{r}', t - t') = \frac{\delta v_{\mathrm{xc}}(\mathbf{r}, t)}
                                                           {\delta n(\mathbf{r}', t')}.
  \label{eq: fxc kernel}
\end{equation}
In prototypical linear-response TDDFT (LR-TDDFT) calculations, the exchange--correlation part of the kernel is either neglected
(leading to the random phase approximation (RPA)) or approximated via the adiabatic local density approximation (ALDA),
\begin{equation}
v_{\mathrm{xc}}(\mathbf{r}, t) \simeq \left.
                                      \frac{\partial[\epsilon_{\mathrm{xc}}(n) n]}{\partial n}
                                      \right|_{n(\mathbf{r}, t)},
\end{equation}
where $\epsilon_{\mathrm{xc}}(n)$ denotes the XC-energy per electron of the homogeneous electron gas of density $n$.
The ALDA is in fact rather restricted as it only ensures a correct description of the
kernel for metals in the long wave length limit. This implies that it cannot account for
excitons in extended systems and furthermore it leads to a divergent XC-hole. The latter
problem can be resolved by a simple renormalization of the ALDA, that regulates the ontop
XC-hole and drastically improves the description of local correlations over the ALDA (and
RPA)~\cite{olsen2012extending}.

\subsubsection{Implementation for periodic systems}\label{sec:lr-tddft for periodic systems}

In crystalline systems, the susceptibility is periodic with respect to translations on the Bravais lattice,
$\chi(\mathbf{r} + \mathbf{R}, \mathbf{r}' + \mathbf{R}, \omega) = \chi(\mathbf{r}, \mathbf{r}', \omega)$.
Consequently, the susceptibility can be Fourier-transformed according to
\begin{equation}
  \chi_{\mathbf{G}\mathbf{G}'}(\mathbf{q}, \omega) = \iint \frac{d\mathbf{r} d\mathbf{r}'}{\Omega} e^{-i(\mathbf{G}+\mathbf{q})\cdot\mathbf{r}} \chi(\mathbf{r}, \mathbf{r}', \omega) e^{i(\mathbf{G}'+\mathbf{q})\cdot\mathbf{r}'},
\end{equation}
translating the Dyson equation \eqref{eq:dyson elec. chi real space} into a plane-wave matrix equation,
which is diagonal in the wave vector $\mathbf{q}$ and can be inverted numerically.

By using a plane-wave representation, the crux of the implementation then
becomes to calculate the reciprocal-space pair densities
\begin{equation}\label{eq:pair_density}
  n_{n\mathbf{k},m\mathbf{k}+\mathbf{q}}(\mathbf{G}+\mathbf{q})
      = \langle \psi_{n \mathbf k}| e^{-i (\mathbf{G}+\mathbf{q})\cdot \mathbf{r}}
                                  |\psi_{m \mathbf{k} + \mathbf{q}} \rangle,
\end{equation}
and to Fourier transform the XC-kernel
\begin{equation}
  f_{\mathrm{LDA}}(\mathbf{G}) = \int_{\Omega_{\mathrm{cell}}} d\mathbf{r}\, e^{-i \mathbf{G} \cdot \mathbf{r}}
                                 \left.
                                 \frac{\partial^2[\epsilon_{\mathrm{xc}}(n) n]}{\partial n^2}
                                 \right|_{n(\mathbf{r})}.
\end{equation}
Crucially, both can be evaluated by adding a PAW correction to the analogous pseudo quantity,
which itself can be evaluated by means of a fast Fourier transform (FFT). Implementational details hereto
are reported in Refs.~\cite{Skovhus2021} and~\cite{yan2011linear}.

The Hartree contribution to the kernel \eqref{eq: Hxc kernel}
is simply given by the bare Coulomb interaction (here given in atomic units)
\begin{equation}
   v_{\mathrm{c},\mathbf{G}\mathbf{G}'}(\mathbf{q}) =
   \frac{4\pi}{|\mathbf{G} + \mathbf{q}|^2} \delta_{\mathbf{G},\mathbf{G}'},
   \label{eq:coulombkernel}
\end{equation}
and can be evaluated analytically when $\mathbf{q}$ is finite.
However, in the optical limit, the $\mathbf{G}=\mathbf{0}$ component diverges and one needs to be careful when inverting the Dyson equation \eqref{eq:dyson elec. chi real space}.
In GPAW, this is handled by expanding the pair densities within $kp$-perturbation theory:
\begin{align}
    \langle \psi_{n \mathbf k} |
  e^{-i (\mathbf{G} + \mathbf{q}) \cdot \mathbf r} | \psi_{m \mathbf k + \mathbf q} \rangle_{\mathbf{q} \rightarrow \mathbf{0}, \mathbf{G}=\mathbf{0}}
  =  -i \mathbf{q} \cdot \frac{ \langle \psi_{n \mathbf k} | \boldsymbol \nabla |\psi_{m \mathbf k} \rangle }{\epsilon_{m \mathbf k} - \epsilon_{n \mathbf k}}.
\end{align}
In the expansion, the diverging Coulomb term is exactly cancelled by the $\mathbf{q}$-dependence of the pair densities,
so that the product $\chi_0 v_{\mathrm{c}}$ remains finite. For additional details, see Ref.~\cite{yan2011linear}.


\subsubsection{Spectral representation}


GPAW offers two different ways of dealing with the frequency dependence of
the Kohn--Sham susceptibility \eqref{eq:chi0 real-space}.
One is to evaluate the expression explicitly for the frequencies of interest.
This is advantageous if one is interested in a few specific frequencies.
The other is to evaluate the associated spectral function
\begin{align}
   S_{0,\mathbf{G} \mathbf{G}^{\prime}}(\mathbf{q}, \omega)
       = &\frac{2}{\Omega}  \sum_{\mathbf{k}} \sum_{n, m}
          (f_{n\mathbf{k}}-f_{m \mathbf{k} + \mathbf{q}})
         \nonumber \\
         &\times
         n_{n\mathbf{k}, m\mathbf{k}+\mathbf{q}}(\mathbf{G}+\mathbf{q})
         n_{m\mathbf{k}+\mathbf{q}, n\mathbf{k}}(-\mathbf{G}'-\mathbf{q})
         \nonumber \\
         &\times
         \delta\hspace{-1pt}\left(\hbar\omega - \left[\epsilon_{m \mathbf{k} + \mathbf{q}} - \epsilon_{n\mathbf{k}}\right]\right)
  \label{eq:chi0 spectrum}
\end{align}
from which the susceptibility can be obtained via a Hilbert transform,
\begin{align}
   \chi_{0,\mathbf{G} \mathbf{G}'}(\mathbf{q}, \omega)
   = \lim_{\eta\rightarrow 0^+} \int_{-\infty}^{\infty} d\omega'\,
     \frac{S_{0,\mathbf{G} \mathbf{G}'}(\mathbf{q}, \omega^{\prime})}
     {\omega' - \omega + i\eta}.
\end{align}
To converge the Hilbert transform, the spectral function is evaluated on a nonlinear frequency grid
spanning the range of eigenenergy differences included in Eq.~\eqref{eq:chi0 spectrum}.
Although the calculation becomes more memory-intensive as a result, it is usually faster to compute
$\chi_0$ via its spectral function. Furthermore, since $S_0$ is linearly interpolatable, the tetrahedron
method can be employed to improve the convergence with respect to $\mathbf{k}$-points.

%% file: linear-response-TDDFT/dielectric-function.tex
The longitudinal part of the dielectric tensor is related to the susceptibility $\chi$ as
\begin{align}
\label{eq:dielectric_matrix}
\varepsilon^{-1}_{\mathbf{G}\mathbf{G}'}(\mathbf{q},\omega) = \delta_{\mathbf{G}\mathbf{G}'} + v_{c,\mathbf{G}\mathbf{G}'}(\mathbf{q})\chi_{\mathbf{G}\mathbf{G}'}(\mathbf{q},\omega).
\end{align}
From the dielectric matrix, the macroscopic dielectric function, including local field corrections, is given by
\begin{align}
\varepsilon_{M}(\mathbf{q},\omega) =
              \frac{1}{\varepsilon^{-1}_{00}(\mathbf{q},\omega)}
\end{align}
from which it is straightforward to extract the optical absorption spectrum
\begin{align}
\mathrm{ABS} = \mathrm{Im}\:
\varepsilon_{M}(\mathbf{q} \rightarrow 0,\omega)
\end{align}
as well as the electron energy-loss spectrum
\begin{align}
\label{eq:eels}
\mathrm{EELS} = -\mathrm{Im}\:
\frac{1}{\varepsilon_{M}(\mathbf{q},\omega)}.
\end{align}

It is also possible to define a symmetrized version of the dielectric matrix as
\begin{align}
\tilde\varepsilon_{\mathbf{GG}'}(\mathbf{q},\omega) = v^{-1/2}_{\mathbf{G}}(\mathbf{q}) \varepsilon_{\mathbf{GG}'}(\omega, \mathbf{q}) v^{1/2}_{\mathbf{G}'}(\mathbf{q}).
\end{align}
The susceptibility can be evaluated at the RPA level (by setting $V_\text{xc}$ in Eq.~\ref{eq:dyson elec. chi real space}
to zero)
or using one of the XC-kernels implemented in GPAW such as the the local ALDA, the non-local rALDA~\cite{olsen2012extending}, or the bootstrap kernel~\cite{sharma2011}.

The evaluation $\chi_0$ is computationally demanding since it involves an integration over the Brillouin zone (BZ) as well as a summation over occupied and unoccupied states.
The $\mathbf{k}$-point convergence can be increased substantially with the tetrahedron method~\cite{macdonald1979}. Contrary to the simple point integration, where the $\delta$-function in Eq.~\ref{eq:chi0 spectrum} is replaced by a smeared-out Lorentzian, the tetrahedron method utilizes linear interpolation of the eigenvalues and matrix elements.

\begin{figure}[tb]
    \centering
    \includegraphics{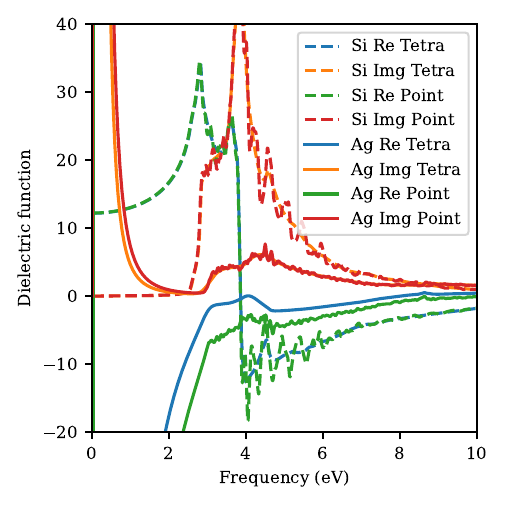}
    \caption{Real and imaginary parts of the dielectric function for Ag (solid) and Si (dashed) using the tetrahedron and point integration methods.}
    \label{fig:df}
\end{figure}

In Fig.~\ref{fig:df} we compare the dielectric function computed using the two different integration methods for two prototypical cases: (semiconducting) Si and (metallic) Ag. Since Si is semiconducting there are no low-energy excitations and consequently the imaginary part of the dielectric function is zero while the real part is flat for low frequencies. The point integration and tetrahedron integration yield the same value for $\omega \rightarrow 0$, but the tetrahedron integration avoids the unphysical oscillations of $\varepsilon$ at higher frequencies, exhibited by the point integration due to the finite $\mathbf{k}$-point sampling. For metals, the $\mathbf{k}$-point convergence is even slower and the difference between the point integration and tetrahedron integration is thus even more pronounced for Ag. By increasing the $\mathbf{k}$-point sampling, the point integration results will eventually approach the results obtained with the tetrahedron method, but at a much higher computational cost.

\subsubsection{Screening in low-dimensional systems}
In a three dimensional (3D) bulk crystal, the
macroscopic dielectric function is related to the macroscopic polarizability $\alpha_M$ of the material as
\begin{align}
\varepsilon_{M}(\mathbf{q},\omega) =
             1+ 4\pi\alpha_M(\mathbf{q},\omega).
\end{align}
Since $\varepsilon_M$ is related to the macroscopic average of the
induced potential, it will depend on the unit cell in
low-dimensional systems and tend to unity when increasing the cell in
the non-periodic direction. In contrast, it is straighforward to
generalize the definition of the polarizability to such that it
measures the induced dipole moment per length or area rather than
volume. The $d$-dimensional polarizability is thus defined as
$\alpha_M^d=\Omega^d\alpha_M$ where $\Omega^d$ is the cell volume with
the periodic directions projected out. For example, for a 2D material
$\Omega^d$ is simply the length of the unit cell in the non-periodic
direction. To improve convergence with respect to the size of the unit
cell, the Coulomb kernel is truncated using the reciprocal-space
method introduced in Ref.~\cite{rozzi2006}. This enables efficient
calculations of dielectric properties of low-dimensional materials
materials~\cite{huser2013dielectric,latini2015excitons}. This does,
however, imply that the polarizability cannot be evaluated directly
from the dielectric constant (which is just one), but one may obtain it directly from the
susceptibility as
\begin{align}
  \alpha_M = -\lim\limits_{\mathbf{q} \to 0}\frac{\Omega^d}{q^2} \chi_{00}(\mathbf{q}).
\end{align}

%% file: acfdt/acfdt.tex
\label{sec:acfdt}
The adiabatic-connection fluctuation-dissipation theorem (ACFDT) is a highly promising approach for constructing accurate correlation functionals with non-local characteristics.
Unlike regular XC-functionals, the ACFDT correlation functionals do not rely on error cancellation between exchange and correlation.
Instead, the ACFDT provides an exact theory for the electronic correlation energy $E_\text{c}$ in terms of the interacting electronic suseptibility, which can be combined with the exact exchange:
\begin{equation}
    E_\text{c} = - \int_0^{\infty} \frac{\mathrm{d} \omega}{2 \pi} \int_0^1 \mathrm{d} \lambda \mathrm{Tr} \left[ v_\text{c} \chi^\lambda \left( w \right) - v_\text{c} \chi^\text{KS} (w) \right].
\end{equation}
The response function can be expressed in terms of the Kohn--Sham response function and the exchange--correlation kernel $f_\text{xc}$
\begin{equation}
\chi^\lambda (\omega) = \chi^\text{KS} (\omega) + \chi^\text{KS} (\omega) \left[ \lambda v_c + f_\text{xc} (\omega) \right] \chi^{\lambda} (\omega).
\end{equation}

The random phase approximation (RPA) can be derived from ACFDT if the XC-kernel $f_\text{xc}$ is neglected. RPA has the strength of capturing non-local correlation effects and provides high accuracy across different bonding types including van der Waals interaction~\cite{Patrick_PRB_2016, Schmidt_JPC_2018, olsen2013random, Olsen_PRL_2011}.

Simple exchange--correlation kernels can also be incorporated into the response function, such as the adiabatic LDA (ALDA) kernel. However, the locality of adiabatic kernels leads to divergent characteristics of the pair-correlation function~\cite{Furche2005}. This issue can be overcome by a renormalization (r) scheme~\cite{olsen2012extending}, which is implemented in GPAW as rALDA. This class of renormalized kernels provides a significantly better description of the short-range correlations and hence also yield highly accurate total correlation energies~\cite{Olsen_PRB_2013_2, Olsen_PRB_2014, Patrick_JCP_2015, Olsen_npj_2019}.

%% file: mag-excitations/mag-excitations.tex
\label{sec:mag_response}
The LR-TDDFT framework described
in Sec.~\ref{sec:LR-TDDFT} can be generalized to include spin degrees of freedom~\cite{Skovhus2021,Buczek2011}.
Using the four-component density as the basic variable, $n^\mu(\mathbf{r})$ (where $\mu\in\{0,x,y,z\}$), one can define the four-component susceptibility tensor:
\begin{equation}
  \chi^{\mu\nu}(\mathbf{r}, \mathbf{r}', t-t') = - \frac{i}{\hbar} \theta(t-t')
      \langle\left[\hat{n}^\mu_0(\mathbf{r},t), \hat{n}^\nu_0(\mathbf{r}',t')\right]\rangle_0.
  \label{eq:four-component kubo formula}
\end{equation}
In a similar fashion to Eq.~\eqref{eq:dyson elec. chi real space}, the many-body $\chi^{\mu\nu}$
can be calculated from the corresponding susceptibility tensor of the Kohn--Sham system,
$\chi^{\mu\nu}_{\mathrm{KS}}$.
In the most general case, the Dyson equation for $\chi^{\mu\nu}$ is a matrix equation in
the $\mu$ and $\nu$ indices, explicitly coupling the charge and spin degrees of freedom.
However, for collinear magnetic systems, the transverse components of the susceptibility
decouple from the rest in the absence of spin--orbit coupling.

\subsubsection{Transverse magnetic susceptibility}

Taking the spins to be polarized along the $z$-direction, the transverse magnetic
susceptibility of collinear nonrelativistic systems may be expressed in terms of
the spin-raising and spin-lowering density operators
$\hat{n}^+(\mathbf{r}) = (\hat{n}^x(\mathbf{r}) + i\hat{n}^y(\mathbf{r}))/2 = \hat{\psi}^\dagger_\uparrow(\mathbf{r}) \hat{\psi}_\downarrow(\mathbf{r})$
and
$\hat{n}^-(\mathbf{r}) = (\hat{n}^x(\mathbf{r}) - i\hat{n}^y(\mathbf{r}))/2 = \hat{\psi}^\dagger_\downarrow(\mathbf{r}) \hat{\psi}_\uparrow(\mathbf{r})$.
In the ALDA, the Dyson equation for $\chi^{+-}$ then becomes a scalar one~\cite{Skovhus2021},
\begin{align}
  \chi^{+-}(\mathbf{r}, \mathbf{r}', \omega) = &\chi^{+-}_{\mathrm{KS}}(\mathbf{r}, \mathbf{r}', \omega)
                                          + \int \mathrm{d}\mathbf{r}_1
                                          \nonumber \\
                                          &\times
                                          \chi^{+-}_{\mathrm{KS}}(\mathbf{r}, \mathbf{r}_1, \omega)
                                          f^{-+}_{\mathrm{LDA}}(\mathbf{r}_1)
                                          \chi^{+-}(\mathbf{r}_1, \mathbf{r}', \omega),
  \label{eq:tms dyson eq}
\end{align}
where the transverse XC-kernel is given by $f^{-+}_{\mathrm{LDA}}=2 W_{\mathrm{LDA}}(\mathbf{r})/n^z(\mathbf{r})$.
The susceptibilities $\chi^{+-}$ and $\chi^{+-}_{\mathrm{KS}}$ are themselves defined
via the Kubo formula \eqref{eq:four-component kubo formula}, where the latter can be
evaluated explicitly in complete analogy to Eq.~\eqref{eq:chi0 real-space}:
\begin{align}
  \chi^{+-}_{\mathrm{KS}}(\mathbf{r}, \mathbf{r}', \omega) = &\lim_{\eta\rightarrow 0^+} \sum_{n\mathbf{k}} \sum_{m\mathbf{k}'}
                                                             \left(f_{n\mathbf{k}\uparrow} - f_{m\mathbf{k}'\downarrow}\right)
                                                             \nonumber \\
                                                             &\times
                                                             \frac{\psi^*_{n\mathbf{k}\uparrow}(\mathbf{r})\psi_{m\mathbf{k}'\downarrow}(\mathbf{r})
                                                                  \psi^*_{m\mathbf{k}'\downarrow}(\mathbf{r}')\psi_{n\mathbf{k}\uparrow}(\mathbf{r}')}
                                                                  {\hbar\omega - (\epsilon_{m\mathbf{k}'\downarrow} - \epsilon_{n\mathbf{k}\uparrow}) + i\hbar\eta}.
  \label{eq:tms chi ks}
\end{align}

In GPAW, the transverse magnetic susceptibility is calculated using a plane-wave
basis as described in Sec. \ref{sec:lr-tddft for periodic systems}, with the notable
exception that no special care is needed to treat the optical limit, since the
Hartree kernel plays no role in the Dyson equation \eqref{eq:tms dyson eq}.
In terms of the temporal representation, it is at the time of this writing only possible
to do a literal evaluation of Eq.~\eqref{eq:tms chi ks} at the frequencies of interest.
For metals, this means that $\eta$ has to be left as a finite broadening parameter,
which has to be carefully chosen depending on the $\mathbf{k}$-point sampling, see~\cite{Skovhus2021}.

\subsubsection{The spectrum of transverse magnetic excitations}

Based on the transverse magnetic susceptibility, one may calculate the corresponding
spectral function~\cite{Skovhus2021}
\begin{align}
  S^{+-}(\mathbf{r},\mathbf{r}',\omega) &= - \frac{1}{2\pi i}
                                           \left[
                                           \chi^{+-}(\mathbf{r},\mathbf{r}',\omega)
                                           -
                                           \chi^{-+}(\mathbf{r}', \mathbf{r}, -\omega)
                                           \right],
\end{align}
which directly governs the energy dissipation in a collinear magnet relating to
induced changes in the spin projection along the $z$-axis $S_z$.
In particular, one can decompose the spectrum into contributions from majority and
minority spin excitations,
$S^{+-}(\mathbf{r},\mathbf{r}',\omega) = A^{+-}(\mathbf{r},\mathbf{r}',\omega) - A^{-+}(\mathbf{r}',\mathbf{r},-\omega)$,
that is, into spectral functions for the excited states where $S_z$ has been lowered
or raised by one unit of spin angular momentum respectively:
\begin{align}
  A^{+-}(\mathbf{r},\mathbf{r}',\omega) = \sum_{\alpha > 0}
                                          &\langle0| \hat{n}^+(\mathbf{r}) |\alpha\rangle
                                          \langle\alpha| \hat{n}^-(\mathbf{r}') |0\rangle
                                          \nonumber \\
                                          &\times
                                          \delta\hspace{-1pt}\left(\hbar\omega - \left[E_\alpha - E_0\right]\right).
\end{align}
Here, $\alpha$ iterates the system eigenstates with $\alpha=0$ denoting the ground state.

\begin{figure}[tb]
    \centering
    \includegraphics{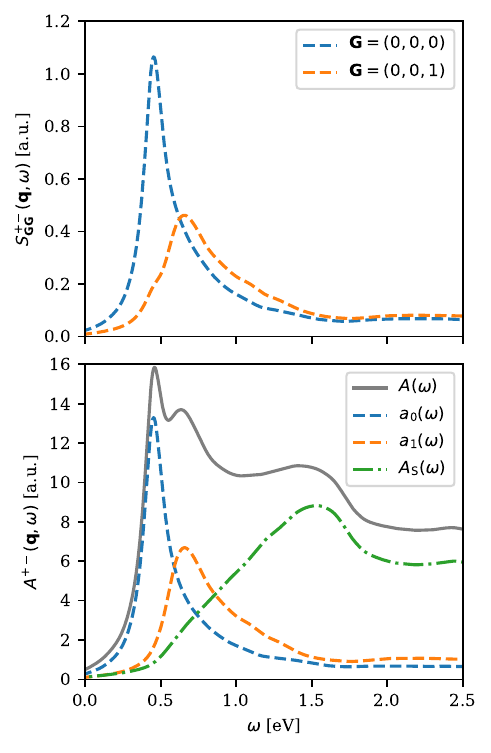}
    \caption{
    Spectrum of transverse magnetic excitations for ferromagnetic hcp-Co evaluated at
    $\mathbf{q}=5\mathbf{q}_{\mathrm{M}}/6$. The spectrum was calculated using
    8 empty-shell bands per atom, a plane-wave cutoff of 800~eV, a $(60, 60, 36)$
    $\mathbf{k}$-point mesh and a spectral broadening of $\eta=50$~meV.
    In the upper panel, the spectrum diagonal is depicted for the 1st and 2nd Brillouin zone out
    of the hexagonal plane, from which the acoustic and optical magnon frequencies can
    be respectively extracted.
    In the lower panel, the spectrum of majority excitations is shown. The full spectral
    weight $A(\omega)$ is calculated as the sum of all positive eigenvalues of $S^{+-}$,
    the acoustic and optical mode lineshapes $a_0(\omega)$ and $a_1(\omega)$ are
    obtained via the two largest eigenvalues (which are significantly larger than the
    rest) and the Stoner spectrum is extracted as the difference
    $A_{\mathrm{S}}(\omega) = A(\omega) - a_0(\omega) - a_1(\omega)$.
    }
    \label{fig:co_magnon_lineshapes}
\end{figure}
For collinear magnetic systems, the spectrum $S^{+-}$ is composed of two types of
excitations: collective spin-wave excitations (referred to as magnons) and excitations
in the Stoner-pair continuum (electron--hole pairs of opposite spin). Since GPAW employs
a plane-wave representation of the spectrum,
$S^{+-}_{\mathbf{G}\mathbf{G}'}(\mathbf{q},\omega)$,
one can directly compare the calculational output to the inelastic neutron scattering
cross-section measured in experiments~\cite{VanHove1954}.
In particular, one can extract the magnon dispersion directly by identifying the position
of the peaks in the spectrum diagonal,
see Fig.~\ref{fig:co_magnon_lineshapes}.
In this way, GPAW allows the user to study various magnon phenomena in magnetic systems
of arbitrary collinear order, such as nonanalytic dispersion effects in itinerant
ferromagnets~\cite{Skovhus2021}, correlation-driven magnetic phase transitions~\cite{Skovhus2022a}
and emergence of distinct collective modes inside the Stoner continuum of an
antiferromagnet~\cite{Skovhus2022b}.

Additionally, one can analyze the spectrum more intricately by extracting the majority
and minority eigenmodes from $S^{+-}$ as the positive and negative eigenvalues
respectively. Contrary to analysis of the plane-wave diagonal, this makes it possible to
completely separate the analysis of each individual magnon lineshape as well as the
many-body Stoner continuum, see Fig.~\ref{fig:co_magnon_lineshapes}.

%% file: mag-force-theorem/mag-force-theorem.tex
Not only can the transverse magnetic susceptibility be used to study magnetic excitations
in a literal sense, but one can also use it to map the spin-degrees of freedom
to a classical Heisenberg model,
\begin{equation}
  E_{\mathrm{H}} = - \frac{1}{2} \sum_{i,j} \sum_{a,b} J_{ij}^{ab}\, \mathbf{u}_{ia} \cdot \mathbf{u}_{jb},
  \label{eq:mft hb model}
\end{equation}
where $i,j$ and $a,b$ denote the indices of the Bravais lattice and the magnetic sublattice,
respectively, $\mathbf{u}_{ia}$ being the direction of spin-polarization of the given
magnetic site. Based on the magnetic force theorem (MFT), the LSDA Heisenberg exchange
parameters $J_{ij}^{ab}$ can be calculated from the reactive part of the static
Kohn--Sham susceptibility~\cite{Durhuus2023} and the effective magnetic field
$B_{\mathrm{xc}}(\mathbf{r}) = \delta E_{\mathrm{xc}}/\delta m(\mathbf{r})$,
using the well-known Liechtenstein MFT formula~\cite{Liechtenstein1987}:
\begin{equation}
J_{ij}^{ab} = - 2 \int_{\Omega_{ia}} \mathrm{d}\mathbf{r} \int_{\Omega_{jb}} \mathrm{d}\mathbf{r}'\,
                B_{\mathrm{xc}}(\mathbf{r}) \chi^{\prime +-}_{\mathrm{KS}}(\mathbf{r}, \mathbf{r}') B_{\mathrm{xc}}(\mathbf{r}').
\end{equation}
Here $\Omega_{ia}$ denotes the site volume, which effectively defines the Heisenberg
model \eqref{eq:mft hb model}.

Using GPAW's plane-wave representation of $\chi^{\prime +-}_{\mathrm{KS}}$, one can directly
compute the lattice Fourier-transformed exchange parameters~\cite{Durhuus2023},
\begin{subequations}
\begin{align}
  \bar{J}^{ab}(\mathbf{q}) &= \sum_i J_{0i}^{ab} e^{i\mathbf{q}\cdot\mathbf{R}_i}
                           \\
                           &= -\frac{2}{\Omega_{\mathrm{cell}}}
                           B_{\mathrm{xc}}^\dagger K^{a\dagger}(\mathbf{q})
                           \chi^{\prime +-}_{\mathrm{KS}}(\mathbf{q})
                           K^b(\mathbf{q}) B_{\mathrm{xc}},
\end{align}
\end{subequations}
where the right-hand side of the second equality is written in a plane-wave basis
and $K_a$ denotes the sublattice site-kernel:
\begin{equation}
  K^a_{\mathbf{G}\mathbf{G}'}(\mathbf{q}) = \frac{1}{\Omega_{\mathrm{cell}}}
  \int_{\Omega_{0a}} \mathrm{}\mathrm{d}\mathbf{r}\, e^{-i(\mathbf{G} - \mathbf{G}' + \mathbf{q})\cdot\mathbf{r}}.
\end{equation}
Since an {\it a priori} definition for the magnetic site volumes does not exist,
GPAW supplies functionality to calculate exchange parameters based on spherical,
cylindrical and/or parallelepipedic site configurations of variable size.

\begin{figure*}[tb]
    \centering
    \includegraphics{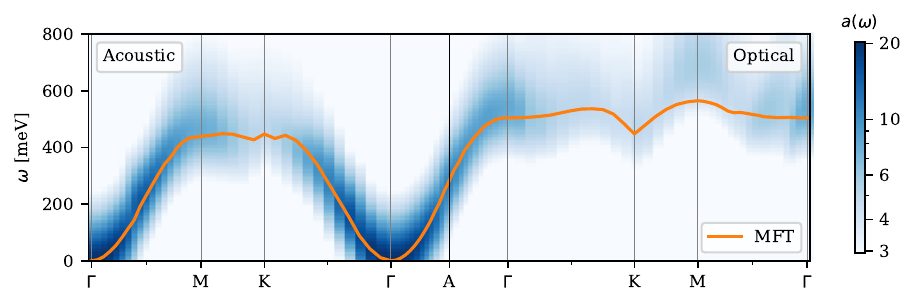}
    \caption{
      Magnon spectrum of ferromagnetic hcp-Co calculated using ALDA LR-TDDFT
      (shown as a heat map), compared to the spin-wave dispersion of Liechtenstein
      MFT.
      The acoustic magnon mode $a_0(\omega)$ is shown to the left of the A-point,
      while the optical magnon mode $a_1(\omega)$ is shown to the right.
      Please note that the two modes are degenerate in both the $\mathrm{A}$ and
      $\mathrm{K}$ high-symmetry points.
      The calculations were carried out using 8 empty-shell bands per atom,
      a plane-wave cutoff of 800~eV and a $(36, 36, 24)$ $\mathbf{k}$-point mesh.
      A finite value of $\eta=100$~meV was used to broaden the ALDA spectrum.
      For the MFT calculations, the dispersion was calculated using linear
      spin-wave theory based on a Heisenberg model of closed-packed spherical
      sites centered at each of the Co atoms.
    }
    \label{fig:co_magnon_dispersion}
\end{figure*}
Upon calculation of the exchange parameters $\bar{J}^{ab}(\mathbf{q})$, it is
straight-forward to compute the magnon dispersion within the classical Heisenberg
model using linear spin-wave theory and to estimate thermal quantities such as
the Curie temperature, see e.g.~\cite{Durhuus2023}.
In Fig.~\ref{fig:co_magnon_dispersion} the MFT magnon dispersion of hcp-Co is
compared to majority magnon spectrum calculated within LR-TDDFT. For Co, the two
are in excellent agreement except for the dispersion of the optical magnon mode
along the $\mathrm{K}-\mathrm{M}-\Gamma$ high-symmetry path, where MFT
underestimates the magnon frequency and neglects the fine structure of the spectrum.
The fine structure of the ALDA spectrum appears due to the overlap between the magnon
mode and the Stoner continuum. This gives rise to so-called Kohn anomalies
(nonanalytical points in the magnon dispersion), which is a trademark of itinerant
electron magnetism. Since the itinerancy is largely ignored in a localized spin model
such as \eqref{eq:mft hb model}, one cannot generally expect to capture such effects.

%% file: gw/gw.tex
GPAW supports standard G$_0$W$_0$ quasiparticle (QP) calculations based
on a first-order perturbative treatment of the linearized QP
equation~\cite{golze2019gw}
\begin{align}\label{eq:QPE}
    E_{n\mathbf{k}\sigma}^\text{QP} =  \varepsilon_{n\mathbf{k}\sigma} + Z_{n\mathbf{k}\sigma}\langle \psi_{n\mathbf{k}\sigma}|\Sigma_{\mathrm{GW}}(\varepsilon_{n\mathbf{k}\sigma})+v_\text{x}-v_
    \text{xc}|\psi_{n\mathbf{k}\sigma}\rangle,
\end{align}
where $\varepsilon_{n\mathbf{k}\sigma}$ and $\psi_{n\mathbf{k}\sigma}$ are Kohn--Sham
eigenvalues and wave functions, and $v_\text{xc}$ and $v_\text{x}$ are the local
XC-potential and nonlocal exchange potential, respectively.
$\Sigma_{\mathrm{GW}}$ is the (dynamical part of) the GW self-energy
whose frequency-dependence is accounted to first order by the
renormalization factor
$Z_{n\mathbf{k}\sigma}=(1-\mathrm{Re}\:\Sigma_{\mathrm{GW}}'(\varepsilon_{n\mathbf{k}\sigma}))^{-1}$.
As indicated by the $\sigma$-index, spin-polarized G$_0$W$_0$
calculations are supported.

The GW self-energy is calculated in a plane-wave basis using full
frequency integration along the real axis to evaluate the convolution
between $G$ and $W$~\cite{huser2013quasiparticle}. Compared to
alternative schemes employing contour deformation techniques or
analytical continuation~\cite{rojas1995space,
friedrich2010efficient,duchemin2020robust}, this approach is time-
consuming but numerically accurate and can provide the full spectral
function. A highly efficient and accurate evaluation of the self-energy
based on a multipole expansion of the screened interaction,
$W$~\cite{leon2021frequency}, is currently being implemented, and a GPU
version of the full GW code is under development.

An important technical issue concerns the treatment of the head and
wings of $W$ ($\mathbf{G}=0$ and/or $\mathbf{G}'=0$, respectively) in
the $\mathbf{q} \rightarrow 0$ limit.
The divergence of the Coulomb interaction appears both in the
evaluation of the inverse dielectric matrix $\varepsilon^{-1}_{\mathbf{GG}'}$
and in the subsequent evaluation of the screened interaction
\begin{align*}
W_{\mathbf{GG}'}(\mathbf{q}, \omega)=\sum_{\mathbf{G}''}
\varepsilon^{-1}_{\mathbf{GG}''}(\mathbf{q}, \omega)v_{c,\mathbf{G}''\mathbf{G}'}(\mathbf{q}).
\end{align*}
For 3D bulk crystals, GPAW obtains these components by evaluating
$\varepsilon^{-1}_{\mathbf{GG}'}$ on a dense $\mathbf{k}$-grid centered at $\mathbf{k}=0$ while
$v_{c,\mathbf{G}\mathbf{G}'}$ can be integrated numerically or analytically around the $\Gamma$-point.

For low-dimensional structures, in particular atomically thin 2D
semiconductors, GPAW can use a truncated Coulomb interaction to avoid
interactions between periodical images when evaluating
$W$~\cite{rozzi2006}. It has been shown that the use of a truncated
Coulomb kernel leads to slower $\mathbf{k}$-point
convergence~\cite{huser2013dielectric}. To mitigate this drawback, a
special 2D treatment of $W(q)$ at $q=0$, can be applied to significantly improve the $\mathbf{k}$-point
convergence~\cite{rasmussen2016efficient}.
A detailed account of the GW implementation in GPAW can be found in
Ref.~\cite{huser2013quasiparticle}.

\begin{figure}[tb]
    \centering
    \includegraphics{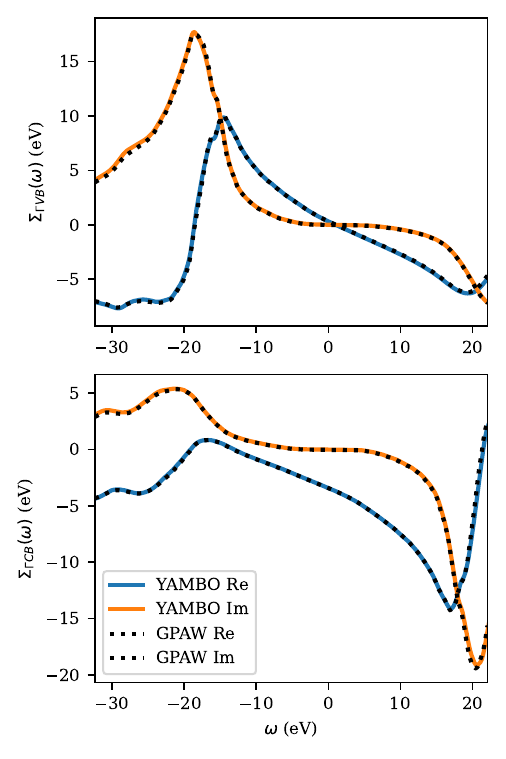}
    \caption{
The real (Re) and imaginary (Im) parts of the self-energy matrix elements
at the Gamma point for valence (top) and conduction (bottom) bands evaluated
with Yambo and GPAW. Both codes are using full
frequency integration with a broadening of 0.1~eV. Yambo is using norm
conserving pseudo potentials, and GPAW its standard PAW setup. The k-point
grid was $12\times 12\times 12$, the plane wave cutoff was 200~eV, and the
number of bands 200 for both codes. The results are virtually indistinguishable.
}
    \label{fig:gw}
\end{figure}

Figure \ref{fig:gw} shows two matrix elements of the dynamical GW
self-energy for the valence and conduction band states at the Gamma point.
As can be seen, the agreement with
the corresponding quantities obtained with the Yambo GW
code~\cite{marini2009yambo} is striking.

%% file: bse/bse.tex
In addition to the LR-TDDFT discussed in Sec. \ref{sec:LR-TDDFT}, the interacting response function may be approximated by solving the Bethe--Salpeter equation (BSE)~\cite{onida2002electronic}. In particular, for a certain wave vector $\mathbf{q}$, one may obtain the two-particle excitations by diagonalizing the Hamiltonian
\begin{align}\label{eq:H_bse}
H_{\substack{\mathbf{k}_2m_1m_2\\\mathbf{k}_4m_3m_4}}(\mathbf{q})&=\delta_{m_1m_3}\delta_{m_2m_4}\delta_{\mathbf{k}_2\mathbf{k}_4}(\varepsilon_{m_1\mathbf{k}_2+\mathbf{q}}-\varepsilon_{m_2\mathbf{k}_2})\\
&-(f_{m_1\mathbf{k}_2+\mathbf{q}}-f_{m_2\mathbf{k}_2})K_{m_2\mathbf{k}_2m_1\mathbf{k}_2+\mathbf{q}m_3\mathbf{k}_4+\mathbf{q}m_4\mathbf{k}_4}\notag,
\end{align}
where $\varepsilon_{\mathbf{k}m}$ are the Kohn--Sham eigenvalues and $f_{\mathbf{k}m}$ the associated occupation numbers. The kernel is defined by $K=v_\mathrm{c}-W$ with
\begin{align}
v_c=&\langle m_2\mathbf{k}_2,m_3\mathbf{k}_4+\mathbf{q}|\hat v_\mathrm{c}|m_1\mathbf{k}_2+\mathbf{q},m_4\mathbf{k}_4\rangle\\
W=&\langle m_2\mathbf{k}_2,m_3\mathbf{k}_4+\mathbf{q}|\hat W|m_4\mathbf{k}_4,m_1\mathbf{k}_2+\mathbf{q}\rangle,
\end{align}
where $\hat W=\varepsilon^{-1}v_\mathrm{c}$ is the static screened Coulomb interaction. The matrix elements of the kernel are evaluated in a plane-wave basis where they are easily expressed in terms of the pair densities \eqref{eq:pair_density} and the reciprocal-space representation of the dielectric matrix \eqref{eq:dielectric_matrix}.

In the Tamm--Dancoff approximation, one only includes states with $\varepsilon_{m_1\mathbf{k}_2+\mathbf{q}}-\varepsilon_{m_2\mathbf{k}_2}>0$ in Eq.~\eqref{eq:H_bse} and the BSE Hamiltonian becomes Hermitian~\cite{onida2002electronic}. The interacting retarded response function may then be written as
\begin{align}\label{eq:susc_bse}
\chi^{\mathbf{G}\mathbf{G}'}(\mathbf{q},\omega)=\frac{1}{\Omega}\sum_{\lambda}\frac{B_{\lambda}(\mathbf{q},\mathbf{G})C_{\lambda}^*(\mathbf{q},\mathbf{G}')}{\omega-E_\lambda(\mathbf{q})+i\eta},
\end{align}
where
\begin{align}
&B_{\lambda}(\mathbf{q},\mathbf{G})=\sum_{\mathbf{k}m_1m_2}n_{m_2\mathbf{k}m_1\mathbf{k}+\mathbf{q}}(\mathbf{G}+\mathbf{q})A^\lambda_{m_1m_2\mathbf{k}}(\mathbf{q}),\\
&C_{\lambda}(\mathbf{q},\mathbf{G})=\sum_{\mathbf{k}m_1m_2}n_{m_2\mathbf{k}m_1\mathbf{k}+\mathbf{q}}(\mathbf{G}+\mathbf{q})A^{\lambda}_{m_1m_2\mathbf{k}}(\mathbf{q})\notag\\
&\qquad\qquad\qquad\qquad\times(f_{m_1\mathbf{k+q}}-f_{m_2\mathbf{k}}),
\end{align}
and $E_\lambda(\mathbf{q})$ denotes the eigenvalue of the Hamiltonian \eqref{eq:H_bse} corresponding to the eigenvector $A^{\lambda}_{m_1m_2\mathbf{k}}(\mathbf{q})$.

In GPAW, the construction of the BSE Hamiltonian \eqref{eq:H_bse} proceeds in two steps~\cite{yan2012optical}. First, the static screened interaction is calculated at all inequivalent $\mathbf{q}$-points in a plane-wave basis and the Kernel is then subsequently expressed in a basis of two-particle KS states. The first step is efficiently parallelized over either states or $\mathbf{k}$-points and the second step is parallelized over pair densities. The Hamiltonian elements are thus distributed over all CPUs and the diagonalization is carried out using ScaLAPACK such that the full Hamiltonian is never collected on a single CPU. The dimension of the BSE Hamiltonian and memory requirements are, therefore, only limited by the number of CPUs used for the calculation. We note that the implementation is not limited to the Tamm--Dancoff approximation, but calculations become more demanding without it. The response function may be calculated for spin-paired as well as spin-polarized systems and spin--orbit coupling can be included non-selfconsistently~\cite{Olsen2016,Gjerding2021}. In low-dimensional systems it is important to eliminate the spurious screening from periodic images of the structure, which is accomplished with the truncated Coulomb interaction of Ref.~\cite{rozzi2006}.

\begin{figure}[tb]
    \centering
    \includegraphics{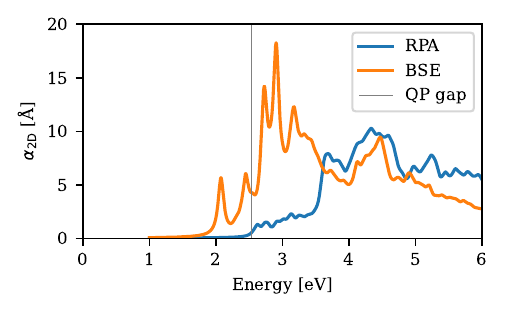}
    \caption{2D polarizability of WS$_2$ calculated from the BSE and the RPA. For this calculation, we included spin--orbit coupling and used the 2D Coulomb truncation to eliminate screening from periodic images.
A gamma-centered uniform $\mathbf{k}$-point grid of $48\times48$ was applied and 8 valence states and 8 conduction states (shifted by 1 eV to match the GW band gap~\cite{Gjerding2021}) was included in the Tamm--Dancoff approximation. This yields a BSE Hamiltonian of size $N\times N$ with $N=147456$, which is easily diagonalized with ScaLAPACK on 240 CPUs.}
    \label{fig:bse}
\end{figure}

The most important application of BSE is arguably the calculation of optical absorption spectra of solids where BSE provides an accurate account of the excitonic effects that are not captured by semi-local approximations for $K_\mathrm{Hxc}$ \eqref{eq: Hxc kernel}. In 2D systems, the excitonic effects are particularly pronounced due to inefficient screening~\cite{huser2013dielectric,Olsen2016} and in Fig.~\ref{fig:bse} we show the 2D polarizability of WS$_2$ calculated from BSE with $\mathbf{q}=\mathbf{0}$. Comparing with Fig.~\ref{fig:WS2_bands}, it is observed that the absorption edge is expected to be located at the $K$ point, where spin--orbit coupling splits the highest valence band by 0.45 eV. This splitting is seen as two excitonic peaks below the band gap, which is interpreted as distinct excitons originating from the highest and next-highest valence bands (the splitting of the lowest conduction band is negligible in this regard). For comparison, we also show the RPA polarizabilty obtained with the BSE module by neglecting the screened interaction in the kernel and this shows the expected absorption edge at the band gap. This yields identical results to the Dyson equation approach of Sec. \ref{sec:LR-TDDFT}, but has the advantage that the eigenvalues and weights of Eq.~\eqref{eq:susc_bse} are obtained directly such that the artificial broadening $\eta$ may be varied without additional computational cost. The eigenstate decomposition also allows one to access ``dark states'' and the BSE calculation reveals two (one for each valley) triplet-like excitons that are situated 70~meV below the lowest bright exciton in Fig.~\ref{fig:bse}.

In addition to optical properties, GPAW allows for solving the BSE at finite $\mathbf{q}$, which can used to obtain plasmon dispersion relations from the EELS \eqref{eq:eels} and magnon dispersions from the transverse magnetic susceptibility of Sec. \ref{sec:mag_response}~\cite{Olsen2021}.

%% file: electron-phonon/elph.tex
\label{sec:elph}
The electron-phonon coupling
is the origin of several important materials properties, ranging from electrical and thermal conductivity to superconductivity. In addition it provides access to the deformation potential, which can be used to obtain transport properties for electrons and holes in solids~\cite{li2021deformation}.

The first-order electron--phonon coupling matrix $g_{mn}^\nu(\mathbf{k}, \mathbf{q})$ measures the strength of the coupling between a phonon branch $\nu$ with wave vector $\mathbf{q}$ and frequency $\omega_\nu$ and the electronic states $m(\mathbf{k}+ \mathbf{q})$ and $n(\mathbf{k})$~\cite{liu1996linear, RevModPhys.89.015003}:
\begin{equation}
g_{mn}^\nu(\mathbf{k}, \mathbf{q}) = \sqrt{  \frac{\hbar}{2 m_0 \omega_\nu}} M_{mn}^\nu(\mathbf{k}, \mathbf{q})
\end{equation}
with
\begin{equation}
	M_{mn}^\nu(\mathbf{k}, \mathbf{q}) = \langle \psi_{m \mathbf{k}+ \mathbf{q}}  \vert \nabla_u v^\text{KS} \cdot \mathbf{e}_\nu \vert \psi_{n\mathbf{k}} \rangle.
\end{equation}
Here $m_0$ is the sum of the masses of all the atoms in the unit cell, $\nabla_u$ denotes the gradient with respect to atomic displacements and $\mathbf{e}_\nu$ projects the gradient onto the direction of the phonon displacements. In the case of the three translational modes at $\vert \mathbf{q}\vert = 0 $, the matrix elements $g_{mn}^\nu$ vanishes as a consequence of the acoustic sum rule~\cite{RevModPhys.89.015003}.

In GPAW, $\nabla_u v^\text{KS}(\mathbf{r})$ is determined using a finite-difference method with a super cell description of the system. This step can be performed in any of the wave function representations available in GPAW. The derivative is then projected onto a set of atomic orbitals from an LCAO basis $\phi_{NM}$, where $N$ denotes the cell index and $M$ the orbital index:
\begin{eqnarray}
	\mathbf{g}_{\substack{N M\\ N^\prime M^\prime}}^{sc} = FT\left[ \langle \phi_{NM}(\mathbf{k}) \vert \nabla_u v^\text{KS} \vert \phi_{N^\prime M^\prime}(\mathbf{k}) \rangle \right].
\end{eqnarray}
The Fourier transform from the Bloch to the real space representation makes it possible to compute $ M_{mn}^\nu$ for arbitrary $\mathbf{q}$. Finally, the electron--phonon coupling matrix is obtained by projecting the matrix corresponding to the supercell into the primitive unit cell bands $m, n$ and phonon modes $\nu$:
\begin{equation}\label{eq:elph}
	M_{mn}^\nu(\mathbf{k}, \mathbf{q}) = \sum_{\substack{N M\\ N^\prime M^\prime}} C_{mM}^{*} C_{nM^\prime} \mathbf{g}_{\substack{N M\\ N^\prime M^\prime}}^{sc} \cdot \mathbf{u}_{q \nu} e^{i [(\mathbf{k}+\mathbf{q})\cdot \mathbf{R}_N - \mathbf{k}\cdot \mathbf{R}_N^\prime]},
\end{equation}
where $C_{nM}$ are the LCAO coefficients and $\mathbf{u}_{q \nu}$ are mass-scaled phonon displacement vectors.

%% file: raman/raman.tex
The Raman effect describes inelastic light scattering, where
vibrational modes are excited within the material.
Resonant and non-resonant Raman spectra of finite systems
such as molecules can be calculated in
various approximations~\cite{walter_ab_2020} using the corresponding interfaces
in ASE. The Stokes Raman intensity is then written as
\begin{equation}
	I(\omega) = I_0 \sum_\nu \frac{n_\nu+1}{\omega_\nu} \vert
	\sum_{\alpha, \beta} u_\mathrm{in}^\alpha R_{\alpha \beta}^\nu u_\mathrm{out}^\beta
	\vert^2 \delta(\omega-\omega_\nu),
\end{equation}
where $\nu$ denotes phonon mode at $\mathbf{q}=0$ with frequency of $\omega_\nu$ and $n_\nu$ is the corresponding Bose-Einstein distribution. Furthermore, $u_\mathrm{in}^\alpha$ and $u_\mathrm{out}^\beta$ are the polarization vectors of the incoming and outgoing light, and $R_{\alpha \beta}^\nu$ denotes the Raman tensor for phonon mode $\nu$.

The predominant approach for calculating $R_{\alpha \beta}^\nu$ involves the use of the Kramers-Heisenberg-Dirac (KHD) method. Within the KHD framework, the Raman tensor is determined by taking the derivative (utilizing a finite-difference method) of the electric polarizability concerning the vibrational normal modes. Alternatively, one can employ time-dependent third-order perturbation theory to compute Raman tensors. These two approaches are equivalent when local field effect are negligible~\cite{taghizadeh2020library}. However, each approach comes with its own set of advantages and drawbacks. The KHD method is computationally more efficient but is limited to computing first-order Raman processes. The perturbative approach can be extended to higher-order Raman processes involving multiple phonons, but it is more computationally demanding, necessitating a greater number of bands and a finer k-mesh grid to achieve convergence. The perturbative approach has been implemented in GPAW and is elaborated below while the KHD method has been implemented in the ASR package~\cite{gjerding2021atomic}, utilizing GPAW as the computational backend.

In the perturbative approach, the Raman tensor $R_{\alpha \beta}^\nu$ is given by~\cite{jorio2011raman, taghizadeh2020library}
\begin{equation}
	\begin{aligned}
		R_{\alpha \beta}^\nu \equiv \sum_{ijmn \mathbf{k}}
		\left[
                  \frac{p_{ij}^\alpha (M_{jm}^\nu \delta_{in} - M_{ni}^\nu \delta_{jm}) p_{mn}^\beta}{(\hbar \omega_\mathrm{in}-\varepsilon_{ji})(\hbar \omega_\mathrm{out}-\varepsilon_{mn})} \right.\\
		+ \frac{p_{ij}^\alpha (p_{jm}^\beta \delta_{in} - p_{ni}^\beta \delta_{jm})M_{mn}^\nu}{(\hbar \omega_\mathrm{in}-\varepsilon_{ji})(\hbar \omega_{\nu}-\varepsilon_{mn})}  \\
		+ \frac{p_{ij}^\beta (M_{jm}^\nu \delta_{in} - M_{ni}^\nu \delta_{jm})p_{mn}^\alpha}{(-\hbar \omega_\mathrm{out}-\varepsilon_{ji})(-\hbar \omega_\mathrm{in}-\varepsilon_{mn})}  \\
		+ \frac{p_{ij}^\beta (p_{jm}^\alpha \delta_{in} - p_{ni}^\alpha \delta_{jm})M_{mn}^\nu}{(-\hbar \omega_\mathrm{out}-\varepsilon_{ji})(\hbar \omega_{\nu}-\varepsilon_{mn})}  \\
		+ \frac{M_{ij}^\nu (p_{jm}^\alpha \delta_{in} - p_{ni}^\alpha \delta_{jm})p_{mn}^\beta}{(-\hbar \omega_{\nu}-\varepsilon_{ji})(\hbar \omega_\mathrm{out}-\varepsilon_{mn})}  \\
		+ \left. \frac{M_{ij}^\nu (p_{jm}^\beta \delta_{in} - p_{ni}^\beta \delta_{jm})p_{mn}^\alpha}{(-\hbar \omega_{\nu}-\varepsilon_{ji})(-\hbar \omega_\mathrm{in}-\varepsilon_{mn})}
		\right] \\
		\times f_i(1-f_j)f_n(1-f_m),
	\end{aligned}
\end{equation}
where the first term is referred to as the resonant part and the remaining
terms represent different time orderings of the interaction in terms of Feynman diagrams. $p_{nm}^\alpha=\langle \psi_{n \mathbf k}| \hat p^\alpha| \psi_{m \mathbf k} \rangle$
is the momentum matrix element between electronic bands $m$ and $n$, with transition energy $\varepsilon_{nm} = E_n - E_m$ in polarization direction $\alpha$ and $M_{nm}^\nu$ is the electron--phonon coupling strength in the optical limit $\vert \mathbf{q}\vert = 0 $ as defined in Eq.~\eqref{eq:elph}.

\begin{figure}[tb]
\centering
\includegraphics{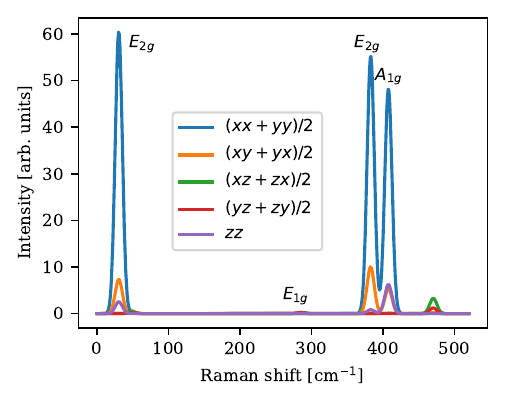}
\caption{
Polarization-resolved Raman spectrum of bulk MoS$_2$ in the 2H phase at
$\omega= 488$~nm. Phonons and potential changes were
computed using a 700 eV plane wave cutoff and $2\times 2\times 2$ $\mathbf{k}$-point
mesh in a $3\times 3\times 2$ supercell.
Each peak has been labeled according to its irreducible representation.}
\label{fig:raman}
\end{figure}

Fig.~\ref{fig:raman} shows the polarization-resolved Raman spectrum of bulk MoS$_2$ in the 2H phase as computed with a laser frequency of $\omega_\mathrm{in}=488$~nm. This example uses only the resonant term, as the other contributions are small in this case.
It's worth noting that we have conducted a comparison of the calculated spectra using the ASR package and observed a high level of agreement for several materials, for example MoS\textsubscript{2}. The results obtained from both methods closely align with each other in terms of peak positions and dominant peaks and minor disagreements between the two methods can be attributed to differences in implementation details and the distinct approximations employed by each. Specifically, within the ASR package, we utilized the phonopy package~\cite{TogoJPCM23} to compute phonon frequencies and eigenvectors, whereas in GPAW, we directly computed phonon frequencies and eigenvectors using ASE's phonon module. Furthermore, in the ASR implementation, we rigorously enforced the symmetry of the polarizability tensor and the ASR results therefore typically exhibit a more accurate adherence to the required symmetry of the Raman tensor compared to the GPAW implementation.

%% file: nlopt/nlopt.tex
The nonlinear optical response of materials can be obtained by going beyond first order perturbation theory. Presently, the GPAW implementation is restricted to second-order response within the dipole approximation and without inclusion of local field effects. We apply the independent particle approximation, which cannot capture collective behavior such as excitonic effects~\cite{taghizadeh2021two}. A spatially homogeneous incident electric field can be written in terms of its Fourier components as
\begin{equation}
    \bm{\mathcal{E}}(t) = \sum_{\alpha,\omega_1} \mathcal{E}_\alpha(\omega_1) \mathbf{e}_\alpha e^{-i\omega_1 t}  \, ,
\end{equation}
where $\omega_1$ runs over positive and negative frequencies, $\mathbf{e}_\alpha$ denotes the unit vector along the $\alpha$-direction, and $\mathcal{E}_\alpha(\omega_1)$ is the electric field at frequency $\omega_1$.
The induced quadratic polarization density $\mathcal{P}_\gamma^{(2)}(t)$ can then be expressed as
\begin{align}
    \mathcal{P}_\gamma^{(2)}(t) = &\epsilon_0 \sum_{\omega_1\omega_2} \sum_{\alpha\beta} \chi_{\gamma\alpha\beta}^{(2)}(\omega_1,\omega_2) \nonumber \\
    & \qquad \qquad \times \mathcal{E}_\alpha(\omega_1) \mathcal{E}_\beta(\omega_2) e^{-i(\omega_1+\omega_2) t} \, ,
\end{align}
where $\chi_{\gamma\alpha\beta}^{(2)}$ is the rank-3 quadratic susceptibility tensor. Due to intrinsic permutation symmetry, i.e. $\chi_{\gamma\alpha\beta}^{(2)}(\omega_1,\omega_2)=\chi_{\gamma\beta\alpha}^{(2)}(\omega_2,\omega_1)$, $\chi_{\gamma\alpha\beta}^{(2)}$ has at most 18 independent elements which may be further reduced by the Neumann principle and point group symmetries~\cite{boyd_nonlinear_2008}. We note that the corresponding quadratic conductivity tensor is readily derived from $\chi_{\gamma\alpha\beta}^{(2)}$ as $\sigma_{\gamma\alpha\beta}^{(2)}(\omega_1,\omega_2) = -i\epsilon_0 \chi_{\gamma\alpha\beta}^{(2)}(\omega_1,\omega_2) (\omega_1+\omega_2)$ due to relationship between current density and polarization density~\cite{taghizadeh_linear_2017}.

Among the various response functions that can be calculated from $\chi_{\gamma\alpha\beta}^{(2)}(\omega_1,\omega_2)$, we have implemented second-harmonics generation (SHG) and the shift current tensor. The implementation currently requires time-reversal symmetry, which limits the application to non-magnetic systems.
For SHG, the susceptibility tensor is separated into a pure interband term $\chi_{\gamma\alpha\beta}^{(2e)}$ and a mixed term $\chi_{\gamma\alpha\beta}^{(2i)}$ that read
\begin{align}
    \label{eq:chiinter}
     & \chi_{\gamma\alpha\beta}^{(2e)} \equiv C_0 \sideset{}{'}\sum_{\mathbf{k},nml} \dfrac{\mathrm{Re} \{ r_{nm}^\gamma (r_{ml}^\alpha r_{ln}^\beta + r_{ml}^\beta r_{ln}^\alpha) \} }{2(\varepsilon_{ln}-\varepsilon_{ml})} \nonumber                \\
     & \qquad \times \bigg( \frac{2f_{nm}}{2\hbar\omega-\varepsilon_{mn}} - \frac{f_{nl}}{\hbar\omega-\varepsilon_{ln}} + \frac{f_{ml}}{\hbar\omega-\varepsilon_{ml}} \bigg) \, ,                                                              \\
    \label{eq:chiintra}
     & \chi_{\gamma\alpha\beta}^{(2i)}  \equiv iC_0 \sideset{}{'}\sum_{\mathbf{k},nm} f_{nm} \Bigg[ \frac{ 2\mathrm{Im} \{ r_{nm}^\gamma (r_{mn;\beta}^\alpha + r_{mn;\alpha}^\beta ) \}}{ \varepsilon_{mn}(2\hbar\omega-\varepsilon_{mn}) } \nonumber \\
     & \qquad +
        \frac{ \mathrm{Im} \{r_{mn}^\alpha r_{nm;\beta}^\gamma + r_{mn}^\beta r_{nm;\alpha}^\gamma \} }{ \varepsilon_{mn}(\hbar\omega-\varepsilon_{mn}) } \nonumber                                                                                    \\
     & \qquad +
        \frac{ \mathrm{Im} \{ r_{nm}^\gamma ( r_{mn}^\alpha \Delta_{mn}^\beta + r_{mn}^\beta \Delta_{mn}^\alpha  ) \} }{ \varepsilon_{mn}^2 } \nonumber                                                                                                \\
     & \qquad \qquad \qquad \times \bigg( \frac{1}{\hbar\omega-\varepsilon_{mn}} - \frac{4}{2\hbar\omega-\varepsilon_{mn}} \bigg) \nonumber                                                                                                    \\
     & \qquad -
        \frac{ \mathrm{Im} \{ r_{mn}^\alpha r_{nm;\gamma}^\beta + r_{mn}^\beta r_{nm;\gamma}^\alpha \} }{ 2\varepsilon_{mn}(\hbar\omega-\varepsilon_{mn}) } \Bigg] \, .
\end{align}
Here, $C_0\equiv e^3/(2\epsilon_0V)$ where $V$ is the crystal volume, $\Delta_{mn}^\alpha\equiv (p_{mm}^\alpha-p_{nn}^\alpha)/m$ denotes the velocity difference between bands $n$ and $m$, and $r_{nm}^\alpha$ is the interband ($n \neq m$) position matrix element, obtained from $im r_{nm}^\alpha=\hbar p_{nm}^\alpha / \varepsilon_{nm}$. All energies, occupations and matrix elements in the preceding expressions depend on the $\mathbf{k}$-vector. Also, the summation over $\mathbf{k}$ implies an integral over the first BZ, i.e. $(2\pi)^3 \sum_\mathbf{k} \rightarrow V \int_\mathrm{BZ} \mathrm{d}^3{\mathbf{k}}$.
The primed summation signs indicate omission of terms with two or more identical indices.
Finally, the generalized derivative $r_{nm;\alpha}^\beta$ (for $n \neq m$) are evaluated from the sum rule~\cite{aversa_nonlinear_1995}
\begin{align}
    \label{eq:sumrule}
    r_{nm;\alpha}^\beta & = \hbar \frac{r_{nm}^\alpha \Delta_{mn}^\beta + r_{nm}^\beta \Delta_{mn}^\alpha }{\varepsilon_{nm}}  \nonumber \\ & +\frac{i\hbar}{\varepsilon_{nm}} \sum_{l\neq n,m} (\varepsilon_{lm} r_{nl}^\alpha r_{lm}^\beta - \varepsilon_{nl} r_{nl}^\beta r_{lm}^\alpha) \, .
\end{align}
Here, infinite sums have been substituted with finite sums over a limited, yet sizable, set of bands. It's important to emphasize that both sides of the equation are dependent on the $\mathbf{k}$-vector, and the summation on the right-hand side pertains exclusively to bands. It should be mentioned that another implementation for the quadratic susceptibility tensor in the velocity gauge is also available in the code, but is not documented here. For sufficiently many bands, the results of the two implementations are identical~\cite{taghizadeh_linear_2017}.

Regarding the shift current, where a DC current is induced in response to an incident AC field, one needs to compute the quadratic conductivity tensor $\sigma_{\gamma\alpha\beta}^{(2)}$:
\begin{align}
    \label{eq:shiftcurrent}
    \sigma_{\gamma\alpha\beta}^{(2)} & = \frac{\pi e^3}{\hbar^2 V} \sideset{}{'}\sum_{\mathbf{k},nm} f_{nm} \mathrm{Im} \{   r_{mn}^\alpha r_{nm;\gamma}^\beta + r_{mn}^\beta r_{nm;\gamma}^\beta \} \nonumber \\
                                     & \qquad \qquad \times \delta(\hbar\omega-\varepsilon_{mn}) \, .
\end{align}
In practice, the delta function is replaced by a Lorentzian with a finite broadeining $\eta$.
To avoid numerical instabilities, the implementation of Eqs.~\eqref{eq:chiinter}-\eqref{eq:shiftcurrent} uses a tolerance, such that terms are neglected if the associated energy differences or differences in Fermi levels are smaller than a tolerance. The default value of the tolerance is $10^{-6}$~eV, and $10^{-4}$ for the energy and Fermi level differences, respectively.

%% file: rt-tddft/rt-tddft.tex
\label{sec:rttddft}
\begin{figure}
    \centering
    \includegraphics{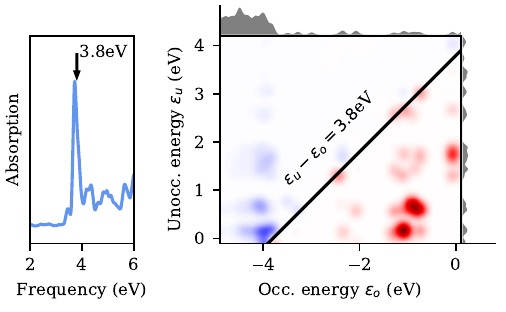}
    \caption{Photoabsorption spectrum for a Ag$_{147}$ icosahedral nanoparticle and the transition contribution map at 3.8\:eV. The map reveals that transitions between KS states near the Fermi level contribute constructively to the plasmon resonance, while transitions from occupied states at the $d$-band edge contribute destructively (screening).
    }
    \label{fig:tddft:tcm}
\end{figure}

The real-time TDDFT (RT-TDDFT) scheme, also known as time-propagation TDDFT, is implemented in FD~\cite{WalHakLeh08} and LCAO~\cite{KuiSakRos15} modes. It requires non-periodic boundary conditions, but is not restricted to the linear regime and can be applied to model molecules in strong fields.
The metod may combined with hybrid quantum-classical modeling to simulate the dynamical interaction between molecules and plasmon resonances at metal surfaces ~\cite{SakRosNie14}.
The LCAO-RT-TDDFT is the more recent implementation supporting versatile analyses and enabling modeling of large systems thanks to its efficiency~\cite{KuiSakRos15, RosLehSak15, RosWinJac17, RosKuiPus17, KumRosKui19, KumRosMar19, RosSheErh19, RosErhKui20, FojRosKui22, SorRinRos23}.
We focus on the capabilities of the LCAO version in this section, but some of the described functionalities are also available in FD mode.

The time-dependent KS equation in the PAW formalism is
\begin{align}
    \label{eq:tddft:td_ks}
    i\hat{S}\frac{\mathrm{d}}{\mathrm{d}t}|\tilde\psi_n(t)\rangle = \left(\hat{H}[n(\mathbf{r})] + \hat{v}(t)\right) |\tilde\psi_n(t)\rangle,
\end{align}
where the Kohn--Sham Hamiltonian $\hat{H}[n(\mathbf{r})]$ is implicitly dependent on time through the time-dependent density and $\hat{v}(t)$ is an explicitly time-dependent external potential.
We have additionally assumed that the overlap matrix $\hat{S}$ is independent of time, i.e. there are no ion dynamics.

Starting from the ground state, Eq.~\eqref{eq:tddft:td_ks} is propagated forward numerically.
After each step, a new density is computed and $\hat{H}[n(\mathbf{r})]$ updated accordingly.
The user can can freely define the external potential and implemented standard potentials include the delta-kick $\hat{v}(t) = \hat{x} \delta(t)$ and a Gaussian pulse $\hat{v}(t) = \hat{x}\exp[-\sigma^2(t-t_0)^2/2] \cos[\omega (t-t_0)]$, where $\hat{x}$ is the dipole operator in the direction $x$.

During the propagation, different time-dependent variables can be recorded, and after the propagation they can be post-processed to quantities of physical and chemical interest.
As a basic example, the time-dependent dipole moment recorded for a delta-kick perturbation can be Fourier-transformed to yield the photoabsorption spectrum~\cite{YabBer96}.
Observables are recorded by attaching observers to the calculation and implemented observers include writers for the dipole moment, magnetic moment, KS density matrix in frequency space and wave functions.

RT-TDDFT calculations can be started from the ground state or continued from the last state of a previous time propagation and the time-limiter feature allows one to limit jobs to a predefined amount of wall time.
Together with the continuation capability, this facilitates time propagation in short chunks, efficiently using shared high-performance resources.

In the LCAO-RT-TDDFT implementation, Eq.~\eqref{eq:tddft:td_ks} is cast into a matrix equation and solved with ScaLAPACK~\cite{KuiSakRos15}.
The intermediate step of updating the Hartree and XC potential is performed on the real-space grid.

\subsection{Kohn--Sham decomposition}

The time-dependent KS density matrix can be written as
\begin{align}
    \begin{split}
    \rho_{nn'}(t)
    = \sum_m f_m
        \times &\int \left(\psi^{(0)}_n(\mathbf{r})\right)^* \psi_m(\mathbf{r}, t) \mathrm{d}{\mathbf{r}} \\
        \times &\int \psi^*_m(\mathbf{r}', t) \psi^{(0)}_{n'}(\mathbf{r'}) \mathrm{d}{\mathbf{r}'},
    \end{split}
\end{align}
where $ \psi^*_m(\mathbf{r}', t)$ are the time-dependent KS orbitals with ground state occupation factors $f_m$. The KS density matrix is a central quantity enabling computation of observables and may be evaluated efficiently in the LCAO mode~\cite{RosKuiPus17}.

The Fourier transform of the induced density matrix $\delta \rho_{nn'}(\omega) = \mathcal{F}[\rho_{nn'}(t) - \rho_{nn'}(0)]$ can be built on the fly during time propagation through the density-matrix observer.
Details on the implementation are described in Ref. ~\cite{RosKuiPus17}.
The KS density matrix in frequency space is related to the Casida eigenvectors and gives similar information as the solution of the Casida equation~\cite{RosKuiPus17}.
Observables such as the polarizability can be decomposed into a sum over the electron--hole part of the KS density matrix $\rho_{nn}$, where $f_{n} > f_{n'}$.
This enables illustrative analyses, e.g., by visualizing $\rho_{nn'}(\omega)$ on energy axes as a transition contribution map~\cite{MalLehEnk13}, from which the nature of the localized surface plasmon resonance can be understood (Fig.~\ref{fig:tddft:tcm}, see Ref.~\cite{RosKuiPus17} for detailed discussion).

\subsection{Hot-carrier analysis}

The KS density matrix is a practical starting point for analyzing hot-carrier generation in plasmonic nanostructures.
In the regime of weak perturbations, the KS density matrix resulting from arbitrary pulses can be obtained from delta-kick calculations by a post-processing routine~\cite{RosErhKui20, FojRosKui22}.
By decomposing the matrix into different spatial and energetic contributions, hot-carrier generation in nanostructures~\cite{RosErhKui20, SorRinRos23} and across nanoparticle-semiconductor~\cite{KumRosMar19} and nanoparticle-molecule~\cite{KumRosKui19, FojRosKui22} interfaces can be studied.
Computational codes and workflows for such hot-carrier analyses are provided in Refs.~\cite{RosErhKui20Code, FojRosKui22Code}.

%% file: rt-tddft/CD.tex
Electronic circular dichroism (ECD) is a powerful spectroscopic method for investigating chiral properties at the molecular level.
The quantity that characterizes the ECD is the rotatory strength, which is defined through the magnetic dipole moment $m(\omega)$ as
\begin{equation}
    R (\omega)=\frac{1}{\pi\kappa} \mathrm{Re}\bigg[\sum_{\alpha} m_{\alpha}^{(\alpha)}(\omega)\bigg],
    \label{RotwithmREAL}
\end{equation}
where the index $\alpha$ enumerates Cartesian directions, the $\alpha$ superscript in parenthesis indicates the $\delta$-kick direction and $\kappa$ is the strength of the kick.
To resolve $R(\omega)$, one needs to perform the $\delta$-kick in all three Cartesian directions using a perturbing electric
field $\mathbf{E}^\alpha(t)=\kappa\hat{\mathbf{\alpha}}\delta(t)$. The frequency components of the dipole moment $m_{i}^{(i)}(\omega)$ are calculated by Fourier transforming $m_{\alpha}^{(\alpha)}(t)$, which is recorded during the propagation. Finally, the time-dependent magnetic dipole moment is obtained
as the expectation value of the operator $\hat{\mathbf{m}} = -\frac{i}{2c} \hat{\mathbf{r}} \times \hat\nabla$,
\begin{equation}
  \mathbf{m}(t) =
  \sum_{n} f_n \int \psi_n^*({\mathbf r},t) \,\hat{\mathbf{m}}\, \psi_n({\mathbf r},t)~{\rm d}\mathbf{r},
  \label{expecgridm2}
\end{equation}
where $f_n$ is the occupation number of KS orbital $n$ and $\psi_n({\bf r},t)$ is the time-evolved KS state.
The current GPAW implementation supports both the FD and LCAO mode and the computational efficiency of the LCAO mode enables calculation of the ECD of nanoscale metal-organic nanoclusters. More details on the implementation can be found in Ref.~\cite{makkonen2021real}.

%% file: rt-tddft/tddft-rr.tex
Plasmonic or collective molecular excitations are strongly susceptible to
any kind of optical interaction. Induced currents will couple via Maxwell's
equation of motion to the optical environment and result in radiative
losses, i.e., decay towards the ground state. It is possible to solve the
Maxwell problem formally by obtaining Green's tensor $\textbf{G}_\perp
(\omega)$. The dipolar interaction between electric field and electronic
dipole can be absorbed into a local potential $
v_\text{rr}(\textbf{r},t)=e\textbf{r}\cdot\textbf{E}_{r,\perp}(t)
=e\textbf{r}\cdot \big[\mathcal{F}_t^{-1}(i\mu_0 \omega \textbf{G}_\perp
(\omega)) \ast \partial_t \textbf{R}(t)\big]$ suitable for Kohn--Sham
TDDFT, where $\textbf{R}(t)$ is the total electronic dipole moment. A
detailed discussion can be found in Ref.~\cite{PhysRevLett.128.156402}.

For many simple structures, such as free-space, one-dimensional wave-guides,
or dielectric spheres, $\textbf{G}_\perp (\omega)$ is analytically
known and radiative losses can then be included in TDDFT without additional
coimputational cost. The tutorials on the GPAW web-page~\cite{tutorials} includes an
example for one-dimensional wave-guides for which the user can specify the
cross sectional area and the polarization of the propagating modes.
Extending the functionality of the radiation-reaction potential to 3D free-space
and collective interaction of large ensembles in Fabry-P\'erot
cavities from first principles~\cite{doi:10.1021/acs.jpclett.2c01169} is
essential for the understanding of polaritonic chemistry. This
functionality is currently under development.

%% file: rt-tddft/ehrenfest.tex
Molecular dynamics (MD) simulations usually rely on the Born-Oppenheimer
approximation, where the electronic system is assumed to react so much faster
than the ionic system that it reaches its ground state at each time step. Thus,
forces for the dynamics are calculated from the DFT ground-state density. While
this approximation is sufficiently valid in most situations, there are cases
where the explicit dynamics of the electronic system can affect the molecular
dynamics, or the movement of the atoms can affect averaged spectral properties.
These cases can be handled using so-called Ehrenfest dynamics, ie. time-
dependent density functional theory molecular dynamics (TDDFT/MD).

Ehrenfest dynamics is implemented in the FD
mode~\cite{ojanpera_nonadiabatic_2012}.  A description of the theory
and a tutorial is available on the GPAW web-page~\cite{tutorials}.
This functionality has been used to model the electronic stopping of ions
including core-electron excitations~\cite{ojanpera_electronic_2014}, study
charge transfer at hybrid interfaces in the presence of
water~\cite{syzgantseva_charge_2015}, simulate the coherent diffraction of
neutral atom beams from graphene~\cite{brand_coherent_2019}, model the
dependence of carbon bond breaking under Ar$^+$-ion irradiation on $sp$
hybridization~\cite{buntov_carbon_2020}, and reveal charge-transfer dynamics
at electrified sulfur cathodes~\cite{aierken_revealing_2021}. An LCAO
implementation, inspired by recent work in the Siesta
code~\cite{artacho_quantum_2017, garcia_siesta_2020}, is currently under
development.

%% file: ivos/ivos.tex
The linear response TDDFT approach generally provides reasonably accurate excitation energies for low-lying valence excited states, where the orbitals associated with the holes and excited electrons overlap significantly. However, it tends to fail for excitations involving spatial rearrangment of the electrons, such as charge transfer~\cite{Dreuw2004, Dreuw2003}, Rydberg~\cite{Seidu2015} and doubly excited states~\cite{Levine2006}.

Some of these problems can be alleviated by using range separated functionals (see section \ref{sec:range_separated}). However, these functionals come with a significantly increased computational cost due to the evaluation of exchange integrals. Moreover, due to the missing cancellation of Coulomb and exchange terms for canonical unoccupied orbitals within Hartree-Fock theory, one obtains spurious unoccupied orbitals. This leads to slow covergence of linear-response TDDFT calculations with respect to the number of unoccuppied orbitals when hybrid and range-separated functionals are used~\cite{wurdemann_berechnung_2016,wurdemann_charge_2018}.

Substantial improvement in convergence with respect to unoccupied orbitals can be obtained using improved virtual orbitals as devised by Huzinaga and Arnau~\cite{huzinaga_virtual_1970,huzinaga_virtual_1971}. In this approach, a modified Fock operator is used for the unoccupied orbitals, which mimics the interaction between a hole arbitrarily chosen among the occupied ground state orbitals and the excited electron. This leads to faster convergence in linear-response TDDFT calculations with hybrid and range-separated functionals, and also makes it possible to evaluate excited state properties. For example, the energetics of long-range charge transfer can be obtained by means of a ground state calculations because the difference between the energy of an improved virtual orbital and a hole tends to approximate the excitation energy. The improved virtual orbitals approach is available in GPAW and details on the implementation are described in Ref.~\cite{wurdemann_charge_2018}.

%% file: variational-excitedstate-calc/variational-excitedstate-calc.tex
\label{sec:variationalexcited}
GPAW also offers the possibility to perform excited-state calculations using an alternative time-independent density functional approach~\cite{Hait2021} (sometimes referred to as the ``$\Delta$SCF'' method), which does not suffer from the limitations of linear-response TDDFT mentioned in the previous section. The method involves variational optimization of the orbitals corresponding to a specific excited state by optimizing the electronic energy to a stationary point other than the ground state. The computational effort is similar to that of ground-state calculations and the variational optimization guarantees that the Hellmann--Feynman theorem is fulfilled. Therefore, all the ground-state machinery available in GPAW to evaluate atomic forces can be used for geometry optimization and for simulating the dynamics of atoms on the excited state. Furthermore, coupling this time-independent, variational approach for excited-state calculations with external MM potentials (see section \ref{sec:ion_dynamics}) does not involve additional implementation efforts compared to ground-state calculations and provides a means for performing excited-state QM/MM molecular dynamics simulations that include the state-specific response of a solvent, i.e. the response due to changes in the electron density of the solute.

Variationally optimized excited states correspond to single Slater determinants of optimal orbitals with a non-aufbau occupation and are typically saddle points on the electronic energy surface. Hence, variational calculations of excited states are prone to collapsing to lower-energy solutions, which preserve the symmetry of the initial guess. A simple maximum overlap method (MOM)~\cite{Barca2018, Gilbert2008} is available in GPAW to address this problem. At each SCF step, the MOM occupies those orbitals that overlap the most with the orbitals of a non-aufbau initial guess, usually obtained from a ground-state calculation. The MOM, however, does not guarantee that variational collapse is avoided, and convergence issues are common when using SCF eigensolvers with density-mixing schemes developed for ground-state calculations.

%% file: do-mom-gmf/do-mom-gmf.tex
To alleviate the issues leading to variational collapse and achieve more robust convergence to excited-state solutions, GPAW contains two alternative strategies that are more reliable than conventional SCF eigensolvers with the MOM. They are based on direct optimization of the orbitals and use saddle point search algorithms akin to those for transition-state searches on potential energy surfaces of atomic rearrangements. These approaches also facilitate variational excited-state calculations of nonunitary invariant functionals, such as self-interaction corrected functionals (see section~\ref{sec:sic}).

The first of these methods is a direct orbital optimization approach supplemented with the MOM (DO-MOM). This method is an extension of the direct minimization approach using the exponential transformation illustrated in section \ref{sec:directmin} where the search is for a generic stationary point of $E[\mathbf{\Psi}]$ instead of a minimum:
\begin{equation}
    \label{eq:dofdpw}
    \underset{\mathbf{\Psi}} {\rm stat\,} E[\mathbf{\Psi}] = \underset{\mathbf{\Psi}_{0}} {\rm min\,} \underset{\mathbf{A}} {\rm stat\,} \mathcal{F}[\mathbf{A}, {\mathbf{\Psi}_{0}}]\,.
\end{equation}
The DO-MOM is available in GPAW for both LCAO~\cite{Levi2020jctc, Levi2020fd}, real-space grid, and plane wave basis sets~\cite{Ivanov2021}. For the LCAO basis, the excited-state optimization only necessitates making the energy stationary with respect to the elements of the anti-Hermitian matrix $\mathbf{A}$ (the orbital rotation angles), while calculations using the real-space grid and plane-wave basis include an outer-loop minimization with respect to the reference orbitals ${\mathbf{\Psi}_{0}}$. The optimization in the linear space of anti-Hermitian matrices uses efficient quasi-Newton algorithms that can handle negative Hessian eigenvalues and therefore converge on saddle points. GPAW implements a novel limited-memory SR1 (L-SR1) algorithm, which has proven to be robust for calculations of excitations in molecules~\cite{Ivanov2021, Levi2020jctc}.

The DO-MOM relies on estimating the degrees of freedom along which the energy needs to be maximized starting from an initial guess. For valence and Rydberg excitations, the initial guess consisting of ground-state canonical orbitals with non-aufbau occupation numbers and preconditioning with a diagonal approximation of the electronic Hessian using the orbital energies~\cite{Head-Gordon1988} can be sufficient. However, if the excitation involves significant charge transfer, large rearrangements of the energy ordering of the orbitals can occur, and DO-MOM can struggle to converge. A second direct optimization method with generalized mode following (DO-GMF)~\cite{Schmerwitz2023} alleviates these problems and is also implemented. In DO-GMF, the components of the energy gradient $\mathbf{g}$ along the modes $\mathbf{v}_{i}$ corresponding to the $n$ lowest eigenvalues of the electronic Hessian are inverted yielding
\begin{equation}
\label{eq:gmfgradient}
    \mathbf{g}^{\mathrm{\,mod}} =
    \mathbf{g} - 2\sum_{i = 1}^{n}\mathbf{v}_{i}\mathbf{v}_{i}^{\mathrm{T}}\mathbf{g}\,,
\end{equation}
and a minimization using the modified gradient $\mathbf{g}^{\mathrm{\,mod}}$ is performed by following the $n$ modes simultaneously. This procedure guarantees convergence to an $n^{\rm th}$-order saddle point, eliminating the risk of variational collapse altogether. While it is computationally more expensive than DO-MOM due to the need for a partial Hessian diagonalization, DO-GMF is more robust. Hence, it is particularly useful in the exploration of potential energy surfaces because it is able to follow an excited state through bond-breaking configurations, where broken-symmetry solutions appear, by targeting the solution that preserves the saddle point order. This important advantage is exemplified by the challenging double-bond twisting in ethylene~\cite{Schmerwitz2023, Schmerwitz2022}, where DO-GMF calculations of the lowest doubly excited state provide an avoided crossing with the ground state whereas other methods fail to do so.

%% file: do-mom-gmf/do-mom-gmf-applications.tex
\begin{figure*}[!tb]
    \centering
    \includegraphics[width=\textwidth]{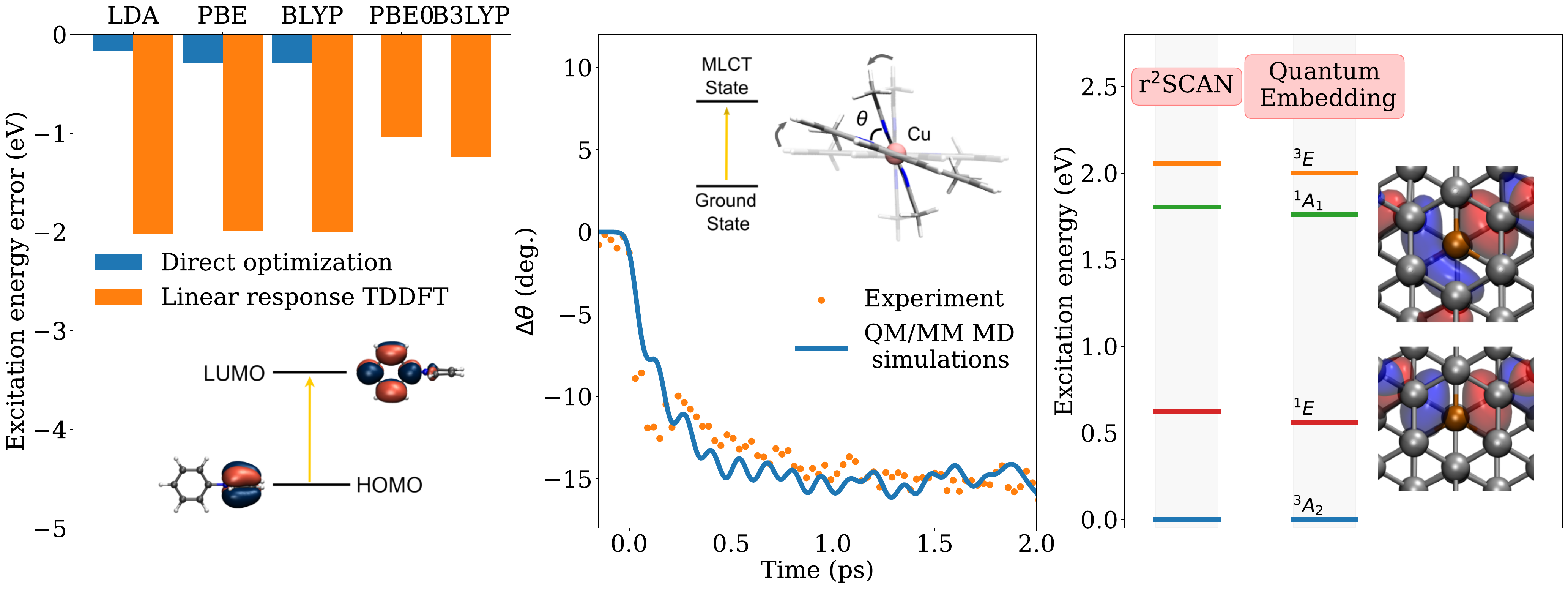}
    \caption{Applications of time-independent, variational calculations of excited states to a molecule in vacuum (left), a molecule in solution (middle) and a solid-state system (right). Left: Deviation of the calculated excitation energy of a charge-transfer excited state in the N-phenylpyrrole molecule from the theoretical best estimate of 5.58 eV ~\cite{Loos2021}. The results of linear-response TDDFT calculations with hybrid functionals are from Ref.~\cite{Loos2021}. Middle: Time-evolution of the interligand angles of the [Cu(dmphen)$_2$]$^+$ complex upon photoexcitation to the lowest metal-to-ligand charge-transfer (MLCT) state in acetonitrile. The experimental results from femtosecond X-ray scattering measurements~\cite{Katayama2023} are compared to the average over excited-state molecular dynamics trajectories obtained using the QM/MM electrostatic embedding scheme in GPAW~\cite{Levi2020pccp} (see section \ref{sec:ion_dynamics}) and convoluted with the experimental instrument-response function~\cite{Katayama2023}. Right: Vertical excitation energy for excitations in the negative nitrogen-vacancy center in diamond obtained with the r\textsuperscript{2}SCAN functional as compared to the results of previous calculations using an advanced quantum embedding approach~\cite{Ma2020}. The orbitals involved in the electronic transitions are visualized in the inset (C atoms are grey and the N atom is orange).}
    \label{fig:excited_state_calc_examples}
\end{figure*}

The efficiency and robustness of the direct-optimization approaches combined with the possibility of choosing different basis-set types make variational calculations of excitated states in GPAW applicable to a great variety of systems ranging from molecules in gas phase or solution to solids.

State-specific orbital relaxation enables the description of challenging excitations characterized by large density rearrangements. Figure~\ref{fig:excited_state_calc_examples} shows the error on the vertical excitation energy of a charge-transfer excitation in the twisted N-phenylpyrrole molecule~\cite{Schmerwitz2023} obtained with direct optimization in GPAW using the LDA, PBE and BLYP functionals and an sz+aug-cc-pVDZ~\cite{RosLehSak15} basis set as compared to the results of linear-response TDDFT calculations with the same basis set and functionals, as well as the hybrid functionals PBE0 and B3LYP (results from Ref.~\cite{Loos2021}). For the variational calculations, the energy of the singlet excited state is computed using the spin-purification formula $E_{\mathrm s}=2E_{\mathrm m}-E_{\mathrm t}$, where $E_{\mathrm m}$ is the energy of a spin-mixed state obtained by promoting an electron in one spin channel and $E_{\mathrm t}$ is the energy of the triplet state with the same character. The variational calculations underestimate the theoretical best-estimate value of the excitation energy (5.58 eV) in Ref.~\cite{Loos2021} by 0.15-0.3 eV, an error that is significantly smaller than that of linear-response TDDFT calculations with the same functionals ($-2.0$ eV) or with the more computationally intensive PBE0 hybrid functional ($-0.85$ eV)~\cite{Loos2021}.

The method has also been used to simulate the photoinduced structural changes of photocatalytic metal complexes and concomitant solvation dynamics~\cite{Katayama2023, Levi2020pccp, Haldrup2019, Abedi2019, Levi2018}. Figure~\ref{fig:excited_state_calc_examples} shows an application to the prototypical copper complex photosensitizer [Cu(dmphen)$_2$]$^+$ (dmphen=2,9-dimethyl-1,10-phenanthroline) in acetonitrile, where QM/MM molecular dynamics simulations elucidated an intricate interplay between deformation of the ligands and rearrangement of the surrounding solvent molecules following a photoexcitation to a metal-to-ligand charge-transfer state~\cite{Katayama2023, Levi2020pccp}.

The last example of an application shown in Figure~\ref{fig:excited_state_calc_examples} is a calculation of the excited states of a solid state system~\cite{Ivanov2023}, the negatively charged nitrogen-vacancy center in diamond, which comprises a prototypical defect for quantum applications. The system is described with a large supercell of up to 511 atoms, and the calculations use a plane-wave basis set. In contrast to previous reports, a range of different density functionals is found to give the correct energy ordering of the excited states, with the r\textsuperscript{2}SCAN functional providing the best agreement with high-level many-body quantum embedding calculations with an error of less than $0.06$~eV~\cite{Ivanov2023, Ma2020}. This example shows that the direct optimization methods in GPAW are promising tools for simulating excited states in extended systems, where alternative approaches are either computationally expensive or lack accuracy.

%% file: berry/static-polarization.tex
The formal polarization of bulk materials may be calculated from the modern theory of polarization~\cite{King-Smith1993,Resta1994} as
\begin{equation}\label{eq:P_F}
\mathbf{P}_\mathrm{F}=\mathbf{P}_\mathrm{F}^\mathrm{el}+\mathbf{P}_\mathrm{F}^\mathrm{n},
\end{equation}
where
\begin{equation}\label{eq:P_el}
\mathbf{P}_\mathrm{F}^\mathrm{el}=\frac{e}{(2\pi)^3}\mathrm{Im}\int_\mathrm{BZ}\mathrm{d}\mathbf{k}\sum_nf_{\mathbf{k}n}\langle u_{\mathbf{k}n}|\nabla_\mathbf{k}|u_{\mathbf{k}n}\rangle
\end{equation}
is the electronic contribution and
\begin{equation}\label{eq:P_n}
\mathbf{P}_\mathrm{F}^\mathrm{n}=\frac{1}{V_\mathrm{cell}}\sum_aZ^a\mathbf{r}^a
\end{equation}
is the contribution from the nuclei. Here the sums run over atoms in the unit cell and $Z^a$ is the charge of nucleus $a$ (including core electrons), situated at position $\mathbf{r}^a$. The electronic contribution can be viewed as a Brillouin-zone integral of $\mathbf{k}$-space Berry phases and may be evaluated from a finite-difference version of Eq.~(\ref{eq:P_el})~\cite{Resta2007}. This involves the overlaps between Bloch orbitals at neighbouring $\mathbf{k}$-points, which are straightforward to evaluate in the PAW formalism~\cite{Olsen2018}.

Eq.~(\ref{eq:P_F}) is only defined modulo $e\mathbf{R_i}/V_\mathrm{cell}$, which follows from the arbitrary choice of unit cell for the atomic positions as well as the choice of phases for $u_{\mathbf{k}n}$, which can shift the Berry phase by $2\pi$. The change in polarization under any adiabatic deformation is, however, well defined and may be calculated as
\begin{equation}\label{eq:dP}
\Delta \mathbf{P}=\int_0^\lambda \mathrm{d}\lambda \frac{\mathrm{d} \mathbf{P}_\mathrm{F}}{\mathrm{d}\lambda},
\end{equation}
where $\lambda$ is some dimensionless variable parameterizing the adiabatic path. In particular, for ferroelectrics the spontaneous polarization $ \mathbf{P}_\mathrm{S}$ can be evaluated by choosing a path that deforms the structure from the polar ground state at $\lambda=1$ to a non-polar structure at $\lambda=0$.

\begin{figure}[tb]
    \centering
    \includegraphics{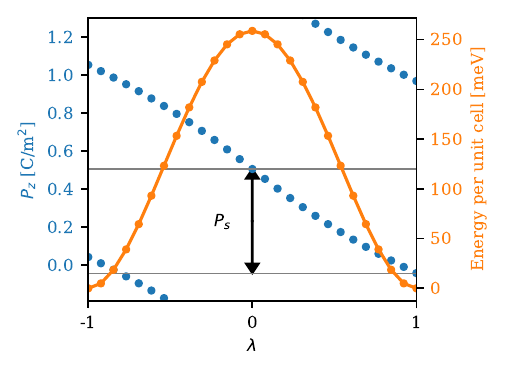}
    \caption{Formal polarization along an adiabatic path connecting two states of polarization and the energy along the path in tetragonal KNbO$_3$. The spontaneous polarization is obtained as the difference in polarization between a polar ground state ($\lambda=1$) and a non-polar reference structure ($\lambda=0$). The energy along the path is also shown.}
    \label{fig:pol_KNbO3}
\end{figure}

In Fig.~\ref{fig:pol_KNbO3}, we show an example of this for tetragonal KNbO$_3$, which is a well-known ferroelectric~\cite{Yoneda2018}. The polar structure was relaxed under the constraint of tetragonal symmetry using PBE ($\lambda=1$) and then linearly interpolated to the inverted structure ($\lambda=-1$) passing through a centrosymmetric point at $\lambda=0$. There are infinitely many polarization branches differing by the polarization quantum $ec/V_\mathrm{cell}$ ($c$ being the lattice constant in the $z$-direction) and the spontaneous polarization is obtained by choosing a single branch and evaluating the difference in formal polarization at $\lambda=1$ and $\lambda=0$. Interestingly, the centrosymmetric point has a non-vanishing polarization given by half the polarization quantum. This is allowed due to the multi-valued nature of the formal polarization and such a ``topological polarization'' in non-polar materials has been shown to yield gapless surface states~\cite{Gibertini2015, Sodequist2023}. Here, however, we merely use the topological polarization to emphasize the importance of evaluating the spontaneous polarization as the change in $\mathbf{P}_\mathrm{F}$ along an adiabatic path.

The expressions (\ref{eq:P_F}-\ref{eq:dP}) can be applied to extract various properties of non-polar materials as well, for example the Born effective charge tensors
\begin{equation}\label{eq:born}
Z^a_{\alpha\beta}=V_\mathrm{cell}\frac{\partial P_\beta}{\partial r_\alpha^a},
\end{equation}
which yield the change in polarization resulting from small shifts in atomic positions. In GPAW, these are obtained by a simple call to a module that introduces a (user-defined) shift of all atoms in the unit cell and calculates the resulting change in polarization from Eq.~(\ref{eq:P_F}-\ref{eq:P_n}). The Born charges may be combined with the atomic force matrix to calculate equilibrium positions under an applied static electric field and the lattice contribution to the dynamic polarizability can be calculated from the eigenvalues and eigenvectors of the force matrix~\cite{Baroni2001}. In addition, the piezoelectric response can be obtained by calculating the change in polarization in response to external strain~\cite{Gjerding2021}. The lattice contribution to the polarizability is typically orders of magnitude smaller than the electronic part, but for ferroelectrics the soft phonon modes associated with spontaneous polarization can give rise to significant lattice polarizabilities.

We exemplify this by the well known case of 2D ferroelectric GeS~\cite{Fei2016, Rangel2017, Wang2017} where we obtain a spontaneous polarization of 490 pC/m~\cite{Petralanda2023, Kruse2023} in excellent agreement with previous calculations~\cite{Wang2017}. The lattice and electronic 2D in-plane polarizabilities are $\alpha^\mathrm{2D}_\mathrm{lat}=4.32$  {\AA}  and $\alpha^\mathrm{2D}_\mathrm{el}=3.75$  {\AA} respectively~\cite{Gjerding2021}. For comparison, the non-polar case of 2D MoS$_2$ yields $\alpha^\mathrm{2D}_\mathrm{lat}=0.09$  {\AA} and  $\alpha^\mathrm{2D}_\mathrm{el}=6.19$  {\AA}~\cite{Gjerding2021}.

%% file: berry/topology.tex
Topological phases such as the quantum spin Hall state and Chern insulator depend crucially on presence of spin--orbit coupling and the band topology may be obtained from the evolution of $\mathbf{k}$-space Berry phases across the Brillouin zone. In particular, for insulators the eigenvalues of the Berry phase matrix of occupied states
\begin{equation}\label{eq:berry_matrix}
\gamma_{mn}^i(\mathbf{k}_\perp)=i\int_0^{2\pi} \mathrm{d}k_i\langle u_m(\mathbf{k})|\partial_{k_i}|u_n(\mathbf{k})\rangle
\end{equation}
must change by an integer multiple of $2\pi$ when a component of $\mathbf{k}_\perp$ (components of $\mathbf{k}$ orthogonal to $k_i$) is cycled through the Brillouin zone~\cite{Taherinejad2014}. Here, $u_n(\mathbf{k})$ are spinor Bloch states, which are typically not smooth functions of $\mathbf{k}$. The evaluation of Eq.~(\ref{eq:berry_matrix}) thus requires the construction of a smooth gauge and in GPAW this is handled by the parallel-transport algorithm of Ref.~\cite{marzari1997maximally}.

\begin{figure}[tb]
    \centering
    \includegraphics{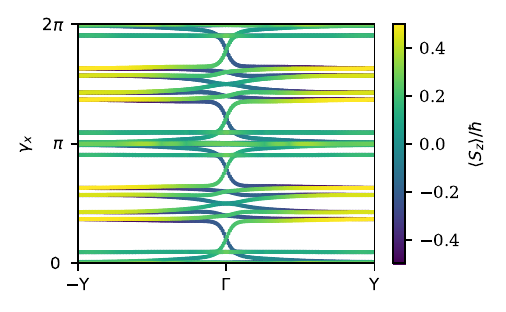}
    \caption{Berry phases of the quantum spin Hall insulator 1T'-MoS$_2$ obtained from PBE with
    non-selfconsistent spin--orbit coupling. The colors indicate the
    expectation value of $S_z$ for each state as defined in Ref. \cite{Olsen2018}
    .}
    \label{fig:berry_MoS2}
\end{figure}

The method has been applied to high throughput search for new topological two-dimensional materials~\cite{Olsen2018} and in Fig.~\ref{fig:berry_MoS2} we show the calculated Berry phases of 1T'-MoS$_2$~\cite{Qian2014}. Due to time-reversal symmetry, the Berry phases at $\Gamma$ and $Y$ are two-fold degenerate, and therefore no perturbation that conserves time-reversal symmetry can open a gap in the Berry-phase spectrum. This property is diagnostic for the quantum spin Hall insulating state and closely related to the presence of gapless edge states~\cite{Taherinejad2014}. 

The eigenvalues of Eq.~(\ref{eq:berry_matrix}) may also be used to calculate the electronic contribution to the formal polarization, since the sum of all individual Berry phases yields the same value as one may obtain from Eq.~(\ref{eq:P_el}). The present approach is, however, more involved since it requires the construction of a smooth gauge, which is not needed in Eq.~(\ref{eq:P_el}).

%% file: wannier/wannier.tex
Wannier functions (WFs) provide a localized representation of the
electronic states of a solid. The WFs are defined by a unitary
transformation of the Bloch eigenstates that minimises the spatial extent
of the resultant orbitals. Specifically, the $n$th Wannier function in
unit cell $i$ is written as
\begin{equation}\label{eq:wannier}
w_{in}(\mathbf r) = \frac{1}{N_\mathbf{k}}\sum_\mathbf{k} e^{-i \mathbf{k}\cdot \mathbf{R}_i} \tilde \psi_{n\mathbf{k}}(\mathbf r),
\end{equation}
where $\tilde \psi_{n\mathbf{k}}$ is a generalized Bloch function (a superposition
of Bloch eigenstates at $\mathbf{k}$).

The minimization of the spatial extent of a set of WFs
$\{w_n(\mathbf{r})\}_{n=1}^{N_w}$ is equivalent to the maximization of the
spread functional~\cite{resta1999electron}
\begin{equation}
  \label{eq:ase-fun}
  \Omega = \sum_{n=1}^{N_w} \sum_{\alpha=1}^{N_q} W_{\alpha} |Z_{\alpha, nn}|^2,
\end{equation}
where
\begin{equation}
  \label{eq:z-mat}
  Z_{\alpha, nn} = \langle w_n | e^{-i \mathbf{q}_{\alpha} \cdot \mathbf{r}} | w_n \rangle.
\end{equation}
The $\{\mathbf{q}_{\alpha}\}$ is a set of at most 6 reciprocal vectors
connecting a $\mathbf{k}$-point to its neighbors and $W_{\alpha}$ are corresponding
weights accounting for the shape of the unit
cell~\cite{berghold2000general}.

The generalized Bloch functions of Eq.~(\ref{eq:wannier}) are determined
by minimizing $\Omega$ using e.g. a conjugate gradient scheme as
implemented in the ASE Wannier module. The inputs to this Wannierization
algorithm are the matrices
\begin{eqnarray}
    Z_{\alpha, ij}^{(0),\mathbf{k}} & = &
    \langle \tilde\psi_{\mathbf{k}i} |
    e^{-i \mathbf{q}_{\alpha} \cdot \mathbf{r}} |
    \tilde\psi_{\mathbf{k}+\mathbf{q}_\alpha,j} \rangle \nonumber\\
    & + & \sum_{aii'}
    e^{-i \mathbf{q}_{\alpha} \cdot \mathbf{R}^a}
    \Delta S_{ii'}^a \nonumber\\
    & & \times \langle \tilde\psi_{\mathbf{k}i} | \tilde p_i^a \rangle
    \langle\tilde p_{i'}^a | \tilde\psi_{\mathbf{k}+\mathbf{q}_\alpha,j}
    \rangle,
\end{eqnarray}
where $\Delta S_{ii'}^a$ are the PAW corrections from Eq.~\ref{eq:ds}.
From these matrices,
the ASE Wannier module can be used to construct partially occupied Wannier
functions~\cite{thygesen2005partly, thygesen2005partly2}, which are a
generalization of maximally-localized Wannier
functions~\cite{marzari1997maximally} to entangled bands and non-periodic
systems. Recently, a further improvement in terms of robustness of the
Wannierisation procedure was achieved using a modified spread functional
containing a penalty term proportional to the variance of the spread
distribution of the WFs, which leads to a more uniform spread
distribution~\cite{fontana2021spread}.

%% file: point_defects/point_defects.tex
Point defects play a crucial role in many applications of
semiconductors\cite{dreyer2018first, park2018point}. First-principles
calculations can be used to determine the atomic structure, formation
energy, and charge-transition levels of points defects. It is well
established that the best description of point defects in
semiconductors/insulators is obtained using range separated hybrids, such
as the HSE06 xc-functional\cite{heyd2003hybrid,weston2018native}. To
illustrate the use of GPAW for point defect calculations, we determine the
formation energy diagrams of the C$_\mathrm{N}$ and C$_\mathrm{B}$ defects
in the hexagonal boron nitride (hBN) crystal with the HSE06 functional.
These defects have been proposed to be responsible for the deep-level
luminescence signal with a zero-phonon line (ZPL) around 4.1
eV.\cite{vuong2016phonon,weston2018native}. The results are compared to
similar results obtained with the VASP software package.

For a point-defect $\mathrm{D}$ in charge state $q$, the formation energy
E$^{\mathrm{f}}$ is calculated from the formula,

\begin{equation}\label{eq:eform}
E^{\mathrm{f}}\left[\mathrm{D}^q\right] = E_{\mathrm{tot}}\left[\mathrm{D}^q\right] - E_{\mathrm{tot}}^\mathrm{bulk} - \sum_i n_{i}\mu_i + qE_\mathrm{F} + E_\mathrm{corr}.
\end{equation}

Here $E_{\mathrm{tot}}\left[\mathrm{D}^q\right]$ and
$E_{\mathrm{tot}}^\mathrm{bulk}$ are the total energies of the crystal with
the point defect in charge state $q$ and of the neutral pristine
crystal, respectively. $\mu_i$
is the chemical potential of the element $i$ while $n_{i}$ is the number
of atoms added to ($n_i > 0$) or removed from ($n_i < 0$) the crystal to
create the defect. $E_\mathrm{F}$ is the chemical potential of the
electrons, i.e. the Fermi level, which is written as
$E_\mathrm{F}=E_{\mathrm{VBM}}+\Delta E_{\mathrm{F}}$, where VBM is the
valence band maximum. Finally, E$_\mathrm{corr}$ is a correction term
which accounts for: (i) the spurious electrostatic interaction between the
periodic images of the defect and their interaction with the compensating
homogeneous background charge and (ii) the potential shift between the
pristine and defect system.





For more details on the methodology of point defect calculations we refer
the reader to the excellent review papers on the topic~\cite{van2004first,
lany2008assessment, freysoldt2014first, zunger2021understanding}.

\begin{figure}
    \centering
    \includegraphics{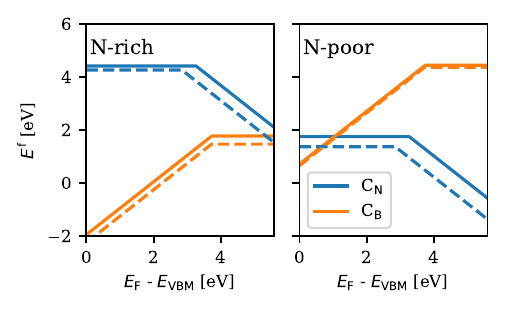}
    \caption{Defect formation energies for C$_{\mathrm N}$ and
             C$_{\mathrm B}$, under N-rich and N-poor conditions,
             respectively. The dashed
             lines are reproduced from Ref. \onlinecite{weston2018native}. }
    \label{fig:hsepd}
\end{figure}
All calculations have been performed using the HSE06 functional with the
default mixing parameter $\alpha$=0.25, plane wave cut-off of 800 eV, and
forces converged to 0.01 eV/{\AA}. The lattice of the hBN crystal was
fixed at the experimental parameters ($a=2.50$ {\AA} and $c=6.64$
{\AA}) \cite{kurakevych2007rhombohedral}. The band gap of the pristine
crystal was determined to be 5.58 eV (using $8\times 8\times 4$
$k$-points) in good agreement with the experimental band gap of 6.08
eV \cite{cassabois2016hexagonal}. The structure of the point defects were
relaxed in a $4\times 4\times 2$ (128 atom) supercell using $\Gamma$-point
$k$-point sampling. For each defect three different charge states ($q=1,
0, -1$) were considered. The corrections ($E_\mathrm{corr}$) due to image
charges and potential alignment are evaluated following Freysoldt,
Neugebauer and Van de Walle \cite{freysoldt2009fully} as implemented in
GPAW.

Figure \ref{fig:hsepd} shows the defect formation energies as a function of
Fermi level for C$_{\mathrm N}$ and C$_{\mathrm B}$, at N-rich and N-poor
conditions, respectively. We can see that under N-rich conditions,
C$_{\mathrm B}$ is energetically lower, whereas C$_{\mathrm N}$ is
favorable under N-poor conditions. C$_{\mathrm B}$ shows a 1/0 charge
transition at 3.73 eV above the VBM, whereas C$_{\mathrm N}$ has a 0/-1
charge transition 3.26 eV deep inside the band gap. We find good agreement
with a similar study \cite{weston2018native} also employing plane waves and
the HSE06 functional (VASP calculations). Minor discrepancies can be
attributed to use of a different supercell size and slightly higher
fraction of the non-local mixing parameter ($\alpha$=0.31).

%% file: pointgroup-reps/pointgroup-reps.tex
GPAW allows for the automatized assignment of point-group symmetry representations for the pre-computed Kohn--Sham wavefunctions.
This can be used for determining the wave-function symmetry representations for both
molecules~\cite{Kaappa2018} and extended structures~\cite{Bertoldo2022} to analyze, for example, the symmetry-involved degeneracy of
the bands and selection rules for dipole transitions.

The analysis follows directly from group theory~\cite{CORNWELL1997}, stating that the solutions to the Schrödinger equation inherit the symmetry group of the respective Hamiltonian, or essentially the external potential invoked by the atomic configuration. The representation matrices $\Gamma$ are computed as
\begin{equation}
    \Gamma_n(T) = \int \phi_n^\dagger (\mathbf r) P(T) \phi_n(\mathbf r) \mathrm d\mathbf r,
\end{equation}
where $\phi$ is a normalized wavefunction, $n$ is the eigenstate label, and $P(T)$ is an operation that corresponds to the
transformation $T$ of the symmetry group of the Hamiltonian. The operations $P(T)$ include rotations that are non-trivial for the
rectangular grid that the computed wavefunctions are projected onto. The wavefunction rotations are performed on the
grid by cubic interpolation. The output of the analysis contains the irreducible representation weights $c_{\alpha,n}$ for each
eigenstate $n$ as solved from
\begin{equation}
    \Gamma_n(T) = \sum_\alpha c_{\alpha,n}\chi_\alpha(T),
\end{equation}
where $\chi_\alpha$ are the character vectors of the group (\textit{i.e.} rows of the character table).

When doing the analysis, the user needs to input the coordinates of the center of symmetry (typically the coordinates of a single atom), and the point group for which the analysis is run. It is possible to analyze only a part of the wave function by selecting a cutoff radius from the center of symmetry beyond which the parts of the wave function are neglected. This enables the investigation of the purity of the local symmetry even if the symmetry of the Hamiltonian is broken far from the center of symmetry~\cite{Kaappa2018,Bertoldo2022}. To date, point groups of C$_2$, C$_\mathrm{2v}$, C$_\mathrm{3v}$, D$_\mathrm{2d}$, D$_\mathrm{3h}$, D$_\mathrm{5}$, D$_\mathrm{5h}$, I, I$_\mathrm{h}$, O$_\mathrm{h}$, T$_\mathrm{d}$, and T$_\mathrm{h}$ are implemented.

%% file: unfolding/unfolding.tex
When studying defect formation, charge-ordered phases or structural phase
transitions, it is often needed to perform DFT calculations on a super-cell.
A super-cell (SC) calculation comes with the cost of having to account for
many more electrons in the unit cell when compared to the primitive cell
(PC). This implies that besides the increased computational effort, the band
structure of a SC contains more bands in a smaller Brillouin zone as compared
to the PC. In order to compare electronic band structures between SC and PC,
unfolding the band structure of the SC into the one of the primitive cell
(PC) becomes convenient.

GPAW features the possibility of performing band-structure unfolding in the
real-space grid, plane wave, and LCAO modes. The implementation allows to
unfold the SC band structure without the explicit calculation of the overlap
between SC and PC wavefunctions, following the procedure described in
Ref.~\cite{popescu2012}. The unfolded band structure is given in terms of
the spectral function
\begin{equation}
  A(\mathbf{k},\epsilon) = \sum_m
  P_{\mathbf{K}m}(\mathbf{k}) \delta(\epsilon_{\mathbf{K}m}-\epsilon),
\end{equation}
with
$\mathbf{k}$, $\mathbf{K}$ momenta in the PC and SC Brillouin zone
respectively, $\epsilon_{\mathbf{K}m}$ the SC eigenvalues obtained for
momentum $\mathbf{K}$ and band index $m$, and $P_{\mathbf{K}m}(\mathbf{k})$
is calculated as
\begin{equation}
  P_{\mathbf{K}m}(\mathbf{k}) = \sum_{\{\mathbf{G}\}}
  |C_{\mathbf{K}m}(\mathbf{G}+\mathbf{k}-\mathbf{K})|^2,
\end{equation}
where
$C_{\mathbf{K}m}$ are the Fourier coefficients of the eigenstate
$|\mathbf{K}m \rangle$ and $\{\mathbf{G}\}$ the subset of reciprocal space
vectors of the SC that match the reciprocal space vectors of the PC. A more
detailed explanation and technical details on how to perform a band-structure
unfolding can be found on the GPAW web-page~\cite{tutorials}.

%% file: qeh/qeh.tex
The quantum-electrostatic heterostructure (QEH)~\cite{qeh_model} model is an add-on GPAW feature for calculating the dielectric response and excitations in vertical stacks of 2D materials, also known as van der Waals (vdW) heterostructures. The QEH model can be used independently from the GPAW code, but it relies on the GPAW implementation for the calculation of the fundamental building blocks used by the model, as elaborated below.

The dielectric screening in 2D materials is particularly sensitive to changes in the environment and depends on the stacking order and thickness of the 2D heterostructure, providing a means to tune the electronic excitations, including quasi-particle band gaps and excitons.  While the dielectric response of freestanding layers can be explicitly represented \emph{ab initio} in GPAW at the linear-response TDDFT, GW, and BSE level of theory, lattice mismatch between different 2D layers often results in large supercells that make these many-body approaches infeasible. Since the interaction between stacked layers is generally governed by van der Waals interactions, the main modification to the non-interacting layers' dielectric response arises from the long-range electrostatic coupling between layers.

Therefore, in the QEH model, the dielectric function of the vdW
heterostructure is obtained through an electrostatic coupling of the quantum
dielectric building blocks of individual 2D
layers~\cite{andersen2015dielectric}. The dielectric building blocks
consist of monopole and dipole components of the density-response function
of the freestanding layers, $\tilde{\chi_{i}}(\mathbf{q}_{\parallel},\omega)$,
calculated from \emph{ab initio} on the RPA level. Subsequently, the full density-response function $\chi_{i,j}(\mathbf{q}_{\parallel},\omega)$ (density perturbation on layer $i$ due to a monopole or dipole
component of the perturbing field acting on layer $j$) is calculated by solving
the Dyson equation
\begin{align}
  \chi_{i,j}(\mathbf{q}_{\parallel},\omega) & =
  \tilde{\chi_{i}}(\mathbf{q}_{\parallel}, \omega) \delta_{i,j}  + \nonumber \\
  &\tilde{\chi_{i}}(\mathbf{q}_{\parallel}, \omega) \sum_{k \neq i}
  V_{i,k}(\mathbf{q}_{\parallel}) \chi_{k,j}(\mathbf{q}_{\parallel}, \omega),
\end{align}
where the Coulomb matrix element is obtained as the real-space overlap over the density, $\rho$, and potential, $\phi$, basis functions on the different layers.
\begin{equation}
V_{i,k}(\mathbf{q}_{\parallel}) = \int \rho_i(z, \mathbf{q}_{\parallel}) \phi_k(z, \mathbf{q}_{\parallel}) \mathrm{d}z.
\end{equation}

From the density-response function, the dielectric function is obtained in the basis of monopole/dipole perturbations on each layer in the heterostructure. While building blocks pre-computed with GPAW for a large variety of 2D materials are provided with the QEH package, GPAW offers the possibility of calculating custom building blocks for any 2D material as explained in Ref.~\cite{qeh_tutorial}.

\begin{figure}
\includegraphics[width=0.95\linewidth]{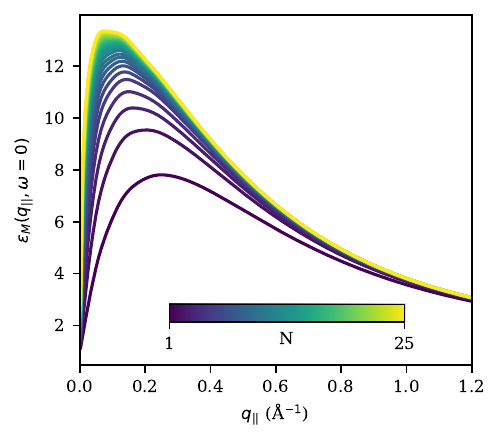}
\llap{\raisebox{5cm}{
      \includegraphics[height=2cm]{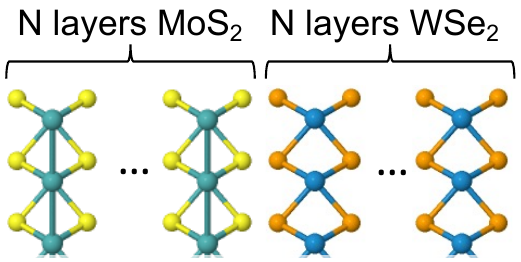} \hspace{0.7cm}}}
\caption{The static macroscopic dielectric function of a vdW heterostructure interface as a function of the number of layers. The heterostructure is made up of $N$ MoS$_2$ layers on one half and $N$ WSe$_2$ layers on the other half (see inset). Increasing the number of layers eventually leads to a bulk-like limit for the chosen stacking configuration.}
\label{fig:qeh}
\end{figure}

As an illustrative example of the QEH model, we show how the static dielectric
function of a heterostructure can be engineered by multi-layer
stacking. Fig.~\ref{fig:qeh} shows the static dielectric function of a vdW
heterostructure made up by stacking $N$ MoS$_2$ layers and $N$ WSe$_2$
layers. We see that the dielectric
function increases significantly as a function of the number of layers, 
eventually approaching a bulk limit. The knowledge of the layer-dependence of
the dielectric response for such a heterostructeres could be further exploited
to investigate inter- and intra-layer excitonic properties and band-edge
renormalization effects.

%% file: pcm/pcm.tex
\label{sec:pcm}
The presence of a solvent has a large effect on the energetics and the
electronic structure of molecules or extended surfaces.
In particular, the arguably most important solvent, water,
is able to stabilize ions or zwitterions that would not form
in the gas phase. The main effect relates to the large
permittivity of water ($\varepsilon_r=78$) that effectively screens
Coulomb interactions.

A convenient and computationally lean method to describe this effect
is the inclusion of a position-dependent
solvent permittivity $\varepsilon(\mathbf{r})$ in the electrostatics
via the Poisson solver~\cite{held_simplified_2014}. The solvent is represented solely as a
polarizable continuum that averages out all movements and re-arrangements
of the solvent molecules and their electrons.
The computational cost is, therefore, practically the same as
a calculation in vacuum.

This implementation allows the calculation of solvation free energies of
neutral and ionic species in solution~\cite{held_simplified_2014}.
Further, it can be applied to periodic surfaces interfaced with an electrolyte 
to reproduce reasonable potential drops within the simulation of
electrochemical reaction processes, as we will elaborate in the following.

%% file: elchem-charging/elchem-charging.tex
\label{sec:Charged electrochemical interfaces}
Simulating atomistic processes at a solid--liquid interface held at a controlled electrode potential is most appropriately performed in the electronically grand-canonical ensemble~\cite{Alavi_JCP_2001, Melander2019, Sakaushi2020, Lindgren2022}. Here, electrons can be exchanged dynamically with an external electron reservoir at a well-defined electrochemical potential.
In a periodic system, a non-zero net charge would lead to divergence of the energy; therefore, any fractional electrons that are added to or removed from the system must be compensated by an equal amount of counter charge.
Several approaches able to account for this change in boundary conditions have recently been brought forward~\cite{Sugino_PRB_2006, Vaspsol_2014,Letchworth-Weaver2012, Hormann2019, Ringe2022implrev}.
In GPAW, this is conveniently accomplished with the introduction of a jellium slab of equal and opposite charge to the needed electronic charge; the jellium is embedded in an implicit solvent localized in the vacuum region above the simulated surface (cf. \ref{sec:pcm}).
As a particular highlight of this \textit{Solvated Jellium Method} (SJM)~\cite{Kastlunger2018}, as it is known in GPAW, we are able to localize the excess charge on only the top side of the simulated atomistic surface, which occurs naturally by introducing the jellium region solely in the top-side vacuum and electrostatically decoupling the two sides of the cell via a dipole correction. Both a purely implicit and hybrid explicit-implicit solvent can be applied.

In the SJM, the simulated electrode potential is a monotonic function of the number of electrons in the simulation; calculations can be run in either a constant-charge or a constant-potential ensemble.
The electrode potential ($\phi_\mathrm{e}$) within SJM is defined as the Fermi level ($\mu$) referenced to an electrostatic potential deep in the solvent (the solution inner potential $\Phi_\mathrm{w}$), where the whole charge on the electrode has been screened and no electric field is present,
\begin{equation}
    \phi_\mathrm{e} = \Phi_\mathrm{w} - \mu .
\end{equation}
%
We can relate $\phi_\mathrm{e}$ to the commonly used reference potentials, for example to the standard hydrogen electrode, by subtracting its absolute potential such as the experimental value of 4.44~V as reported by Trasatti~\cite{Trasatti1986}. In practice, the reference potentials depend on the used solvent model~\cite{10.1063/1.4976971}, and the reference can be calibrated using computed and measured potentials of zero charge~\cite{10.1063/1.5047829}.

The energy used in the analysis of electrode reactions is the grand-potential energy $\Omega$,
\begin{equation}
 \Omega \equiv E_\mathrm{tot} + \phi_\mathrm{e} N_\mathrm{e},
\end{equation}
where $N_\mathrm{e}$ are the excess electrons in the simulation.
This allows for energetic comparisons between calculations with different numbers of electrons, and is the default energy returned to atomistic methods by SJM.
While $E_\mathrm{tot}$ is consistent with the forces in traditional electronic-structure calculations, the grand-potential energy $\Omega$ is consistent with the forces in constant-potential simulations~\cite{PhysRevB.45.13709, Lindgren2022}.
This means that relaxations that follow forces will correctly find local minima in $\Omega$, and any kind of structure optimization or dynamical routine can be performed on the grand-canonical potential energy surface, such as the search for saddle points~\cite{Lindgren2020HER, Kastlunger2022pH, Kastlunger2023selectivity} or molecular-dynamics simulations~\cite{melander_wu_honkala_2023}.

In constant-potential mode, the potential is controlled by a damped iterative technique that varies $N_\mathrm{e}$ to find the target $\phi_\mathrm{e}$; in practice, a trajectory (such as a relaxation or nudged elastic band) is run, where in a first series of SCF cycles the potential equilibrates. Upon achieving a target potential within the given threshold, the code will conduct the chosen geometry-optimization routine under constant potential control.

%% file: cdft/cdft.tex
Constrained DFT (cDFT)~\cite{PhysRevA.72.024502, 10.1063/1.2360263, doi:10.1021/cr200148b} is a computationally efficient method for constructing diabatic or charge/spin-localized states. GPAW includes a real-space implementation of cDFT~\cite{doi:10.1021/acs.jctc.6b00815}, which can be used in both the FD and LCAO modes. Compared to most cDFT implementations, in GPAW the periodicity can be chosen flexibly between isolated molecules and periodic systems in one-, two-, or three-dimensions with $\mathbf{k}$-point sampling.

The key difference of cDFT compared to normal DFT is the introduction of an auxiliary potential to force a certain region (in real space around a molecule, molecular fragment, or atom) to carry a predefined charge or spin. This leads to a modified energy functional
\begin{equation} \label{eq:cdft_energy}
\begin{split}
F[n(\mathbf{r}), \{V_i\}] &= E^\mathrm{KS}[n(\mathbf{r})] + \\
&\sum_i  V_i\sum_{\sigma}  \left(\int \mathrm{d}\mathbf{r}w^{\sigma}_i(\mathbf{r}) n^{\sigma}(\mathbf{r})-N_i\right),
\end{split}
\end{equation}
where $E^\mathrm{KS}$ is the Kohn--Sham energy functional, $\sigma$ denotes the spin, $n^{\sigma}(\mathbf{r})$ is the spin-dependent electron density, and $N_i$ is the predefined charge or spin constraint. $V_i$ acts as Lagrange multiplier, which determines the strength of the auxiliary potential and needs to be determined self-consistently as discussed below. $w^{\sigma}_i(\mathbf{r})$ is the weight function that defines how the charge or spin are to be partitioned, i.e. the regions where charge/spin is to be localised. This will be discussed in some detail below.

Introducing the constraining term in Eq.~\eqref{eq:cdft_energy} leads to a new localized, spin-dependent external potential
\begin{equation}
v_\text{eff}^{\sigma}(\mathbf{r})=
\dfrac{\delta F[n(\mathbf{r})]}{\delta n(\mathbf{r})} =
\dfrac{\delta E^\mathrm{KS}[n(\mathbf{r})]}{\delta n(\mathbf{r})} + \sum_i  V_i\sum_{\sigma} w^{\sigma}_i(\mathbf{r}).
\end{equation}
The constraint is further enforced by demanding that the ${V_i}$ need to satisfy the chosen constraints
\begin{equation} \label{eq:convergence}
C \geq \bigg\lvert \sum_{\sigma} \int \mathrm{d}\mathbf{r} w^{\sigma}_i(\mathbf{r})n^{\sigma}(\mathbf{r}) -N_i\bigg\rvert.
\end{equation}
In practice, the strength of the constraining potential ${V_i}$ is found through a self-consistent two-stage optimization of both $\{V_i\}$ and $n(\mathbf{r})$. As the derivatives of $F[n(\mathbf{r}), \{V_i\}]$ with respect to ${V_i}$ are readily available~\cite{doi:10.1021/acs.jctc.6b00815}, gradient-based optimization algorithms in SciPy~\cite{2020SciPy-NMeth} are used for optimizing $\{V_i\}$. The weight functions are defined by a Hirshfeld-type partitioning scheme with Gaussian atomic densities, and $w^{\sigma}_i(\mathbf{r})$ and the resulting external potential are presented on the grid. With these definitions, the forces resulting from the cDFT external potential can be analytically computed and used in e.g. geometry optimization or molecular-dynamics simulations.

cDFT has been widely used for computing transfer rates within Marcus theory, which depends on the reorganization and reaction (free) energies and the diabatic coupling matrix element; the GPAW-cDFT implementation includes all the needed tools for obtaining these parameters for bulk, surface, and molecular systems~\cite{doi:10.1021/acs.jctc.6b00815, doi:10.1021/acs.chemmater.7b04618}. Recently, the cDFT approach has been combined with molecular-dynamics methods to compute the reorganization energy at electrochemical interfaces~\cite{kumeda2022cations} as well as with grand-canonical DFT methods (see section \ref{sec:Charged electrochemical interfaces}) to construct fixed electron-potential diabatic states~\cite{Melander_2020}.

%% file: ofdft/ofdft.tex
Orbital-free DFT (OFDFT) approximates the DFT energy functional by modelling 
the kinetic energy as a direct functional of the density
\begin{equation}
\begin{split}
E_\text{OF} [n] = \underbrace{\int \text{d}\textbf{r} \, 
n^{1/2} (\textbf{r}) \left( - \frac{1}{2} \nabla^2 \right) \, 
n^{1/2} (\textbf{r})}_{T_\text{W} [n]} + J[n] + V[n] \\+ E_\text{xc} 
[n] + T_\text{s} [n] - T_\text{W} [n].
\end{split}
\end{equation}
Levy and colleagues showed that a Kohn--Sham-like equation derived 
variationally from the equation above holds for the square-root of
the density~\cite{LevyOFDFT}
\begin{equation}
\left( - \frac{1}{2} \nabla^2 + V_\text{eff}(\textbf{r}) \right) \,
 n^{1/2} (\textbf{r}) = \mu \, n^{1/2} (\textbf{r}).
\end{equation}
OFDFT approximately enforces the Pauli principle, partially 
accounting for quantum effects in an averaged way. 

The OFDFT scheme implemented in GPAW offers the advantage of accessing 
all-electron values while maintaining computational linear-scaling time 
with respect to system size. To achieve this, we employ the PAW method 
in conjunction with 
real-space methods obtaining a mean absolute error of 10~meV per atom when 
compared to reference all-electron values~\cite{Lehtomaki2014}. 

While OFDFT functionals perform better using local pseudopotentials in bulk materials, the OFDFT PAW implementation can be interesting for assessing density functionals. For example, in studying large-$Z$ limits or semiclassical limits of density functionals, the all-electron values allow us to find highly-performing OFDFT functionals~\cite{Lehtomaki2019}.

%% file: zfs/zfs.tex
The zero-field splitting (ZFS) refers to the energetic splitting of the
magnetic sub-levels of a localized triplet state in the absence of a
magnetic field~\cite{ivady2014pressure}. The origin of the ZFS is the
magnetic dipole-dipole interactions between the two electrons of the
triplet. This interaction is described by a spin Hamiltonian of the form
($\alpha,\beta=x,y,z$)~\cite{rayson2008first,biktagirov2020spin}
\begin{equation}\label{eq:ZFS}
\hat    H_{\mathrm{ZFS}} = \sum_{\alpha\beta}\hat S_aD_{\alpha\beta}\hat S_\beta,
\end{equation}
where $\mathbf{\hat S}$ is the total spin operator and $\mathbf{D}$ is the
ZFS tensor given by
\begin{equation}
 \begin{aligned}
 D_{\alpha\beta} &= \frac{1}{2} \frac{\mu_0}{4\pi}g_e^2 \mu_{\mathrm{B}}^2
    \int \rho_2(\mathbf {r}_{1},\mathbf {r}_{2})
    \frac{\delta_{\alpha\beta}{r}^2 - 3 r_{\alpha} r_{\beta}}{r^5}
    \mathrm{d}\mathbf{r}_1 \mathrm{d} \mathbf{r}_2.
    \end{aligned}
\end{equation}
where $r_\alpha$ and $r_\beta$ denote the Cartesian components of
$\mathbf r = \mathbf r_1-\mathbf r_2$,
$\rho_2$ is the two-particle density matrix of the
Kohn--Sham ground-state Slater determinant,
$\mu_{\mathrm{B}}$ is the Bohr magneton and
$g_e$ is the Landé splitting factor.
GPAW computes the $\mathbf{D}$-tensor by
evaluating the double integral in reciprocal space using the pseudo density
including compensation charges~\cite{Bertoldo2022}.

%% file: hyperfine/hyperfine.tex
The hyperfine coupling describes the interaction between a magnetic dipole
of a nuclear spin, $\mathbf {\hat {I}}^N$, and the magnetic dipole of the
electron-spin distribution, $\mathbf{\hat {S}}(\mathbf r)$. The interaction
is described by the spin Hamiltonian ($\alpha,\beta=x,y,z$)
\begin{equation}
\hat H_{\mathrm {HF}}^N = \sum_{\alpha\beta}
\hat {S}_\alpha A_{\alpha\beta}^{N} \hat {I}^{N}_\beta
\end{equation}
where the hyperfine tensor of nucleus $N$ at $\mathbf R= \mathbf 0$ is
given by~\cite{blochl2000first}
\begin{equation}
 \begin{aligned}
 A^N_{\alpha\beta} &= \frac{2\mu_0}{3} g_e\mu_B g_N\mu_N
    \int \delta_T(\mathbf{r}) \rho_s(\mathbf{r}) \mathrm{d}\mathbf{r}\\
        &  + \frac{\mu_0}{4\pi}g_e\mu_Bg_N\mu_N \int \frac{3 r_\alpha r_\beta - \delta_{\alpha\beta} r^2}{r^5}
    \rho_s(\mathbf{r}) \mathrm{d}\mathbf{r}.\end{aligned}
\end{equation}
The first term is the isotropic Fermi contact term, which is proportional
to the spin density, $\rho_s(\mathbf{r})$, at the centre of the nucleus.
$\delta_T(\mathbf{r})$ is a smeared-out $\delta$-function. $g_e$ and $g_N$
are the gyromagnetic ratios for the electron and nucleus, and $\mu_N$ is
the nuclear magneton. The second term represents the anisotropic part of
the hyperfine coupling tensor and results from dipole-dipole interactions
between nuclear and electronic magnetic moments. GPAW evaluates $A^N$ using
the pseudo spin density with compensation charges~\cite{Bertoldo2022}.

%% file: outlook/outlook.tex
As described in this review, GPAW is a highly versatile code that is both
maintenance-, user-, and developer-friendly at the same time. The continued
expansion of the code requires substantial efforts and can be lifted only
because of the dedicated team of developers contributing at all levels.
There are currently a number of ongoing as well as planned developments for
GPAW, which will further improve performance and applicability of the code.
We are currently finishing a major refactoring of the code, which will make
it even more developer friendly and facilitate easier implementation of new
functionality.

Another priority is to improve the parallelization of hybrid functional
calculations in plane wave mode by enabling parallelization over bands and
$\mathbf{k}$-points. In the same vein, there is
ongoing work to support LCAO-based hybrid functional calculations using a
resolution-of-identity approach. A natural next step would then be
LCAO-based GW calculations. Such a method could potentially be very efficient
compared to plane wave calculations, but it is currently unclear if the
accuracy can be maintained with the limited LCAO basis. In relation to
quasiparticle calculations, there are plans to implement (quasiparticle)
self-consistent GW and vertex corrected GW using nonlocal xc-kernels from
TDDFT. Constrained RPA calculations that provide a partially screened
Coulomb interaction useful for ab initio calculation of interaction
parameters in low-energy model Hamiltonians is currently being implemented.

Underlying any GPAW calculation are the PAW potentials. The current
potentials date back to 2009. A new set of potentials, including both soft
and norm-conserving potentials (for response function calculations), is being
worked on.

As described herein, GPAW already has an efficient implementation of real
time TDDFT in the LCAO basis while Ehrenfest dynamics is supported only the
comparatively slower in grid mode. Work to enable Ehrenfest dynamics in LCAO
mode is ongoing.

The current version of GPAW supports GPU acceleration only for standard
ground state calculations. The CuPy library greatly simplifies the task of
porting GPAW to GPU and we foresee that large parts of the code, including
more advanced features such as linear response and GW calculations, will
become GPU compatible.

%% file: acknowledgements.tex
K. S. T. acknowledges funding from the European Research Council (ERC) under
the European Union’s Horizon 2020 research and innovation program Grant No.
773122 (LIMA) and Grant agreement No. 951786 (NOMAD CoE). K. S. T. is a
Villum Investigator supported by VILLUM FONDEN (grant no. 37789).
Funding for A. O. D. and G. L. was provided by the Icelandic Research Fund
(grant nos. 196279 and 217734, respectively).
F. N has received funding from the European Union’s Horizon 2020 research and
innovation program under the Marie Skłodowska-Curie grant agreement No.
899987.
M. M. M. was supported by the Academy of Finland (grant \#338228).
T. O. acknowledges support from the Villum foundation Grant No. 00029378.
S. K. and K. W. J. acknowledge support from the VILLUM Center for Science of
Sustainable Fuels and Chemicals, which is funded by the VILLUM Fonden research
grant (9455).
T. B. was funded by the Danish National Research Foundation (DNRF 146).
J. S. acknowledges funding from the Independent Research Fund Denmark (DFF-
FTP) through grant no. 9041-00161B.
C. S. acknowledges support from the Swedish Research Council (VR) through Grant
No. 2016-06059 and funding from the Horizon Europe research and innovation
program of the European Union under the Marie Sk{\l}odowska-Curie grant
agreement no.\ 101065117. Partially funded by the European Union. Views and
opinions expressed are, however, those of the author(s) only and do not
necessarily reflect those of the European Union or REA. Neither the European
Union nor the granting authority can be held responsible for them.
T. S. received funding from the European Research Council (ERC) under the
European Union’s Horizon 2020 research and innovation programme (Grant
agreement No. 756277-ATMEN).
O. L.-A. have been supported by Minciencias and University of Antioquia
(Colombia).
K. T. W. was supported by the U.S. Department of Energy, Office of Science,
Office of Basic Energy Sciences, Chemical Sciences, Geosciences, and
Biosciences Division, Catalysis Science Program to the SUNCAT Center for
Interface Science and Catalysis.
G. K. acknowledges funding from V-Sustain: The VILLUM Centre for the
Science of Sustainable Fuels and Chemicals (grant no. 9455).
Additional funding:
Knut and Alice Wallenberg Foundation (2019.0140; J. F. and P. E.),
the Swedish Research Council (2020-04935; J. F. and P. E.),
the European Union’s Horizon 2020 research and innovation programme under
the Marie Sk{\l}odowska-Curie grant agreement
No~838996 (T. R.) and 101065117 (C. S.).
Computations were enabled by resources provided by the National Academic
Infrastructure for Supercomputing in Sweden (NAISS) at NSC, PDC, and C3SE
partially funded by the Swedish Research Council through grant agreement
no. 2022-06725.

%% file: main.bbl
\begin{thebibliography}{100}

\bibitem{hohenberg1964inhomogeneous}
Pierre Hohenberg and Walter Kohn.
\newblock Inhomogeneous electron gas.
\newblock {\em Phys. Rev.}, 136(3B):B864, 1964.

\bibitem{kohn1965self}
Walter Kohn and Lu~Jeu Sham.
\newblock Self-consistent equations including exchange and correlation effects.
\newblock {\em Phys. Rev.}, 140(4A):A1133, 1965.

\bibitem{perdew1996generalized}
John~P Perdew, Kieron Burke, and Matthias Ernzerhof.
\newblock Generalized gradient approximation made simple.
\newblock {\em Phys. Rev. Lett.}, 77(18):3865, 1996.

\bibitem{becke1993density}
Axel~D Becke.
\newblock Density-functional thermochemistry. iii. the role of exact exchange.
\newblock {\em The Journal of Chemical Physics}, 98:5648--5652, 1993.

\bibitem{heyd2003hybrid}
Jochen Heyd, Gustavo~E Scuseria, and Matthias Ernzerhof.
\newblock Hybrid functionals based on a screened coulomb potential.
\newblock {\em The Journal of Chemical Physics}, 118(18):8207--8215, 2003.

\bibitem{torrie1977nonphysical}
Glenn~M Torrie and John~P Valleau.
\newblock Nonphysical sampling distributions in monte carlo free-energy
  estimation: Umbrella sampling.
\newblock {\em J. Comput. Phys.}, 23(2):187--199, 1977.

\bibitem{souvatzis2008entropy}
Petros Souvatzis, Olle Eriksson, MI~Katsnelson, and SP~Rudin.
\newblock Entropy driven stabilization of energetically unstable crystal
  structures explained from first principles theory.
\newblock {\em Phys. Rev. Lett.}, 100(9):095901, 2008.

\bibitem{hybertsen1986electron}
Mark~S. Hybertsen and Steven~G. Louie.
\newblock Electron correlation in semiconductors and insulators: Band gaps and
  quasiparticle energies.
\newblock {\em Phys. Rev. B}, 34:5390--5413, Oct 1986.

\bibitem{golze2019gw}
Dorothea Golze, Marc Dvorak, and Patrick Rinke.
\newblock The {{GW}} compendium: {{A}} practical guide to theoretical
  photoemission spectroscopy.
\newblock {\em Front. Chem.}, 7:377, 2019.

\bibitem{casida1998molecular}
Mark~E Casida, Christine Jamorski, Kim~C Casida, and Dennis~R Salahub.
\newblock Molecular excitation energies to high-lying bound states from
  time-dependent density-functional response theory: Characterization and
  correction of the time-dependent local density approximation ionization
  threshold.
\newblock {\em The Journal of Chemical Physics}, 108(11):4439--4449, 1998.

\bibitem{onida2002electronic}
Giovanni Onida, Lucia Reining, and Angel Rubio.
\newblock Electronic excitations: density-functional versus many-body
  green's-function approaches.
\newblock {\em Rev. Mod. Phys.}, 74(2):601, 2002.

\bibitem{norskov2009towards}
Jens~Kehlet N{\o}rskov, Thomas Bligaard, Jan Rossmeisl, and Claus~Hviid
  Christensen.
\newblock Towards the computational design of solid catalysts.
\newblock {\em Nat. Chem.}, 1(1):37--46, 2009.

\bibitem{jain2013commentary}
Anubhav Jain, Shyue~Ping Ong, Geoffroy Hautier, Wei Chen, William~Davidson
  Richards, Stephen Dacek, Shreyas Cholia, Dan Gunter, David Skinner, Gerbrand
  Ceder, et~al.
\newblock Commentary: The materials project: A materials genome approach to
  accelerating materials innovation.
\newblock {\em APL Mater.}, 1(1):011002, 2013.

\bibitem{curtarolo2013high}
Stefano Curtarolo, Gus~LW Hart, Marco~Buongiorno Nardelli, Natalio Mingo,
  Stefano Sanvito, and Ohad Levy.
\newblock The high-throughput highway to computational materials design.
\newblock {\em Nat. Mater.}, 12(3):191--201, 2013.

\bibitem{haastrup2018computational}
Sten Haastrup, Mikkel Strange, Mohnish Pandey, Thorsten Deilmann, Per~S
  Schmidt, Nicki~F Hinsche, Morten~N Gjerding, Daniele Torelli, Peter~M Larsen,
  Anders~C Riis-Jensen, et~al.
\newblock The computational 2d materials database: high-throughput modeling and
  discovery of atomically thin crystals.
\newblock {\em 2D Mater.}, 5(4):042002, 2018.

\bibitem{castelli2012computational}
Ivano~E Castelli, Thomas Olsen, Soumendu Datta, David~D Landis, S{\o}ren Dahl,
  Kristian~S Thygesen, and Karsten~W Jacobsen.
\newblock Computational screening of perovskite metal oxides for optimal solar
  light capture.
\newblock {\em Energ. Environ. Sci.}, 5(2):5814--5819, 2012.

\bibitem{marek2014elpa}
Andreas Marek, Volker Blum, Rainer Johanni, Ville Havu, Bruno Lang, Thomas
  Auckenthaler, Alexander Heinecke, Hans-Joachim Bungartz, and Hermann Lederer.
\newblock The elpa library: scalable parallel eigenvalue solutions for
  electronic structure theory and computational science.
\newblock {\em Journal of Physics: Condensed Matter}, 26(21):213201, 2014.

\bibitem{liu2016cubic}
Peitao Liu, Merzuk Kaltak, Ji{\v{r}}{\'\i} Klime{\v{s}}, and Georg Kresse.
\newblock Cubic scaling {{GW}}: Towards fast quasiparticle calculations.
\newblock {\em Phys. Rev. B}, 94(16):165109, 2016.

\bibitem{jain2015fireworks}
Anubhav Jain, Shyue~Ping Ong, Wei Chen, Bharat Medasani, Xiaohui Qu, Michael
  Kocher, Miriam Brafman, Guido Petretto, Gian-Marco Rignanese, Geoffroy
  Hautier, et~al.
\newblock Fireworks: a dynamic workflow system designed for high-throughput
  applications.
\newblock {\em Concurr. Comp-pract. E.}, 27(17):5037--5059, 2015.

\bibitem{pizzi2016aiida}
Giovanni Pizzi, Andrea Cepellotti, Riccardo Sabatini, Nicola Marzari, and Boris
  Kozinsky.
\newblock Aiida: automated interactive infrastructure and database for
  computational science.
\newblock {\em Nato. Sc. S. Ss. Iii. C. S.}, 111:218--230, 2016.

\bibitem{myqueue}
Jens~Jørgen Mortensen, Morten Gjerding, and Kristian~Sommer Thygesen.
\newblock {MyQueue}: Task and workflow scheduling system.
\newblock {\em J. Open Source Softw.}, 5(45):1844, 2020.

\bibitem{gjerding2021atomic}
Morten Gjerding, Thorbj{\o}rn Skovhus, Asbj{\o}rn Rasmussen, Fabian Bertoldo,
  Ask~Hjorth Larsen, Jens~J{\o}rgen Mortensen, and Kristian~Sommer Thygesen.
\newblock Atomic simulation recipes: A python framework and library for
  automated workflows.
\newblock {\em Nato. Sc. S. Ss. Iii. C. S.}, 199:110731, 2021.

\bibitem{schmidt2019recent}
Jonathan Schmidt, M{\'a}rio~RG Marques, Silvana Botti, and Miguel~AL Marques.
\newblock Recent advances and applications of machine learning in solid-state
  materials science.
\newblock {\em npj Comput. Mater.}, 5(1):83, 2019.

\bibitem{bartok2010gaussian}
Albert~P Bart{\'o}k, Mike~C Payne, Risi Kondor, and G{\'a}bor Cs{\'a}nyi.
\newblock Gaussian approximation potentials: The accuracy of quantum mechanics,
  without the electrons.
\newblock {\em Phys. Rev. Lett.}, 104(13):136403, 2010.

\bibitem{rupp2012fast}
Matthias Rupp, Alexandre Tkatchenko, Klaus-Robert M{\"u}ller, and O~Anatole
  Von~Lilienfeld.
\newblock Fast and accurate modeling of molecular atomization energies with
  machine learning.
\newblock {\em Phys. Rev. Lett.}, 108(5):058301, 2012.

\bibitem{teale2022dft}
Andrew~M Teale, Trygve Helgaker, Andreas Savin, Carlo Adamo, B{\'a}lint Aradi,
  Alexei~V Arbuznikov, Paul~W Ayers, Evert~Jan Baerends, Vincenzo Barone,
  Patrizia Calaminici, et~al.
\newblock {{DFT}} exchange: sharing perspectives on the workhorse of quantum
  chemistry and materials science.
\newblock {\em Phys. Chem. Chem. Phys.}, 24(47):28700--28781, 2022.

\bibitem{ren2012random}
Xinguo Ren, Patrick Rinke, Christian Joas, and Matthias Scheffler.
\newblock Random-phase approximation and its applications in computational
  chemistry and materials science.
\newblock {\em J. Mater. Sci.}, 47:7447--7471, 2012.

\bibitem{sun2015strongly}
Jianwei Sun, Adrienn Ruzsinszky, and John~P Perdew.
\newblock Strongly constrained and appropriately normed semilocal density
  functional.
\newblock {\em Phys. Rev. Lett.}, 115(3):036402, 2015.

\bibitem{olsen2013random}
Thomas Olsen and Kristian~S Thygesen.
\newblock Random phase approximation applied to solids, molecules, and
  graphene-metal interfaces: {{From}} van der {{Waals}} to covalent bonding.
\newblock {\em Phys. Rev. B}, 87(7):075111, 2013.

\bibitem{tkatchenko2009accurate}
Alexandre Tkatchenko and Matthias Scheffler.
\newblock Accurate molecular van der {{Waals}} interactions from ground-state
  electron density and free-atom reference data.
\newblock {\em Phys. Rev. Lett.}, 102(7):073005, 2009.

\bibitem{dion2004van}
Max Dion, Henrik Rydberg, Elsebeth Schr{\"o}der, David~C Langreth, and Bengt~I
  Lundqvist.
\newblock Van der {{Waals}} density functional for general geometries.
\newblock {\em Phys. Rev. Lett.}, 92(24):246401, 2004.

\bibitem{gpaw1}
J.~J. Mortensen, L.~B. Hansen, and K.~W. Jacobsen.
\newblock Real-space grid implementation of the projector augmented wave
  method.
\newblock {\em Phys. Rev. B}, 71:035109, Jan 2005.

\bibitem{gpaw2}
J~Enkovaara, C~Rostgaard, J~J Mortensen, J~Chen, M~Dułak, L~Ferrighi,
  J~Gavnholt, C~Glinsvad, V~Haikola, H~A Hansen, H~H Kristoffersen, M~Kuisma,
  A~H Larsen, L~Lehtovaara, M~Ljungberg, O~Lopez-Acevedo, P~G Moses, J~Ojanen,
  T~Olsen, V~Petzold, N~A Romero, J~Stausholm-Møller, M~Strange, G~A
  Tritsaris, M~Vanin, M~Walter, B~Hammer, H~Häkkinen, G~K~H Madsen, R~M
  Nieminen, J~K Nørskov, M~Puska, T~T Rantala, J~Schiøtz, K~S Thygesen, and
  K~W Jacobsen.
\newblock Electronic structure calculations with {GPAW}: a real-space
  implementation of the projector augmented-wave method.
\newblock {\em Journal of Physics: Condensed Matter}, 22(25):253202, jun 2010.

\bibitem{WalHakLeh08}
Michael Walter, Hannu H{\"a}kkinen, Lauri Lehtovaara, Martti Puska, Jussi
  Enkovaara, Carsten Rostgaard, and Jens~J{\o}rgen Mortensen.
\newblock Time-dependent density-functional theory in the projector
  augmented-wave method.
\newblock {\em The Journal of Chemical Physics}, 128(24):244101, June 2008.

\bibitem{larsen2009localized}
A.~H. Larsen, M.~Vanin, J.~J. Mortensen, K.~S. Thygesen, and K.~W. Jacobsen.
\newblock Localized atomic basis set in the projector augmented wave method.
\newblock {\em Phys. Rev. B}, 80:195112, Nov 2009.

\bibitem{yan2011linear}
Jun Yan, Jens~J Mortensen, Karsten~W Jacobsen, and Kristian~S Thygesen.
\newblock Linear density response function in the projector augmented wave
  method: Applications to solids, surfaces, and interfaces.
\newblock {\em Phys. Rev. B}, 83(24):245122, 2011.

\bibitem{olsen2012extending}
Thomas Olsen and Kristian~S Thygesen.
\newblock Extending the random-phase approximation for electronic correlation
  energies: The renormalized adiabatic local density approximation.
\newblock {\em Phys. Rev. B}, 86(8):081103, 2012.

\bibitem{huser2013quasiparticle}
Falco H{\"u}ser, Thomas Olsen, and Kristian~S Thygesen.
\newblock Quasiparticle gw calculations for solids, molecules, and
  two-dimensional materials.
\newblock {\em Phys. Rev. B}, 87(23):235132, 2013.

\bibitem{yan2012optical}
Jun Yan, Karsten~W Jacobsen, and Kristian~S Thygesen.
\newblock Optical properties of bulk semiconductors and graphene/boron nitride:
  The bethe-salpeter equation with derivative discontinuity-corrected density
  functional energies.
\newblock {\em Phys. Rev. B}, 86(4):045208, 2012.

\bibitem{Sodequist2023a}
Joachim Sødequist and Thomas Olsen.
\newblock Type ii multiferroic order in two-dimensional transition metal
  halides from first principles spin-spiral calculations.
\newblock {\em 2D Mater.}, 10(3):035016, may 2023.

\bibitem{Torelli2018}
Daniele Torelli and Thomas Olsen.
\newblock {Calculating critical temperatures for ferromagnetic order in
  two-dimensional materials}.
\newblock {\em 2D Mater.}, 6:015028, 2018.

\bibitem{Durhuus2023}
Frederik~L Durhuus, Thorbjørn Skovhus, and Thomas Olsen.
\newblock Plane wave implementation of the magnetic force theorem for magnetic
  exchange constants: application to bulk fe, co and ni.
\newblock {\em Journal of Physics: Condensed Matter}, 35(10):105802, jan 2023.

\bibitem{Skovhus2021}
Thorbj{\o}rn Skovhus and Thomas Olsen.
\newblock {Dynamic transverse magnetic susceptibility in the projector
  augmented-wave method: Application to Fe, Ni, and Co}.
\newblock {\em Phys. Rev. B}, 103:245110, 2021.

\bibitem{Kruse2023}
Mads Kruse, Urko Petralanda, Morten~N. Gjerding, Karsten~W. Jacobsen,
  Kristian~S. Thygesen, and Thomas Olsen.
\newblock {Two-dimensional ferroelectrics from high throughput computational
  screening}.
\newblock {\em npj Comput. Mater.}, 9:45, 2023.

\bibitem{Gjerding2021}
Morten~Niklas Gjerding, Alireza Taghizadeh, Asbj{\o}rn Rasmussen, Sajid Ali,
  Fabian Bertoldo, Thorsten Deilmann, Nikolaj~R{\o}rb{\ae}k Kn{\o}sgaard, Mads
  Kruse, Ask~Hjorth Larsen, Simone Manti, Thomas~Garm Pedersen, Urko
  Petralanda, Thorbj{\o}rn Skovhus, Mark~Kamper Svendsen, Jens~J{\o}rgen
  Mortensen, Thomas Olsen, and Kristian~Sommer Thygesen.
\newblock {Recent progress of the computational 2D materials database (C2DB)}.
\newblock {\em 2D Mater.}, 8:044002, 2021.

\bibitem{Olsen2018}
Thomas Olsen, Erik Andersen, Takuya Okugawa, Daniele Torelli, Thorsten
  Deilmann, and Kristian~S. Thygesen.
\newblock {Discovering two-dimensional topological insulators from
  high-throughput computations}.
\newblock {\em Phys. Rev. Mater.}, 3:024005, 2019.

\bibitem{larsen2017atomic}
Ask~Hjorth Larsen, Jens~Jørgen Mortensen, Jakob Blomqvist, Ivano~E Castelli,
  Rune Christensen, Marcin Dułak, Jesper Friis, Michael~N Groves, Bjørk
  Hammer, Cory Hargus, Eric~D Hermes, Paul~C Jennings, Peter~Bjerre Jensen,
  James Kermode, John~R Kitchin, Esben~Leonhard Kolsbjerg, Joseph Kubal,
  Kristen Kaasbjerg, Steen Lysgaard, Jón~Bergmann Maronsson, Tristan Maxson,
  Thomas Olsen, Lars Pastewka, Andrew Peterson, Carsten Rostgaard, Jakob
  Schiøtz, Ole Schütt, Mikkel Strange, Kristian~S Thygesen, Tejs Vegge, Lasse
  Vilhelmsen, Michael Walter, Zhenhua Zeng, and Karsten~W Jacobsen.
\newblock The atomic simulation environment—a python library for working with
  atoms.
\newblock {\em Journal of Physics: Condensed Matter}, 29(27):273002, jun 2017.

\bibitem{mostofi2008wannier90}
Arash~A Mostofi, Jonathan~R Yates, Young-Su Lee, Ivo Souza, David Vanderbilt,
  and Nicola Marzari.
\newblock wannier90: A tool for obtaining maximally-localised wannier
  functions.
\newblock {\em Comput. Phys. Commun.}, 178(9):685--699, 2008.

\bibitem{Schmerwitz2023}
Yorick L.~A. Schmerwitz, Gianluca Levi, and Hannes J{\'{o}}nsson.
\newblock {Calculations of Excited Electronic States by Converging on Saddle
  Points Using Generalized Mode Following}.
\newblock {\em J. Chem. Theory Comput.}, 19(12):3634--3651, 2023.

\bibitem{Ivanov2021}
Aleksei~V. Ivanov, Gianluca Levi, Elvar~\"O. J\'onsson, and Hannes J\'onsson.
\newblock Method for calculating excited electronic states using density
  functionals and direct orbital optimization with real space grid or
  plane-wave basis set.
\newblock {\em J. Chem. Theory Comput.}, 17:5034--5049, 8 2021.

\bibitem{Levi2020jctc}
Gianluca Levi, Aleksei~V. Ivanov, and Hannes J\'onsson.
\newblock Variational density functional calculations of excited states via
  direct optimization.
\newblock {\em J. Chem. Theory Comput.}, 16:6968--6982, 11 2020.

\bibitem{KuiSakRos15}
M.~Kuisma, A.~Sakko, T.~P. Rossi, A.~H. Larsen, J.~Enkovaara, L.~Lehtovaara,
  and T.~T. Rantala.
\newblock Localized surface plasmon resonance in silver nanoparticles:
  {{Atomistic}} first-principles time-dependent density-functional theory
  calculations.
\newblock {\em Phys. Rev. B}, 91(11):115431, 2015.

\bibitem{RosLehSak15}
Tuomas~P. Rossi, Susi Lehtola, Arto Sakko, Martti~J. Puska, and Risto~M.
  Nieminen.
\newblock Nanoplasmonics simulations at the basis set limit through
  completeness-optimized, local numerical basis sets.
\newblock {\em The Journal of Chemical Physics}, 142(9):094114, 2015.

\bibitem{RosKuiPus17}
Tuomas~P. Rossi, Mikael Kuisma, Martti~J. Puska, Risto~M. Nieminen, and Paul
  Erhart.
\newblock Kohn\textendash{{Sham Decomposition}} in {{Real-Time Time-Dependent
  Density-Functional Theory}}: {{An Efficient Tool}} for {{Analyzing Plasmonic
  Excitations}}.
\newblock {\em J. Chem. Theory Comput.}, 13(10):4779--4790, 2017.

\bibitem{makkonen2021real}
Esko Makkonen, Tuomas~P Rossi, Ask~Hjorth Larsen, Olga Lopez-Acevedo, Patrick
  Rinke, Mikael Kuisma, and Xi~Chen.
\newblock Real-time time-dependent density functional theory implementation of
  electronic circular dichroism applied to nanoscale metal--organic clusters.
\newblock {\em The Journal of Chemical Physics}, 154(11):114102, 2021.

\bibitem{taghizadeh2020library}
Alireza Taghizadeh, Ulrik Leffers, Thomas~G Pedersen, and Kristian~S Thygesen.
\newblock A library of ab initio raman spectra for automated identification of
  2d materials.
\newblock {\em Nat. Commun.}, 11(1):3011, 2020.

\bibitem{taghizadeh2021two}
Alireza Taghizadeh, Kristian~S Thygesen, and Thomas~G Pedersen.
\newblock Two-dimensional materials with giant optical nonlinearities near the
  theoretical upper limit.
\newblock {\em Acs Nano}, 15(4):7155--7167, 2021.

\bibitem{sauer2023shift}
Mikkel~Ohm Sauer, Alireza Taghizadeh, Urko Petralanda, Martin Ovesen,
  Kristian~Sommer Thygesen, Thomas Olsen, Horia Cornean, and Thomas~Garm
  Pedersen.
\newblock Shift current photovoltaic efficiency of 2d materials.
\newblock {\em npj Comput. Mater.}, 9(1):35, 2023.

\bibitem{gitlab}
{GPAW}'s git-repository: \url{https://gitlab.com/gpaw/gpaw/}.

\bibitem{pytest}
Holger Krekel, Bruno Oliveira, Ronny Pfannschmidt, Floris Bruynooghe, Brianna
  Laugher, and Florian Bruhin.
\newblock pytest, 2004.

\bibitem{tutorials}
{GPAW} tutorials and exercises:
  \url{https://wiki.fysik.dtu.dk/gpaw/tutorialsexercises/tutorialsexercises.html}.

\bibitem{fftw}
Matteo Frigo and Steven~G. Johnson.
\newblock The design and implementation of {FFTW3}.
\newblock {\em P. Ieee}, 93(2):216--231, 2005.
\newblock Special issue on ``Program Generation, Optimization, and Platform
  Adaptation''.

\bibitem{scalapack}
L.~S. Blackford, J.~Choi, A.~Cleary, E.~D'Azevedo, J.~Demmel, I.~Dhillon,
  J.~Dongarra, S.~Hammarling, G.~Henry, A.~Petitet, K.~Stanley, D.~Walker, and
  R.~C. Whaley.
\newblock {\em {ScaLAPACK} Users' Guide}.
\newblock Society for Industrial and Applied Mathematics, Philadelphia, PA,
  1997.

\bibitem{libxc}
Miguel~A.L. Marques, Micael~J.T. Oliveira, and Tobias Burnus.
\newblock Libxc: A library of exchange and correlation functionals for density
  functional theory.
\newblock {\em Comput. Phys. Commun.}, 183(10):2272--2281, 2012.

\bibitem{lehtola_recent_2018}
Susi Lehtola, Conrad Steigemann, Micael J.~T. Oliveira, and Miguel A.~L.
  Marques.
\newblock Recent developments in libxc \textemdash{} {{A}} comprehensive
  library of functionals for density functional theory.
\newblock {\em SoftwareX}, 7:1--5, January 2018.

\bibitem{libvdwxc}
Ask~Hjorth Larsen, Mikael Kuisma, Joakim Löfgren, Yann Pouillon, Paul Erhart,
  and Per Hyldgaard.
\newblock libvdwxc: a library for exchange–correlation functionals in the
  vdw-df family.
\newblock {\em Model. Simul. Mater. Sc.}, 25(6):065004, jun 2017.

\bibitem{numpy}
Charles~R. Harris, K.~Jarrod Millman, St{\'{e}}fan~J. van~der Walt, Ralf
  Gommers, Pauli Virtanen, David Cournapeau, Eric Wieser, Julian Taylor,
  Sebastian Berg, Nathaniel~J. Smith, Robert Kern, Matti Picus, Stephan Hoyer,
  Marten~H. van Kerkwijk, Matthew Brett, Allan Haldane, Jaime~Fern{\'{a}}ndez
  del R{\'{i}}o, Mark Wiebe, Pearu Peterson, Pierre G{\'{e}}rard-Marchant,
  Kevin Sheppard, Tyler Reddy, Warren Weckesser, Hameer Abbasi, Christoph
  Gohlke, and Travis~E. Oliphant.
\newblock Array programming with {NumPy}.
\newblock {\em Nature}, 585(7825):357--362, September 2020.

\bibitem{2020SciPy-NMeth}
Pauli Virtanen, Ralf Gommers, Travis~E. Oliphant, Matt Haberland, Tyler Reddy,
  David Cournapeau, Evgeni Burovski, Pearu Peterson, Warren Weckesser, Jonathan
  Bright, St{\'e}fan~J. {van der Walt}, Matthew Brett, Joshua Wilson, K.~Jarrod
  Millman, Nikolay Mayorov, Andrew R.~J. Nelson, Eric Jones, Robert Kern, Eric
  Larson, C~J Carey, {\.I}lhan Polat, Yu~Feng, Eric~W. Moore, Jake
  {VanderPlas}, Denis Laxalde, Josef Perktold, Robert Cimrman, Ian Henriksen,
  E.~A. Quintero, Charles~R. Harris, Anne~M. Archibald, Ant{\^o}nio~H. Ribeiro,
  Fabian Pedregosa, Paul {van Mulbregt}, and {SciPy 1.0 Contributors}.
\newblock {{SciPy} 1.0: Fundamental Algorithms for Scientific Computing in
  Python}.
\newblock {\em Nat. Methods}, 17:261--272, 2020.

\bibitem{Blochl1994}
P.~E. Bl\"ochl.
\newblock Projector augmented-wave method.
\newblock {\em Phys. Rev. B}, 50:17953--17979, Dec 1994.

\bibitem{susi_efficient_2019}
Toma Susi, Jacob Madsen, Ursula Ludacka, Jens~Jørgen Mortensen, Timothy~J.
  Pennycook, Zhongbo Lee, Jani Kotakoski, Ute Kaiser, and Jannik~C. Meyer.
\newblock Efficient first principles simulation of electron scattering factors
  for transmission electron microscopy.
\newblock {\em Ultramicroscopy}, 197:16--22, February 2019.

\bibitem{madsen_ab_2021}
Jacob Madsen, Timothy~J. Pennycook, and Toma Susi.
\newblock ab initio description of bonding for transmission electron
  microscopy.
\newblock {\em Ultramicroscopy}, 231:113253, December 2021.

\bibitem{madsen_abtem_2021}
Jacob Madsen and Toma Susi.
\newblock {abTEM}: transmission electron microscopy from first principles.
\newblock {\em Open Research Europe}, 1(24):13015, 2021.

\bibitem{Kresse1996}
G.~Kresse and J.~Furthm\"uller.
\newblock Efficient iterative schemes for ab initio total-energy calculations
  using a plane-wave basis set.
\newblock {\em Phys. Rev. B}, 54:11169--11186, Oct 1996.

\bibitem{cold-smearing}
Nicola Marzari, David Vanderbilt, Alessandro De~Vita, and M.~C. Payne.
\newblock Thermal contraction and disordering of the al(110) surface.
\newblock {\em Phys. Rev. Lett.}, 82:3296--3299, Apr 1999.

\bibitem{tetra}
Peter~E. Bl\"ochl, O.~Jepsen, and O.~K. Andersen.
\newblock Improved tetrahedron method for brillouin-zone integrations.
\newblock {\em Phys. Rev. B}, 49:16223--16233, Jun 1994.

\bibitem{Lehtola2020}
Susi Lehtola, Frank Blockhuys, and Christian {Van Alsenoy}.
\newblock {An overview of self-consistent field calculations within finite
  basis sets}.
\newblock {\em Molecules}, 25(5):1--23, 2020.

\bibitem{Voorhis2002}
Troy~Van Voorhis and Martin Head-Gordon.
\newblock {A geometric approach to direct minimization}.
\newblock {\em Mol. Phys.}, 100:1713--1721, 2002.

\bibitem{Payne1992}
M.~C. Payne, M.~P. Teter, D.~C. Allan, T.~A. Arias, and J.~D. Joannopoulos.
\newblock {Iterative minimization techniques for ab initio total-energy
  calculations: Molecular dynamics and conjugate gradients}.
\newblock {\em Rev. Mod. Phys.}, 64(4):1045--1097, 1992.

\bibitem{Head-Gordon1988}
Martin Head-Gordon and John~A. Pople.
\newblock {Optimization of wave function and geometry in the finite basis
  Hartree-Fock method}.
\newblock {\em J. Phys. Chem-us.}, 92(11):3063--3069, 1988.

\bibitem{ivanov_directmin_2021}
Aleksei~V. Ivanov, Elvar~\"O. J\'onsson, Tejs Vegge, and Hannes J\'onsson.
\newblock Direct energy minimization based on exponential transformation in
  density functional calculations of finite and extended systems.
\newblock {\em Comput. Phys. Commun.}, 267:108047, 2021.

\bibitem{paw-xml}
{PAW-XML} specification.
\newblock \url{https://esl.cecam.org/data/paw-xml/}.

\bibitem{abinit}
Xavier Gonze, Bernard Amadon, Gabriel Antonius, Frédéric Arnardi, Lucas
  Baguet, Jean-Michel Beuken, Jordan Bieder, François Bottin, Johann Bouchet,
  Eric Bousquet, Nils Brouwer, Fabien Bruneval, Guillaume Brunin, Théo
  Cavignac, Jean-Baptiste Charraud, Wei Chen, Michel Côté, Stefaan Cottenier,
  Jules Denier, Grégory Geneste, Philippe Ghosez, Matteo Giantomassi, Yannick
  Gillet, Olivier Gingras, Donald~R. Hamann, Geoffroy Hautier, Xu~He, Nicole
  Helbig, Natalie Holzwarth, Yongchao Jia, François Jollet, William
  Lafargue-Dit-Hauret, Kurt Lejaeghere, Miguel~A.L. Marques, Alexandre Martin,
  Cyril Martins, Henrique~P.C. Miranda, Francesco Naccarato, Kristin Persson,
  Guido Petretto, Valentin Planes, Yann Pouillon, Sergei Prokhorenko, Fabio
  Ricci, Gian-Marco Rignanese, Aldo~H. Romero, Michael~Marcus Schmitt, Marc
  Torrent, Michiel~J. {van Setten}, Benoit {Van Troeye}, Matthieu~J.
  Verstraete, Gilles Zérah, and Josef~W. Zwanziger.
\newblock The {{ABINIT}} project: Impact, environment and recent developments.
\newblock {\em Comput. Phys. Commun.}, 248:107042, 2020.

\bibitem{JTH}
François Jollet, Marc Torrent, and Natalie Holzwarth.
\newblock Generation of projector augmented-wave atomic data: {{A}} 71 element
  validated table in the {{XML}} format.
\newblock {\em Comput. Phys. Commun.}, 185(4):1246--1254, 2014.

\bibitem{hgh}
C.~Hartwigsen, S.~Goedecker, and J.~Hutter.
\newblock Relativistic separable dual-space gaussian pseudopotentials from h to
  rn.
\newblock {\em Phys. Rev. B}, 58:3641--3662, Aug 1998.

\bibitem{sg15}
Martin Schlipf and François Gygi.
\newblock Optimization algorithm for the generation of oncv pseudopotentials.
\newblock {\em Comput. Phys. Commun.}, 196:36--44, 2015.

\bibitem{cupy}
{CuPy}.
\newblock \url{https://cupy.dev/}.

\bibitem{hop}
Header only porting.
\newblock \url{https://github.com/mlouhivu/hop/}.

\bibitem{Hakala2013}
Samuli Hakala, Ville Havu, Jussi Enkovaara, and Risto~M. Nieminen.
\newblock Parallel electronic structure calculations using multiple graphics
  processing units (gpus).
\newblock {\em Lect. Notes Comput. Sc.}, 7782:63--76, 2013.

\bibitem{Hakala2015}
Samuli Hakala, Jussi Enkovaara, Ville Havu, Jun Yan, Lin Li, Chris O'Grady, and
  Risto~M. Nieminen.
\newblock Grid-based projector-augmented wave method.
\newblock {\em Method in Electronic Structure Calculations on Graphics
  Processing Units}, 2015.

\bibitem{perdew_jacobs_2001}
John~P. Perdew and Karla Schmidt.
\newblock Jacob's ladder of density functional approximations for the
  exchange-correlation energy.
\newblock In {\em {{AIP Conference Proceedings}}}, volume 577, pages 1--20.
  {AIP Publishing}, July 2001.

\bibitem{tao_climbing_2003}
Jianmin Tao, John~P. Perdew, Viktor~N. Staroverov, and Gustavo~E. Scuseria.
\newblock Climbing the {{Density Functional Ladder}}: {{Nonempirical
  Meta}}-{{Generalized Gradient Approximation Designed}} for {{Molecules}} and
  {{Solids}}.
\newblock {\em Phys. Rev. Lett.}, 91(14):146401, September 2003.

\bibitem{perdew_accurate_1992}
John~P. Perdew and Yue Wang.
\newblock Accurate and simple analytic representation of the electron-gas
  correlation energy.
\newblock {\em Phys. Rev. B}, 45(23):13244--13249, June 1992.

\bibitem{perez2009}
Guillermo Rom\'an-P\'erez and Jos\'e~M. Soler.
\newblock Efficient implementation of a van der waals density functional:
  Application to double-wall carbon nanotubes.
\newblock {\em Phys. Rev. Lett.}, 103:096102, Aug 2009.

\bibitem{GriLeeLen95}
Oleg Gritsenko, Robert van Leeuwen, Erik van Lenthe, and Evert~Jan Baerends.
\newblock Self-consistent approximation to the {{Kohn-Sham}} exchange
  potential.
\newblock {\em Phys. Rev. A}, 51(3):1944, 1995.

\bibitem{KuiOjaEnk10}
M.~Kuisma, J.~Ojanen, J.~Enkovaara, and T.~T. Rantala.
\newblock Kohn-{{Sham}} potential with discontinuity for band gap materials.
\newblock {\em Phys. Rev. B}, 82(11):115106, 2010.

\bibitem{YanJacThy11}
Jun Yan, Karsten~W. Jacobsen, and Kristian~S. Thygesen.
\newblock First-principles study of surface plasmons on {{Ag}}(111) and
  {{H}}/{{Ag}}(111).
\newblock {\em Phys. Rev. B}, 84(23):235430, 2011.

\bibitem{YanJacThy12}
Jun Yan, Karsten~W. Jacobsen, and Kristian~S. Thygesen.
\newblock Conventional and acoustic surface plasmons on noble metal surfaces:
  {{A}} time-dependent density functional theory study.
\newblock {\em Phys. Rev. B}, 86(24):241404, 2012.

\bibitem{RahTibRos20}
J.~Magnus Rahm, Christopher Tiburski, Tuomas~P. Rossi, Ferry Anggoro~Ardy
  Nugroho, Sara Nilsson, Christoph Langhammer, and Paul Erhart.
\newblock A library of late transition metal alloy dielectric functions for
  nanophotonic applications.
\newblock {\em Adv. Funct. Mater.}, 30(35):2002122, 2020.

\bibitem{Bashyal_2018}
Keshab Bashyal, Christopher~K Pyles, Sajjad Afroosheh, Aneer Lamichhane, and
  Alexey~T Zayak.
\newblock Empirical optimization of {{DFT}}+{{U}} and {{HSE}} for the band
  structure of {{ZnO}}.
\newblock {\em Journal of Physics: Condensed Matter}, 30(6):065501, jan 2018.

\bibitem{PhysRevLett.105.196403}
M.~K.~Y. Chan and G.~Ceder.
\newblock Efficient band gap prediction for solids.
\newblock {\em Phys. Rev. Lett.}, 105:196403, Nov 2010.

\bibitem{https://doi.org/10.1002/qua.560280846}
John~P. Perdew.
\newblock Density functional theory and the band gap problem.
\newblock {\em Int. J. Quantum Chem.}, 28(S19):497--523, 1985.

\bibitem{PhysRevB.101.165117}
Kevin~J. May and Alexie~M. Kolpak.
\newblock Improved description of perovskite oxide crystal structure and
  electronic properties using self-consistent {{Hubbard}} $u$ corrections from
  {{ACBN0}}.
\newblock {\em Phys. Rev. B}, 101:165117, Apr 2020.

\bibitem{PhysRevB.57.1505}
S.~L. Dudarev, G.~A. Botton, S.~Y. Savrasov, C.~J. Humphreys, and A.~P. Sutton.
\newblock Electron-energy-loss spectra and the structural stability of nickel
  oxide: {{An LSDA+U study}}.
\newblock {\em Phys. Rev. B}, 57:1505--1509, Jan 1998.

\bibitem{PhysRevB.52.R5467}
A.~I. Liechtenstein, V.~I. Anisimov, and J.~Zaanen.
\newblock Density-functional theory and strong interactions: Orbital ordering
  in mott-hubbard insulators.
\newblock {\em Phys. Rev. B}, 52:R5467--R5470, Aug 1995.

\bibitem{aryasetiawan2011constrained}
F~Aryasetiawan, T~Miyake, and R~Sakuma.
\newblock The constrained {{RPA}} method for calculating the {{Hubbard U}} from
  first-principles.
\newblock {\em The {{LDA+ DMFT}} approach to strongly correlated materials},
  2011.

\bibitem{Cococcioni_2005}
Matteo Cococcioni and Stefano de~Gironcoli.
\newblock Linear response approach to the calculation of the effective
  interaction parameters in the {{LDA+U}} method.
\newblock {\em Phys. Rev. B}, 71(3), jan 2005.

\bibitem{moore2022highthroughput}
Guy~C. Moore, Matthew~K. Horton, Alexander~M. Ganose, Martin Siron, Edward
  Linscott, David~D. O'Regan, and Kristin~A. Persson.
\newblock High-throughput determination of {{Hubbard U and Hund J}} values for
  transition metal oxides via linear response formalism, 2022.

\bibitem{PhysRevB.73.195107}
Lei Wang, Thomas Maxisch, and Gerbrand Ceder.
\newblock Oxidation energies of transition metal oxides within the
  $\mathrm{GGA}+\mathrm{U}$ framework.
\newblock {\em Phys. Rev. B}, 73:195107, May 2006.

\bibitem{Taib_2019}
M~F~M Taib, D~T Mustaffa, N~H Hussin, M~H Samat, A~M~M Ali, O~H Hassan, and
  M~Z~A Yahya.
\newblock {First principles study on Zn doped MgO using Hubbard U correction}.
\newblock {\em Materials Research Express}, 6(9):094012, jul 2019.

\bibitem{yu2020machine}
Maituo Yu, Shuyang Yang, Chunzhi Wu, and Noa Marom.
\newblock Machine learning the {{Hubbard U}} parameter in {{DFT}}+{{U}} using
  {{Bayesian}} optimization.
\newblock {\em npj Comput. Mater.}, 6(1):180, 2020.

\bibitem{tawada_long-range-corrected_2004}
Yoshihiro Tawada, Takao Tsuneda, Susumu Yanagisawa, Takeshi Yanai, and Kimihiko
  Hirao.
\newblock A long-range-corrected time-dependent density functional theory.
\newblock {\em The Journal of Chemical Physics}, 120(18):8425--8433, May 2004.

\bibitem{cohen_insights_2008}
Aron~J. Cohen, Paula {Mori-S{\'a}nchez}, and Weitao Yang.
\newblock Insights into {{Current Limitations}} of {{Density Functional
  Theory}}.
\newblock {\em Science}, 321(5890):792--794, August 2008.

\bibitem{Dreuw2003}
Andreas Dreuw, Jennifer~L. Weisman, and Martin Head-Gordon.
\newblock {Long-range charge-transfer excited states in time-dependent density
  functional theory require non-local exchange}.
\newblock {\em J. Chem. Phys.}, 119(6):2943--2946, 2003.

\bibitem{baerends_kohnsham_2013}
E.~J. Baerends, O.~V. Gritsenko, and R.~van Meer.
\newblock The {{Kohn}}\textendash{{Sham}} gap, the fundamental gap and the
  optical gap: The physical meaning of occupied and virtual
  {{Kohn}}\textendash{{Sham}} orbital energies.
\newblock {\em Phys. Chem. Chem. Phys.}, 15(39):16408--16425, September 2013.

\bibitem{kummel_charge-transfer_2017}
Stephan K{\"u}mmel.
\newblock Charge-{{Transfer Excitations}}: {{A Challenge}} for {{Time-Dependent
  Density Functional Theory That Has Been Met}}.
\newblock {\em Adv. Energy Mater.}, 7(16):1700440, August 2017.

\bibitem{baer_tuned_2010}
Roi Baer, Ester Livshits, and Ulrike Salzner.
\newblock Tuned {{Range-Separated Hybrids}} in {{Density Functional Theory}}.
\newblock {\em Annu. Rev. Phys. Chem.}, 61(1):85--109, 2010.

\bibitem{adamo_toward_1998}
Carlo Adamo and Vincenzo Barone.
\newblock Toward chemical accuracy in the computation of {{NMR}} shieldings:
  The {{PBE0}} model.
\newblock {\em Chem. Phys. Lett.}, 298(1\textendash 3):113--119, December 1998.

\bibitem{yanai_new_2004}
Takeshi Yanai, David~P Tew, and Nicholas~C Handy.
\newblock A new hybrid exchange\textendash correlation functional using the
  {{Coulomb-attenuating}} method ({{CAM-B3LYP}}).
\newblock {\em Chem. Phys. Lett.}, 393(1\textendash 3):51--57, July 2004.

\bibitem{akinaga_range-separation_2008}
Yoshinobu Akinaga and Seiichiro {Ten-no}.
\newblock Range-separation by the {{Yukawa}} potential in long-range corrected
  density functional theory with {{Gaussian-type}} basis functions.
\newblock {\em Chem. Phys. Lett.}, 462(4\textendash 6):348--351, September
  2008.

\bibitem{seth_range-separated_2012}
Michael Seth and Tom Ziegler.
\newblock Range-{{Separated Exchange Functionals}} with {{Slater-Type
  Functions}}.
\newblock {\em J. Chem. Theory Comput.}, 8(3):901--907, March 2012.

\bibitem{livshits_well-tempered_2007}
Ester Livshits and Roi Baer.
\newblock A well-tempered density functional theory of electrons in molecules.
\newblock {\em Phys. Chem. Chem. Phys.}, 9(23):2932--2941, June 2007.

\bibitem{wurdemann_berechnung_2016}
Rolf W{\"u}rdemann.
\newblock {\em {Berechnung optischer Spektren und Grundzustandseigenschaften
  neutraler und geladener Molek\"ule mittels Dichtefunktionaltheorie}}.
\newblock PhD thesis, Universit\"at Freiburg, {Freiburg, Germany}, July 2016.

\bibitem{wurdemann_charge_2018}
Rolf W{\"u}rdemann and Michael Walter.
\newblock Charge {{Transfer Excitations}} with {{Range Separated Functionals
  Using Improved Virtual Orbitals}}.
\newblock {\em J. Chem. Theory Comput.}, 14(7):3667--3676, July 2018.

\bibitem{perdew_zunger_1981}
Perdew~J. P. and A.~Zunger.
\newblock Self-interaction correction to density-functional approximations for
  many-electron systems.
\newblock {\em Phys. Rev. B}, 23:5048, 1981.

\bibitem{Kluepfel_2011}
S.~Kl\"upfel, P.~J. Kl\"upfel, and H.~J\'onsson.
\newblock Importance of complex orbitals in calculating the self-interaction
  corrected ground state of atoms.
\newblock {\em Phys. Rev. A}, 84:050501(R), 2011.

\bibitem{Lehtola_2014}
S.~Lehtola and H.~J\'onsson.
\newblock Variational, self-consistent implementation of the perdew-zunger
  self-interaction correction with complex optimal orbitals.
\newblock {\em J. Chem. Theory Comput.}, 10:5324, 2014.

\bibitem{Lehtola_2016}
S.~Lehtola, M.~Head-Gordon, and H.~J\'onsson.
\newblock Complex orbitals, multiple local minima, and symmetry breaking in
  perdew-zunger self-interaction corrected density functional theory
  calculations.
\newblock {\em J. Chem. Theory Comput.}, 12:3195, 2016.

\bibitem{Jonsson_2011}
H.~J\'onsson.
\newblock Simulation of surface processes.
\newblock {\em Proceedings of the National Academy of Sciences}, 108:944, 2011.

\bibitem{Kluepfel_2012}
S.~Kl\"upfel, P.~Kl\"upfel, and H.~J\'onsson.
\newblock The effect of the perdew-zunger self-interaction correction to
  density functionals on the energetics of small molecules.
\newblock {\em J. Chem. Phys.}, 137:124102, 2012.

\bibitem{Gudmundsdottir_2013}
H.~Gudmundsd\'ottir, Y.~Zhang, P.~M. Weber, and H.~J\'onsson.
\newblock Self-interaction corrected density functional calculations of
  molecular rydberg states.
\newblock {\em J. Chem. Phys.}, 139:194102, 2013.

\bibitem{IvanovMnDimer_2021}
A.~V. Ivanov, T.~Ghosh, E.~\"O. J\'onsson, and H.~J\'onsson.
\newblock Mn dimer can be described accurately with density functional
  calculations when self-interaction correction is applied.
\newblock {\em J. Phys. Chem. Lett.}, 12:4240, 2021.

\bibitem{Gudmundsdottir_2015}
H.~Gudmundsd\'ottir, E.~\"O. J\'onsson, and H.~J\'onsson.
\newblock Calculations of al dopant in alpha-quartz using a variational
  implementation of the perdew-zunger self-interaction correction.
\newblock {\em New J. Phys.}, 17:083006, 2015.

\bibitem{Schmerwitz2022}
Yorick L.~A. Schmerwitz, Aleksei~V. Ivanov, Elvar~\"O. J\'onsson, Hannes
  J\'onsson, and Gianluca Levi.
\newblock Variational density functional calculations of excited states:
  Conical intersection and avoided crossing in ethylene bond twisting.
\newblock {\em J. Phys. Chem. Lett.}, 13:3990--3999, 2022.

\bibitem{mortensen_bayesian_2005}
J.~J. Mortensen, K.~Kaasbjerg, S.~L. Frederiksen, J.~K. N{\o}rskov, J.~P.
  Sethna, and K.~W. Jacobsen.
\newblock Bayesian {{Error Estimation}} in {{Density-Functional Theory}}.
\newblock {\em Phys. Rev. Lett.}, 95(21):216401, November 2005.

\bibitem{BEEF}
Jess Wellendorff, Keld~T. Lundgaard, Andreas Møgelhøj, Vivien Petzold,
  David~D. Landis, Jens~K. Nørskov, Thomas Bligaard, and Karsten~W. Jacobsen.
\newblock Density functionals for surface science: {E}xchange-correlation model
  development with {B}ayesian error estimation.
\newblock {\em Phys. Rev. B}, 85:235149, 2012.

\bibitem{wellendorffMBEEFAccurateSemilocal2014}
Jess Wellendorff, Keld~T. Lundgaard, Karsten~W. Jacobsen, and Thomas Bligaard.
\newblock {mBEEF}: {An} accurate semi-local {Bayesian} error estimation density
  functional.
\newblock {\em The Journal of Chemical Physics}, 140(14):144107, April 2014.

\bibitem{mBEEF-vdW}
Keld~T. Lundgaard, Jess Wellendorff, J.~Voss, Karsten~W. Jacobsen, and Thomas
  Bligaard.
\newblock {mBEEF-vdW: Robust fitting of error estimation density functionals}.
\newblock {\em Phys. Rev. B}, 93(23):235162, 2016.

\bibitem{BEEFammonia}
Andrew~J. Medford, Jess Wellendorff, Aleksandra Vojvodic, Felix Studt, Frank
  Abild-Pedersen, Karsten~W. Jacobsen, Thomas Bligaard, and Jens~K. Nørskov.
\newblock Assessing the reliability of calculated catalytic ammonia synthesis
  rates.
\newblock {\em Science}, 345(6193):197--200, 2014.

\bibitem{Ciccotti1982}
G~Ciccotti, M~Ferrario, and J~P Ryckaert.
\newblock Molecular dynamics of rigid systems in cartesian coordinates a
  general formulation.
\newblock {\em Mol. Phys.}, 47:1253--1264, 1982.

\bibitem{Andy2014}
Andrew~A Peterson.
\newblock Global optimization of adsorbate--surface structures while preserving
  molecular identity.
\newblock {\em Top. Catal.}, 57:40--53, 2014.

\bibitem{Kaxiras1999}
E~B Tadmor, G~S Smith, N~Bernstein, and E~Kaxiras.
\newblock Mixed finite element and atomistic formulation for complex crystals.
\newblock {\em Phys. Rev. B}, 59:235, 1999.

\bibitem{Broyden}
C~G Broyden.
\newblock The convergence of a class of double-rank minimization algorithms.
\newblock {\em J. I. Math. Appl.}, 6:76--90, 1970.

\bibitem{Fletcher}
R~Fletcher.
\newblock A new approach to variable metric algorithms.
\newblock {\em Comput. J.}, 13:317--322, 1970.

\bibitem{Goldfarb}
D~Goldfarb.
\newblock A family of variable metric updates derived by variational means.
\newblock {\em Math. Comput.}, 24:23--26, 1970.

\bibitem{Shanno}
D~F Shanno.
\newblock Conditioning of quasi-newton methods for function minimization.
\newblock {\em Math. Comput.}, 24:647--656, 1970.

\bibitem{LBFGS}
D~Liu and J~Nocedal.
\newblock On the limited memory method for large scale optimization.
\newblock {\em Mathematical Programming B}, 45:503--528, 1989.

\bibitem{FIRE}
Erik Bitzek, Pekka Koskinen, Franz Gahler, Michael Moseler, and Peter Gumbsch.
\newblock Structural relaxation made simple.
\newblock {\em Phys. Rev. Lett.}, 97:170201, 2006.

\bibitem{Packwood2016}
D~Packwood, J~R Kermode, L~Mones, N~Bernstein, J~Woolley, N~Gould, C~Ortner,
  and G~Csanyi.
\newblock A universal preconditioner for simulating condensed phase materials.
\newblock {\em J. Chem. Phys.}, 144:164109, 2016.

\bibitem{GPMin}
Estefania~Garijo del Rio, Jens~J Mortensen, and Karsten~W Jacobsen.
\newblock Local bayesian optimizer for atomic structures.
\newblock {\em Phys. Rev. B}, 100:104103, 2019.

\bibitem{DimerMethod}
G~Henkelman and H~Jonsson.
\newblock A dimer method for finding saddle points on high dimensional
  potential surfaces using only first derivatives.
\newblock {\em J. Chem. Phys.}, 111:7010, 1999.

\bibitem{NEB}
H~Jonsson, G~Mills, and K~W Jacobsen.
\newblock in ‘classical and quantum dynamics in condensed phase systems’,
  edited by b. j. berne, g. cicotti, and d. f. coker.
\newblock {\em World Scientific}, 1998.

\bibitem{improvedtangent}
G~Henkelman and H~Jonsson.
\newblock Improved tangent estimate in the neb method for finding minimum
  energy paths and saddle points.
\newblock {\em J. Chem. Phys.}, 113:9978, 2000.

\bibitem{CI}
G~Henkelman, B~P Uberuaga, and H~Jonsson.
\newblock A climbing-image neb method for finding saddle points and minimum
  energy paths.

\bibitem{II}
S~Smidstrup, A~Pedersen, K~Stokbro, and H~Jonsson.
\newblock Improved initial guess for minimum energy path calculations.
\newblock {\em J. Chem. Phys.}, 140:214106, 2014.

\bibitem{KolsbergNEB}
E~L Kolsbjerg, M~N Groves, and B~Hammer.
\newblock An automated nudged elastic band method.
\newblock {\em J. Chem. Phys.}, 145:094107, 2016.

\bibitem{LindgrenNEB}
P~Lindgren, G~Kastlunger, and A~A Peterson.
\newblock Scaled and dynamic optimizations of nudged elastic bands.
\newblock {\em J. Chem. Theory Comput.}, 15:5787--5793, 2019.

\bibitem{MakriPrecon}
S~Makri, C~Ortner, and J~R Kermode.
\newblock A preconditioning scheme for minimum energy path finding methods.
\newblock {\em J. Chem. Phys.}, 150:094109, 2019.

\bibitem{AndyNEB}
A~A Peterson.
\newblock Acceleration of saddle-point searches with machine learning.
\newblock {\em J. Chem. Phys.}, 145:074106, 2016.

\bibitem{OlipekkaNEB}
O~Koistinen, F~B Dagbjartsdottir, V~Asgeirsson, A~Vehtari, and H~Jonsson.
\newblock Nudged elastic band calculations accelerated with gaussian process
  regression.
\newblock {\em J. Chem. Phys.}, 147:152720, 2017.

\bibitem{JoseNEB}
J~A~G Torres, P~C Jennings, M~H Hansen, J~R Boes, and T~Bligaard.
\newblock Low--scaling algorithm for nudged elastic band calculations using a
  surrogate machine learning model.
\newblock {\em Phys. Rev. Lett.}, 122:156001, 2019.

\bibitem{ari2018}
Manuel {\v{S}}ari{\'{c}}, Jan Rossmeisl, and Poul~Georg Moses.
\newblock Modeling the adsorption of sulfur containing molecules and their
  hydrodesulfurization intermediates on the {{Co}}-promoted {{Mo}S}2 catalyst
  by {{DFT}}.
\newblock {\em J. Catal.}, 358:131--140, February 2018.

\bibitem{stergaard2018}
Thomas~M. {\O}stergaard, Livia Giordano, Ivano~E. Castelli, Filippo Maglia,
  Byron~K. Antonopoulos, Yang Shao-Horn, and Jan Rossmeisl.
\newblock Oxidation of ethylene carbonate on li metal oxide surfaces.
\newblock {\em The Journal of Physical Chemistry C}, 122(19):10442--10449,
  April 2018.

\bibitem{Arnarson2018}
Logi Arnarson, Per~S. Schmidt, Mohnish Pandey, Alexander Bagger, Kristian~S.
  Thygesen, Ifan E.~L. Stephens, and Jan Rossmeisl.
\newblock Fundamental limitation of electrocatalytic methane conversion to
  methanol.
\newblock {\em Phys. Chem. Chem. Phys.}, 20(16):11152--11159, 2018.

\bibitem{Wan2020}
Hao Wan, Anders~W. Jensen, Mar{\'{\i}}a Escudero-Escribano, and Jan Rossmeisl.
\newblock Insights in the oxygen reduction reaction: From metallic
  electrocatalysts to diporphyrins.
\newblock {\em ACS Catal.}, 10(11):5979--5989, April 2020.

\bibitem{Levi2020pccp}
Gianluca Levi, Elisa Biasin, Asmus~Ougaard Dohn, and Hannes J{\'{o}}nsson.
\newblock {On the interplay of solvent and conformational effects in simulated
  excited-state dynamics of a copper phenanthroline photosensitizer}.
\newblock {\em Phys. Chem. Chem. Phys.}, 22:748--757, 2020.

\bibitem{BasinHopping}
D~A Wales and J~P~K Doye.
\newblock Global optimization of clusters, crystals, and biomolecules.
\newblock {\em Science}, 285:1368, 1999.

\bibitem{MinimaHopping}
S~Goedecker.
\newblock Minima hopping: An efficient search method for the global minimum of
  the potential energy surface of complex molecular systems.
\newblock {\em J. Chem. Phys.}, 120:9911, 2004.

\bibitem{HammerGA}
L~B Vilhelmsen and B~Hammer.
\newblock A genetic algorithm for first principles global structure
  optimization of supported nano structures.
\newblock {\em The Journal of Chemical Physics}, 141:044711, 2014.

\bibitem{Christiansen2022}
Mads-Peter~V. Christiansen, Nikolaj R{\o}nne, and Bj{\o}rk Hammer.
\newblock Atomistic global optimization x: A python package for optimization of
  atomistic structures.
\newblock {\em The Journal of Chemical Physics}, 157(5):054701, August 2022.

\bibitem{Jennings2019}
Paul~C. Jennings, Steen Lysgaard, Jens~Strabo Hummelsh{\o}j, Tejs Vegge, and
  Thomas Bligaard.
\newblock Genetic algorithms for computational materials discovery accelerated
  by machine learning.
\newblock {\em npj Comput. Mater.}, 5(1), April 2019.

\bibitem{gofee}
Malthe~K. Bisbo and Bjørk Hammer.
\newblock Efficient {Global} {Structure} {Optimization} with a
  {Machine}-{Learned} {Surrogate} {Model}.
\newblock {\em Phys. Rev. Lett.}, 124(8):086102, February 2020.

\bibitem{Kaappa2021}
Sami Kaappa, Casper Larsen, and Karsten~Wedel Jacobsen.
\newblock Atomic structure optimization with machine-learning enabled
  interpolation between chemical elements.
\newblock {\em Phys. Rev. Lett.}, 127(16), October 2021.

\bibitem{Kaappa2021b}
Sami Kaappa, Estefan{\'{\i}}a~Garijo del R{\'{\i}}o, and Karsten~Wedel
  Jacobsen.
\newblock Global optimization of atomic structures with gradient-enhanced
  gaussian process regression.
\newblock {\em Phys. Rev. B}, 103(17), May 2021.

\bibitem{larsen2022machinelearning}
Casper Larsen, Sami Kaappa, Andreas~Lynge Vishart, Thomas Bligaard, and
  Karsten~Wedel Jacobsen.
\newblock Machine-learning enabled optimization of atomic structures using
  atoms with fractional existence, 2022.

\bibitem{Wanzenbck2022}
Ralf Wanzenb\"{o}ck, Marco Arrigoni, Sebastian Bichelmaier, Florian Buchner,
  Jes{\'{u}}s Carrete, and Georg K.~H. Madsen.
\newblock Neural-network-backed evolutionary search for {SrTiO}$\_3$(110)
  surface reconstructions.
\newblock {\em Digital Discovery}, 1(5):703--710, 2022.

\bibitem{Mgelhj2011}
Andreas M{\o}gelh{\o}j, Andr{\'{e}}~K. Kelkkanen, K.~Thor Wikfeldt, Jakob
  Schi{\o}tz, Jens~J{\o}rgen Mortensen, Lars G.~M. Pettersson, Bengt~I.
  Lundqvist, Karsten~W. Jacobsen, Anders Nilsson, and Jens~K. N{\o}rskov.
\newblock Ab initio van der {{Waals}} interactions in simulations of water
  alter structure from mainly tetrahedral to high-density-like.
\newblock {\em The Journal of Physical Chemistry B}, 115(48):14149--14160,
  August 2011.

\bibitem{Hansen2016}
Martin~Hangaard Hansen and Jan Rossmeisl.
\newblock {pH} in grand canonical statistics of an electrochemical interface.
\newblock {\em The Journal of Physical Chemistry C}, 120(51):29135--29143,
  December 2016.

\bibitem{Dohn2017}
A.~O. Dohn, E.~\"{O}. J{\'{o}}nsson, G.~Levi, J.~J. Mortensen,
  O.~Lopez-Acevedo, K.~S. Thygesen, K.~W. Jacobsen, J.~Ulstrup, N.~E.
  Henriksen, K.~B. M{\o}ller, and H.~J{\'{o}}nsson.
\newblock Grid-based projector augmented wave ({GPAW}) implementation of
  quantum mechanics/molecular mechanics ({QM}/{MM}) electrostatic embedding and
  application to a solvated diplatinum complex.
\newblock {\em J. Chem. Theory Comput.}, 13(12):6010--6022, 2017.

\bibitem{Dohn2014}
Asmus~Ougaard Dohn, Elvar \"{O}rn J{\'{o}}nsson, Kasper~Skov Kj{\ae}r,
  Tim~Brandt van Driel, Martin~Meedom Nielsen, Karsten~Wedel Jacobsen,
  Niels~Engholm Henriksen, and Klaus~Braagaard M{\o}ller.
\newblock Direct dynamics studies of a binuclear metal complex in solution: The
  interplay between vibrational relaxation, coherence, and solvent effects.
\newblock {\em J. Phys. Chem. Lett.}, 5(14):2414--2418, June 2014.

\bibitem{vanDriel2016}
Tim~B. van Driel, Kasper~S. Kj{\ae}r, Robert~W. Hartsock, Asmus~O. Dohn, Tobias
  Harlang, Matthieu Chollet, Morten Christensen, Wojciech Gawelda, Niels~E.
  Henriksen, Jong~Goo Kim, Kristoffer Haldrup, Kyung~Hwan Kim, Hyotcherl Ihee,
  Jeongho Kim, Henrik Lemke, Zheng Sun, Villy Sundstr\"{o}m, Wenkai Zhang,
  Diling Zhu, Klaus~B. M{\o}ller, Martin~M. Nielsen, and Kelly~J. Gaffney.
\newblock Atomistic characterization of the active-site solvation dynamics of a
  model photocatalyst.
\newblock {\em Nat. Commun.}, 7(1), 2016.

\bibitem{Levi2018}
Gianluca Levi, M{\'{a}}ty{\'{a}}s P{\'{a}}pai, Niels~E. Henriksen, Asmus~O.
  Dohn, and Klaus~B. M{\o}ller.
\newblock Solution structure and ultrafast vibrational relaxation of the ptpop
  complex revealed by $\delta$scf-qm/mm direct dynamics simulations.
\newblock {\em J. Phys. Chem. C}, 122:7100--7119, 4 2018.

\bibitem{Haldrup2019}
Kristoffer Haldrup, Gianluca Levi, Elisa Biasin, Peter Vester, Mads~Goldschmidt
  Laursen, Frederik Beyer, Kasper~Skov Kj{\ae}r, Tim {Brandt Van Driel}, Tobias
  Harlang, Asmus~O. Dohn, Robert~J. Hartsock, Silke Nelson, James~M. Glownia,
  Henrik~T. Lemke, Morten Christensen, Kelly~J. Gaffney, Niels~E. Henriksen,
  Klaus~B. M{\o}ller, and Martin~M. Nielsen.
\newblock {Ultrafast X-Ray Scattering Measurements of Coherent Structural
  Dynamics on the Ground-State Potential Energy Surface of a Diplatinum
  Molecule}.
\newblock {\em Phys. Rev. Lett.}, 122(6):63001, 2019.

\bibitem{Dohn2016}
Asmus~O. Dohn, Kasper~S. Kj{\ae}r, Tobias~B. Harlang, Sophie~E. Canton,
  Martin~M. Nielsen, and Klaus~B. M{\o}ller.
\newblock Electron transfer and solvent-mediated electronic localization in
  molecular photocatalysis.
\newblock {\em Inorg. Chem.}, 55(20):10637--10644, October 2016.

\bibitem{Jnsson2019}
Elvar \"{O}rn J{\'{o}}nsson, Asmus~Ougaard Dohn, and Hannes J{\'{o}}nsson.
\newblock Polarizable embedding with a transferable {H$\_2$O} potential
  function {I}: Formulation and tests on dimer.
\newblock {\em J. Chem. Theory Comput.}, 15(12):6562--6577, 2019.

\bibitem{Dohn2019}
Asmus~Ougaard Dohn, Elvar \"{O}rn J{\'{o}}nsson, and Hannes J{\'{o}}nsson.
\newblock Polarizable embedding with a transferable {H$\_2$O} potential
  function {II}: Application to {H$\_2$O} clusters and liquid water.
\newblock {\em J. Chem. Theory Comput.}, 15(12):6578--6587, 2019.

\bibitem{Olsen2016a}
Thomas Olsen.
\newblock Designing in-plane heterostructures of quantum spin {{Hall}}
  insulators from first principles: 1{{T}}'-{{MoS}}$\_2$ with adsorbates.
\newblock {\em Phys. Rev. B}, 94:235106, 2016.

\bibitem{Liechtenstein1987}
A.I. Liechtenstein, M.I. Katsnelson, V.P Antropov, and V.A. Gubanov.
\newblock {Local spin density functional approach to the theory of exchange
  interactions in ferromagnetic metals and alloys}.
\newblock {\em J. Magn. Magn. Mater.}, 67:65--74, 1987.

\bibitem{Mermin1966}
N.~D. Mermin and H.~Wagner.
\newblock {Absence of Ferromagnetism or Antiferromagnetism in One- or
  Two-Dimensional Isotropic Heisenberg Models}.
\newblock {\em Phys. Rev. Lett.}, 17:1133--1136, 1966.

\bibitem{Lado2017}
J.~L. Lado and J.~Fern{\'{a}}ndez-Rossier.
\newblock {On the origin of magnetic anisotropy in two dimensional CrI 3}.
\newblock {\em 2D Mater.}, 4:035002, 2017.

\bibitem{Torelli2019b}
Daniele Torelli, Kristian~Sommer Thygesen, and Thomas Olsen.
\newblock {High throughput computational screening for 2D ferromagnetic
  materials: the critical role of anisotropy and local correlations}.
\newblock {\em 2D Mater.}, 6:045018, 2019.

\bibitem{Torelli2020}
Daniele Torelli, Hadeel Moustafa, Karsten~W. Jacobsen, and Thomas Olsen.
\newblock {High-throughput computational screening for two-dimensional magnetic
  materials based on experimental databases of three-dimensional compounds}.
\newblock {\em npj Comput. Mater.}, 6:158, 2020.

\bibitem{Barth1972}
U~von Barth and L.~Hedin.
\newblock {A local exchange-correlation potential for the spin polarized case.
  i}.
\newblock {\em J. Phys. C Solid State Phys.}, 5:1629--1642, 1972.

\bibitem{Scalmani2012}
Giovanni Scalmani and Michael~J. Frisch.
\newblock {A New Approach to Noncollinear Spin Density Functional Theory beyond
  the Local Density Approximation}.
\newblock {\em J. Chem. Theory Comput.}, 8:2193--2196, 2012.

\bibitem{Kim2021}
Tae~Yun Kim and Cheol-Hwan Park.
\newblock {Magnetic Anisotropy and Magnetic Ordering of Transition-Metal
  Phosphorus Trisulfides}.
\newblock {\em Nano Lett.}, 21:10114--10121, 2021.

\bibitem{Ceresoli2010}
D.~Ceresoli, U.~Gerstmann, A.~P. Seitsonen, and F.~Mauri.
\newblock First-principles theory of orbital magnetization.
\newblock {\em Phys. Rev. B}, 81(6):060409, 2010.

\bibitem{Meyer1961}
A.~J.~P. Meyer and G.~Asch.
\newblock Experimental g' and g values of {{Fe}}, {{Co}}, {{Ni}}, and their
  alloys.
\newblock {\em J. Appl. Phys.}, 32(3):S330--S333, 1961.

\bibitem{PhysRev.130.183}
Simon Foner.
\newblock High-field antiferromagnetic resonance in {{Cr}}$\_{2}${{O}}$\_{3}$.
\newblock {\em Phys. Rev.}, 130:183--197, 1963.

\bibitem{Kaplan1959}
T.~A. Kaplan.
\newblock {Classical Spin-Configuration Stability in the Presence of Competing
  Exchange Forces}.
\newblock {\em Phys. Rev.}, 116(4):888--889, nov 1959.

\bibitem{Sandratskii1986}
L.~M. Sandratskii.
\newblock Energy band structure calculations for crystals with spiral magnetic
  structure.
\newblock {\em Phys. Status Solidi B.}, 136(1):167--180, 1986.

\bibitem{sandratskii2017insight}
LM~Sandratskii.
\newblock Insight into the dzyaloshinskii-moriya interaction through
  first-principles study of chiral magnetic structures.
\newblock {\em Phys. Rev. B}, 96(2):024450, 2017.

\bibitem{Kubo1957}
Ryogo Kubo.
\newblock {Statistical-Mechanical Theory of Irreversible Processes. I. General
  Theory and Simple Applications to Magnetic and Conduction Problems}.
\newblock {\em J. Phys. Soc. Jpn.}, 12(6):570--586, 6 1957.

\bibitem{Gross1985}
E.~K~U Gross and Walter Kohn.
\newblock Local density-functional theory of frequency-dependent linear
  response.
\newblock {\em Phys. Rev. Lett.}, 55:2850--2852, 1985.

\bibitem{Runge1984}
Erich Runge and E.~K.~U. Gross.
\newblock {Density-Functional Theory for Time-Dependent Systems}.
\newblock {\em Phys. Rev. Lett.}, 52:997--1000, 1984.

\bibitem{sharma2011}
S.~Sharma, J.~K. Dewhurst, A.~Sanna, and E.~K.~U. Gross.
\newblock Bootstrap approximation for the exchange-correlation kernel of
  time-dependent density-functional theory.
\newblock {\em Phys. Rev. Lett.}, 107:186401, Oct 2011.

\bibitem{macdonald1979}
AH~MacDonald, SH~Vosko, and PT~Coleridge.
\newblock Extensions of the tetrahedron method for evaluating spectral
  properties of solids.
\newblock {\em Journal of Physics C: Solid State Physics}, 12(15):2991, 1979.

\bibitem{rozzi2006}
Carlo~A. Rozzi, Daniele Varsano, Andrea Marini, Eberhard K.~U. Gross, and Angel
  Rubio.
\newblock Exact coulomb cutoff technique for supercell calculations.
\newblock {\em Phys. Rev. B}, 73:205119, May 2006.

\bibitem{huser2013dielectric}
Falco H{\"u}ser, Thomas Olsen, and Kristian~S Thygesen.
\newblock How dielectric screening in two-dimensional crystals affects the
  convergence of excited-state calculations: Monolayer {{MoS}}$\_2$.
\newblock {\em Phys. Rev. B}, 88(24):245309, 2013.

\bibitem{latini2015excitons}
S.~Latini, T.~Olsen, and K.~S. Thygesen.
\newblock Excitons in van der waals heterostructures: The important role of
  dielectric screening.
\newblock {\em Phys. Rev. B}, 92:245123, Dec 2015.

\bibitem{Patrick_PRB_2016}
Christopher~E. Patrick and Kristian~S. Thygesen.
\newblock Hubbard-{{U}} corrected {{Hamiltonians}} for non-self-consistent
  random-phase approximation total-energy calculations: A study of {{ZnS,
  TiO}}$\_{2}$, and {{NiO}}.
\newblock {\em Phys. Rev. B}, 93:035133, Jan 2016.

\bibitem{Schmidt_JPC_2018}
Per~S. Schmidt and Kristian~S. Thygesen.
\newblock Benchmark database of transition metal surface and adsorption
  energies from many-body perturbation theory.
\newblock {\em The Journal of Physical Chemistry C}, 122(8):4381--4390, 2018.

\bibitem{Olsen_PRL_2011}
Thomas Olsen, Jun Yan, Jens~J. Mortensen, and Kristian~S. Thygesen.
\newblock Dispersive and covalent interactions between graphene and metal
  surfaces from the random phase approximation.
\newblock {\em Phys. Rev. Lett.}, 107:156401, Oct 2011.

\bibitem{Furche2005}
Filipp Furche and Troy Van~Voorhis.
\newblock {Fluctuation-dissipation theorem density-functional theory}.
\newblock {\em The Journal of Chemical Physics}, 122(16), 04 2005.
\newblock 164106.

\bibitem{Olsen_PRB_2013_2}
Thomas Olsen and Kristian~S. Thygesen.
\newblock Beyond the random phase approximation: Improved description of
  short-range correlation by a renormalized adiabatic local density
  approximation.
\newblock {\em Phys. Rev. B}, 88:115131, Sep 2013.

\bibitem{Olsen_PRB_2014}
Thomas Olsen and Kristian~S. Thygesen.
\newblock Accurate ground-state energies of solids and molecules from
  time-dependent density-functional theory.
\newblock {\em Phys. Rev. Lett.}, 112:203001, May 2014.

\bibitem{Patrick_JCP_2015}
Christopher~E. Patrick and Kristian~S. Thygesen.
\newblock {Adiabatic-connection fluctuation-dissipation {{DFT}} for the
  structural properties of solids—{{The}} renormalized {{ALDA}} and electron
  gas kernels}.
\newblock {\em The Journal of Chemical Physics}, 143(10), 04 2015.
\newblock 102802.

\bibitem{Olsen_npj_2019}
Thomas Olsen, Christopher~E. Patrick, Jefferson~E. Bates, Adrienn Ruzsinszky,
  and Kristian~S. Thygesen.
\newblock Beyond the {{RPA}} and {{GW}} methods with adiabatic xc-kernels for
  accurate ground state and quasiparticle energies.
\newblock {\em npj Comput. Mater.}, 5(1):106, Nov 2019.

\bibitem{Buczek2011}
Pawe{\l} Buczek, Arthur Ernst, and Leonid~M. Sandratskii.
\newblock {Different dimensionality trends in the Landau damping of magnons in
  iron, cobalt, and nickel: Time-dependent density functional study}.
\newblock {\em Phys. Rev. B}, 84:174418, 2011.

\bibitem{VanHove1954}
L{\'{e}}on {Van Hove}.
\newblock {Time-Dependent Correlations between Spins and Neutron Scattering in
  Ferromagnetic Crystals}.
\newblock {\em Phys. Rev.}, 95(6):1374--1384, 9 1954.

\bibitem{Skovhus2022a}
Thorbjørn Skovhus, Thomas Olsen, and Henrik~M Rønnow.
\newblock Influence of static correlation on the magnon dynamics of an
  itinerant ferromagnet with competing exchange interactions: First-principles
  study of mnbi.
\newblock {\em Phys. Rev. Mater.}, 6:054402, 5 2022.

\bibitem{Skovhus2022b}
Thorbj\o{}rn Skovhus and Thomas Olsen.
\newblock Magnons in antiferromagnetic bcc cr and
  ${\mathrm{cr}}\_{2}{\mathrm{o}}\_{3}$ from time-dependent density functional
  theory.
\newblock {\em Phys. Rev. B}, 106:085131, Aug 2022.

\bibitem{rojas1995space}
HN~Rojas, Rex~William Godby, and RJ~Needs.
\newblock Space-time method for ab initio calculations of self-energies and
  dielectric response functions of solids.
\newblock {\em Phys. Rev. Lett.}, 74(10):1827, 1995.

\bibitem{friedrich2010efficient}
Christoph Friedrich, Stefan Bl{\"u}gel, and Arno Schindlmayr.
\newblock Efficient implementation of the g w approximation within the
  all-electron flapw method.
\newblock {\em Phys. Rev. B}, 81(12):125102, 2010.

\bibitem{duchemin2020robust}
Ivan Duchemin and Xavier Blase.
\newblock Robust analytic-continuation approach to many-body gw calculations.
\newblock {\em J. Chem. Theory Comput.}, 16(3):1742--1756, 2020.

\bibitem{leon2021frequency}
Dario~A Leon, Claudia Cardoso, Tommaso Chiarotti, Daniele Varsano, Elisa
  Molinari, and Andrea Ferretti.
\newblock Frequency dependence in g w made simple using a multipole
  approximation.
\newblock {\em Phys. Rev. B}, 104(11):115157, 2021.

\bibitem{rasmussen2016efficient}
Filip~A Rasmussen, Per~S Schmidt, Kirsten~T Winther, and Kristian~S Thygesen.
\newblock Efficient many-body calculations for two-dimensional materials using
  exact limits for the screened potential: Band gaps of {{MoS}}$\_2$,
  {{H}}-{{BN}}, and phosphorene.
\newblock {\em Phys. Rev. B}, 94(15):155406, 2016.

\bibitem{marini2009yambo}
Andrea Marini, Conor Hogan, Myrta Gr{\"u}ning, and Daniele Varsano.
\newblock Yambo: an ab initio tool for excited state calculations.
\newblock {\em Comput. Phys. Commun.}, 180(8):1392--1403, 2009.

\bibitem{Olsen2016}
Thomas Olsen, Simone Latini, Filip Rasmussen, and Kristian~S. Thygesen.
\newblock {Simple Screened Hydrogen Model of Excitons in Two-Dimensional
  Materials}.
\newblock {\em Phys. Rev. Lett.}, 116:056401, 2016.

\bibitem{Olsen2021}
Thomas Olsen.
\newblock {Unified Treatment of Magnons and Excitons in Monolayer CrI$\_3$ from
  Many-Body Perturbation Theory}.
\newblock {\em Phys. Rev. Lett.}, 127:166402, 2021.

\bibitem{li2021deformation}
Zhen Li, Patrizio Graziosi, and Neophytos Neophytou.
\newblock Deformation potential extraction and computationally efficient
  mobility calculations in silicon from first principles.
\newblock {\em Phys. Rev. B}, 104(19):195201, 2021.

\bibitem{liu1996linear}
Amy~Y Liu and Andrew~A Quong.
\newblock Linear-response calculation of electron-phonon coupling parameters.
\newblock {\em Phys. Rev. B}, 53(12):R7575, 1996.

\bibitem{RevModPhys.89.015003}
Feliciano Giustino.
\newblock Electron-phonon interactions from first principles.
\newblock {\em Rev. Mod. Phys.}, 89:015003, Feb 2017.

\bibitem{walter_ab_2020}
Michael Walter and Michael Moseler.
\newblock Ab {Initio} {Wavelength}-{Dependent} {Raman} {Spectra}: {Placzek}
  {Approximation} and {Beyond}.
\newblock {\em J. Chem. Theory Comput.}, 16(1):576--586, January 2020.

\bibitem{jorio2011raman}
Ado Jorio, Mildred~S Dresselhaus, Riichiro Saito, and Gene Dresselhaus.
\newblock {\em Raman spectroscopy in graphene related systems}.
\newblock John Wiley \& Sons, 2011.

\bibitem{TogoJPCM23}
Atsushi Togo, Laurent Chaput, Terumasa Tadano, and Isao Tanaka.
\newblock Implementation strategies in phonopy and phono3py.
\newblock {\em J. Phys. Condens. Matter}, 35(35):353001, 2023.

\bibitem{boyd_nonlinear_2008}
Robert~W. Boyd.
\newblock {\em Nonlinear Optics}.
\newblock Elsevier Science Publishing Co Inc., Amsterdam, 3rd edition, 2008.

\bibitem{taghizadeh_linear_2017}
Alireza Taghizadeh, F.~Hipolito, and T.~G. Pedersen.
\newblock Linear and nonlinear optical response of crystals using length and
  velocity gauges: Effect of basis truncation.
\newblock {\em Phys. Rev. B}, 96:195413, 2017.

\bibitem{aversa_nonlinear_1995}
Claudio Aversa and J.~E. Sipe.
\newblock Nonlinear optical susceptibilities of semiconductors: Results with a
  length-gauge analysis.
\newblock {\em Phys. Rev. B}, 52:14636--14645, 1995.

\bibitem{SakRosNie14}
Arto Sakko, Tuomas~P Rossi, and Risto~M Nieminen.
\newblock Dynamical coupling of plasmons and molecular excitations by hybrid
  quantum/classical calculations: time-domain approach.
\newblock {\em Journal of Physics: Condensed Matter}, 26(31):315013, jul 2014.

\bibitem{RosWinJac17}
Tuomas~P. Rossi, Kirsten~T. Winther, Karsten~W. Jacobsen, Risto~M. Nieminen,
  Martti~J. Puska, and Kristian~S. Thygesen.
\newblock Effect of edge plasmons on the optical properties of {{MoS}}$\_{2}$
  monolayer flakes.
\newblock {\em Phys. Rev. B}, 96:155407, Oct 2017.

\bibitem{KumRosKui19}
Priyank~V. Kumar, Tuomas~P. Rossi, Mikael Kuisma, Paul Erhart, and David~J.
  Norris.
\newblock Direct hot-carrier transfer in plasmonic catalysis.
\newblock {\em Faraday Discuss.}, 214:189--197, 2019.

\bibitem{KumRosMar19}
Priyank~V. Kumar, Tuomas~P. Rossi, Daniel {Marti-Dafcik}, Daniel Reichmuth,
  Mikael Kuisma, Paul Erhart, Martti~J. Puska, and David~J. Norris.
\newblock Plasmon-{{Induced Direct Hot-Carrier Transfer}} at
  {{Metal}}\textendash{{Acceptor Interfaces}}.
\newblock {\em Acs Nano}, 13(3):3188--3195, 2019.

\bibitem{RosSheErh19}
Tuomas~P. Rossi, Timur Shegai, Paul Erhart, and Tomasz~J. Antosiewicz.
\newblock Strong plasmon-molecule coupling at the nanoscale revealed by
  first-principles modeling.
\newblock {\em Nat. Commun.}, 10(1):3336, 2019.

\bibitem{RosErhKui20}
Tuomas~P. Rossi, Paul Erhart, and Mikael Kuisma.
\newblock Hot-{{Carrier Generation}} in {{Plasmonic Nanoparticles}}: {{The
  Importance}} of {{Atomic Structure}}.
\newblock {\em Acs Nano}, 14(8):9963--9971, 2020.

\bibitem{FojRosKui22}
Jakub Fojt, Tuomas~P. Rossi, Mikael Kuisma, and Paul Erhart.
\newblock Hot-{{Carrier Transfer}} across a
  {{Nanoparticle}}\textendash{{Molecule Junction}}: {{The Importance}} of
  {{Orbital Hybridization}} and {{Level Alignment}}.
\newblock {\em Nano Lett.}, 22(21):8786--8792, 2022.

\bibitem{SorRinRos23}
Daniel Sorvisto, Patrick Rinke, and Tuomas~P. Rossi.
\newblock Single-atom dopants in plasmonic nanocatalysts.
\newblock {\em The Journal of Physical Chemistry C}, 127(18):8585--8590, 2023.

\bibitem{YabBer96}
K.~Yabana and G.~F. Bertsch.
\newblock Time-dependent local-density approximation in real time.
\newblock {\em Phys. Rev. B}, 54(7):4484--4487, 1996.

\bibitem{MalLehEnk13}
Sami Malola, Lauri Lehtovaara, Jussi Enkovaara, and Hannu Häkkinen.
\newblock Birth of the localized surface plasmon resonance in
  monolayer-protected gold nanoclusters.
\newblock {\em Acs Nano}, 7(11):10263--10270, 2013.

\bibitem{RosErhKui20Code}
Tuomas Rossi, Paul Erhart, and Mikael Kuisma.
\newblock Code for "hot-carrier generation in plasmonic nanoparticles: The
  importance of atomic structure".
\newblock {\em Zenodo}.

\bibitem{FojRosKui22Code}
Jakub Fojt, Tuomas Rossi, Mikael Kuisma, and Paul Erhart.
\newblock Code for "hot-carrier transfer across a nanoparticle-molecule
  junction: The importance of orbital hybridization and level alignment".
\newblock {\em Zenodo}.

\bibitem{PhysRevLett.128.156402}
Christian Sch\"afer and G\"oran Johansson.
\newblock Shortcut to self-consistent light-matter interaction and realistic
  spectra from first principles.
\newblock {\em Phys. Rev. Lett.}, 128:156402, Apr 2022.

\bibitem{doi:10.1021/acs.jpclett.2c01169}
Christian Schäfer.
\newblock Polaritonic chemistry from first principles via embedding radiation
  reaction.
\newblock {\em J. Phys. Chem. Lett.}, 13(30):6905--6911, 2022.
\newblock PMID: 35866694.

\bibitem{ojanpera_nonadiabatic_2012}
Ari Ojanperä, Ville Havu, Lauri Lehtovaara, and Martti Puska.
\newblock Nonadiabatic {Ehrenfest} molecular dynamics within the projector
  augmented-wave method.
\newblock {\em The Journal of Chemical Physics}, 136(14):144103, April 2012.
\newblock Publisher: American Institute of Physics.

\bibitem{ojanpera_electronic_2014}
Ari Ojanperä, Arkady~V. Krasheninnikov, and Martti Puska.
\newblock Electronic stopping power from first-principles calculations with
  account for core electron excitations and projectile ionization.
\newblock {\em Phys. Rev. B}, 89(3):035120, 2014.

\bibitem{syzgantseva_charge_2015}
Olga~A. Syzgantseva, Martti Puska, and Kari Laasonen.
\newblock Charge {Transfer} at the {Hybrid} {Interfaces} in the {Presence} of
  {Water}: {A} {Theoretical} {Study}.
\newblock {\em The Journal of Physical Chemistry C}, 119(51):28347--28352,
  December 2015.

\bibitem{brand_coherent_2019}
Christian Brand, Maxime Debiossac, Toma Susi, François Aguillon, Jani
  Kotakoski, Philippe Roncin, and Markus Arndt.
\newblock Coherent diffraction of hydrogen through the 246 pm lattice of
  graphene.
\newblock {\em New J. Phys.}, 21(3):033004, March 2019.

\bibitem{buntov_carbon_2020}
Evgeny~A. Buntov and Anatoly~F. Zatsepin.
\newblock Carbon {Bond} {Breaking} under {Ar}+-{Ion} {Irradiation} in
  {Dependence} on sp {Hybridization}: {Car}–{Parrinello}, {Ehrenfest}, and
  {Classical} {Dynamics} {Study}.
\newblock {\em J. Phys. Chem. A}, 124(44):9128--9132, November 2020.

\bibitem{aierken_revealing_2021}
Yierpan Aierken, Ankit Agrawal, Meiling Sun, Marko Melander, Ethan~J. Crumlin,
  Brett~A. Helms, and David Prendergast.
\newblock Revealing {Charge}-{Transfer} {Dynamics} at {Electrified} {Sulfur}
  {Cathodes} {Using} {Constrained} {Density} {Functional} {Theory}.
\newblock {\em J. Phys. Chem. Lett.}, 12(2):739--744, January 2021.
\newblock Publisher: American Chemical Society.

\bibitem{artacho_quantum_2017}
Emilio Artacho and David~D. O'Regan.
\newblock Quantum mechanics in an evolving {Hilbert} space.
\newblock {\em Phys. Rev. B}, 95(11):115155, March 2017.
\newblock Publisher: American Physical Society.

\bibitem{garcia_siesta_2020}
Siesta: {Recent} developments and applications.
\newblock {\em The Journal of Chemical Physics}, 152(20):204108, May 2020.
\newblock Publisher: American Institute of Physics.

\bibitem{Dreuw2004}
Andreas Dreuw and Martin Head-Gordon.
\newblock {Failure of Time-Dependent Density Functional Theory for Long-Range
  Charge-Transfer Excited States: The Zincbacteriochlorin-Bacteriochlorin and
  Bacteriochlorophyll-Spheroidene Complexes}.
\newblock {\em J. Am. Chem. Soc.}, 126(12):4007--4016, 2004.

\bibitem{Seidu2015}
Issaka Seidu, Mykhaylo Krykunov, and Tom Ziegler.
\newblock {Applications of time-dependent and time-independent density
  functional theory to Rydberg transitions}.
\newblock {\em J. Phys. Chem. A}, 119(21):5107--5116, 2015.

\bibitem{Levine2006}
Benjamin~G. Levine, Chaehyuk Ko, Jason Quenneville, and Todd~J. Martínez.
\newblock Conical intersections and double excitations in time-dependent
  density functional theory.
\newblock {\em Mol. Phys.}, 104:1039--1051, 3 2006.

\bibitem{huzinaga_virtual_1970}
Sigeru Huzinaga and Catalina Arnau.
\newblock Virtual {{Orbitals}} in {{Hartree-Fock Theory}}.
\newblock {\em Phys. Rev. A}, 1(5):1285--1288, May 1970.

\bibitem{huzinaga_virtual_1971}
Sigeru Huzinaga and Catalina Arnau.
\newblock Virtual {{Orbitals}} in {{Hartree}}\textendash{{Fock Theory}}.
  {{II}}.
\newblock {\em The Journal of Chemical Physics}, 54(5):1948--1951, March 1971.

\bibitem{Hait2021}
Diptarka Hait and Martin Head-Gordon.
\newblock Orbital optimized density functional theory for electronic excited
  states.
\newblock {\em J. Phys. Chem. Lett.}, 12:4517--4529, 5 2021.

\bibitem{Barca2018}
Giuseppe~M.J. Barca, Andrew~T.B. Gilbert, and Peter~M.W. Gill.
\newblock Simple models for difficult electronic excitations.
\newblock {\em J. Chem. Theory Comput.}, 14:1501--1509, 3 2018.

\bibitem{Gilbert2008}
Andrew~T.B. Gilbert, Nicholas~A. Besley, and Peter~M.W. Gill.
\newblock Self-consistent field calculations of excited states using the
  maximum overlap method (mom).
\newblock {\em J. Phys. Chem. A}, 112:13164--13171, 12 2008.

\bibitem{Levi2020fd}
Gianluca Levi, Aleksei~V. Ivanov, and Hannes Jónsson.
\newblock Variational calculations of excited states via direct optimization of
  orbitals in {{DFT}}.
\newblock {\em Faraday Discuss.}, 224:448--466, 12 2020.

\bibitem{Loos2021}
Pierre~François Loos, Massimiliano Comin, Xavier Blase, and Denis Jacquemin.
\newblock Reference energies for intramolecular charge-transfer excitations.
\newblock {\em J. Chem. Theory Comput.}, 17:3666--3686, 6 2021.

\bibitem{Katayama2023}
Tetsuo Katayama, Tae~Kyu Choi, Dmitry Khakhulin, Asmus~O Dohn, Christopher~J.
  Milne, Gy{\"{o}}rgy Vank{\'{o}}, Zolt{\'{a}}n N{\'{e}}meth, Frederico~A.
  Lima, Jakub Szlachetko, Tokushi Sato, Shunsuke Nozawa, Shin~Ichi Adachi,
  Makina Yabashi, Thomas~J. Penfold, Wojciech Gawelda, and Gianluca Levi.
\newblock {Atomic-scale observation of solvent reorganization influencing
  photoinduced structural dynamics in a copper complex photosensitizer}.
\newblock {\em Chem. Sci.}, 14:2572--2584, 2023.

\bibitem{Ma2020}
He~Ma, Marco Govoni, and Giulia Galli.
\newblock Quantum simulations of materials on near term quantum computers.
\newblock {\em npj Comput. Mater.}, 6(1):85, December 2020.

\bibitem{Abedi2019}
Mostafa Abedi, Gianluca Levi, Diana~B Zederkof, Niels~Engholm Henriksen,
  M{\'{a}}ty{\'{a}}s P{\'{a}}pai, and Klaus~B. M{\o}ller.
\newblock {Excited-State Solvation Structure of Transition Metal Complexes from
  Molecular Dynamics Simulations and Assessment of Partial Atomic Charge
  Methods}.
\newblock {\em Phys. Chem. Chem. Phys.}, 21:4082--4095, 2019.

\bibitem{Ivanov2023}
Aleksei~V. Ivanov, Yorick L.~A. Schmerwitz, Gianluca Levi, and Hannes Jónsson.
\newblock {Electronic excitations of the charged nitrogen-vacancy center in
  diamond obtained using time-independent variational density functional
  calculations}.
\newblock {\em SciPost Phys.}, 15:009, 2023.

\bibitem{King-Smith1993}
{Theory of polarization of crystalline solids}.
\newblock {\em Phys. Rev. B}, 47:1651--1654, 1993.

\bibitem{Resta1994}
Raffaele Resta.
\newblock {Macroscopic polarization in crystalline dielectrics: the geometric
  phase approach}.
\newblock {\em Rev. Mod. Phys.}, 66:899--915, 1994.

\bibitem{Resta2007}
Raffaele Resta and David Vanderbilt.
\newblock {Theory of Polarization: A Modern Approach}.
\newblock In {\em Phys. Ferroelectr.}, volume 105, pages 31--68. 2007.

\bibitem{Yoneda2018}
Yasuhiro Yoneda, Koji Ohara, and Hajime Nagata.
\newblock {Local structure and phase transitions of KNbO 3}.
\newblock {\em Jpn. J. Appl. Phys.}, 57:11UB07, 2018.

\bibitem{Gibertini2015}
Marco Gibertini and Nicola Marzari.
\newblock {Emergence of One-Dimensional Wires of Free Carriers in
  Transition-Metal-Dichalcogenide Nanostructures}.
\newblock {\em Nano Lett.}, 15:6229--6238, 2015.

\bibitem{Sodequist2023}
Joachim S{\o}dequist, Urko Petralanda, and Thomas Olsen.
\newblock {Abundance of second order topology in C 3 symmetric two-dimensional
  insulators}.
\newblock {\em 2D Mater.}, 10:015009, 2023.

\bibitem{Baroni2001}
Stefano Baroni, Stefano {De Gironcoli}, Andrea {Dal Corso}, and Paolo
  Giannozzi.
\newblock {Phonons and related crystal properties from density-functional
  perturbation theory}.
\newblock {\em Rev. Mod. Phys.}, 73:515--562, 2001.

\bibitem{Fei2016}
Ruixiang Fei, Wei Kang, and Li~Yang.
\newblock {Ferroelectricity and Phase Transitions in Monolayer Group-IV
  Monochalcogenides}.
\newblock {\em Phys. Rev. Lett.}, 117:097601, 2016.

\bibitem{Rangel2017}
Tonatiuh Rangel, Benjamin~M Fregoso, Bernardo~S Mendoza, Takahiro Morimoto,
  Joel~E Moore, and Jeffrey~B Neaton.
\newblock {Large Bulk Photovoltaic Effect and Spontaneous Polarization of
  Single-Layer Monochalcogenides}.
\newblock {\em Phys. Rev. Lett.}, 119:067402, 2017.

\bibitem{Wang2017}
Hua Wang and Xiaofeng Qian.
\newblock {Two-dimensional multiferroics in monolayer group IV
  monochalcogenides}.
\newblock {\em 2D Mater.}, 4:015042, 2017.

\bibitem{Petralanda2023}
Urko Petralanda and Thomas Olsen.
\newblock {Polarization switching induced by domain wall sliding in
  two-dimensional ferroelectric monochalcogenides}.
\newblock {\em 2D Mater.}, 10:015001, 2023.

\bibitem{Taherinejad2014}
Maryam Taherinejad, Kevin~F. Garrity, and David Vanderbilt.
\newblock {Wannier center sheets in topological insulators}.
\newblock {\em Phys. Rev. B}, 89:115102, 2014.

\bibitem{marzari1997maximally}
Nicola Marzari and David Vanderbilt.
\newblock Maximally localized generalized wannier functions for composite
  energy bands.
\newblock {\em Phys. Rev. B}, 56(20):12847, 1997.

\bibitem{Qian2014}
Xiaofeng Qian, Junwei Liu, Liang Fu, and Ju~Li.
\newblock {Quantum spin Hall effect in two-dimensional transition metal
  dichalcogenides}.
\newblock {\em Science}, 346:1344--1347, 2014.

\bibitem{resta1999electron}
Raffaele Resta and Sandro Sorella.
\newblock Electron localization in the insulating state.
\newblock {\em Phys. Rev. Lett.}, 82(2):370, 1999.

\bibitem{berghold2000general}
Gerd Berghold, Christopher~J Mundy, Aldo~H Romero, J{\"u}rg Hutter, and Michele
  Parrinello.
\newblock General and efficient algorithms for obtaining maximally localized
  wannier functions.
\newblock {\em Phys. Rev. B}, 61(15):10040, 2000.

\bibitem{thygesen2005partly}
Kristian~S Thygesen, Lars~Bruno Hansen, and Karsten~Wedel Jacobsen.
\newblock Partly occupied wannier functions.
\newblock {\em Phys. Rev. Lett.}, 94(2):026405, 2005.

\bibitem{thygesen2005partly2}
Kristian~Sommer Thygesen, Lars~Bruno Hansen, and Karsten~Wedel Jacobsen.
\newblock Partly occupied wannier functions: Construction and applications.
\newblock {\em Phys. Rev. B}, 72(12):125119, 2005.

\bibitem{fontana2021spread}
Pietro~F Fontana, Ask~H Larsen, Thomas Olsen, and Kristian~S Thygesen.
\newblock Spread-balanced wannier functions: Robust and automatable orbital
  localization.
\newblock {\em Phys. Rev. B}, 104(12):125140, 2021.

\bibitem{dreyer2018first}
Cyrus~E Dreyer, Audrius Alkauskas, John~L Lyons, Anderson Janotti, and Chris~G
  Van~de Walle.
\newblock First-principles calculations of point defects for quantum
  technologies.
\newblock {\em Annu. Rev. Mater. Res.}, 48:1--26, 2018.

\bibitem{park2018point}
Ji~Sang Park, Sunghyun Kim, Zijuan Xie, and Aron Walsh.
\newblock Point defect engineering in thin-film solar cells.
\newblock {\em Nature Reviews Materials}, 3(7):194--210, 2018.

\bibitem{weston2018native}
L~Weston, D~Wickramaratne, M~Mackoit, A~Alkauskas, and CG~Van~de Walle.
\newblock Native point defects and impurities in hexagonal boron nitride.
\newblock {\em Phys. Rev. B}, 97(21):214104, 2018.

\bibitem{vuong2016phonon}
TQP Vuong, Guillaume Cassabois, Pierre Valvin, Abdelkarim Ouerghi, Yannick
  Chassagneux, Christophe Voisin, and Bernard Gil.
\newblock Phonon-photon mapping in a color center in hexagonal boron nitride.
\newblock {\em Phys. Rev. Lett.}, 117(9):097402, 2016.

\bibitem{van2004first}
Chris~G Van~de Walle and J{\"o}rg Neugebauer.
\newblock First-principles calculations for defects and impurities:
  Applications to iii-nitrides.
\newblock {\em J. Appl. Phys.}, 95(8):3851--3879, 2004.

\bibitem{lany2008assessment}
Stephan Lany and Alex Zunger.
\newblock Assessment of correction methods for the band-gap problem and for
  finite-size effects in supercell defect calculations: Case studies for zno
  and gaas.
\newblock {\em Phys. Rev. B}, 78(23):235104, 2008.

\bibitem{freysoldt2014first}
Christoph Freysoldt, Blazej Grabowski, Tilmann Hickel, J{\"o}rg Neugebauer,
  Georg Kresse, Anderson Janotti, and Chris~G Van~de Walle.
\newblock First-principles calculations for point defects in solids.
\newblock {\em Rev. Mod. Phys.}, 86(1):253, 2014.

\bibitem{zunger2021understanding}
Alex Zunger and Oleksandr~I Malyi.
\newblock Understanding doping of quantum materials.
\newblock {\em Chem. Rev.}, 121(5):3031--3060, 2021.

\bibitem{kurakevych2007rhombohedral}
Oleksandr~O Kurakevych and Vladimir~L Solozhenko.
\newblock Rhombohedral boron subnitride, b13n2, by x-ray powder diffraction.
\newblock {\em Acta Crystallogr. C.}, 63(9):i80--i82, 2007.

\bibitem{cassabois2016hexagonal}
Guillaume Cassabois, Pierre Valvin, and Bernard Gil.
\newblock Hexagonal boron nitride is an indirect bandgap semiconductor.
\newblock {\em Nat. Photonics}, 10(4):262--266, 2016.

\bibitem{freysoldt2009fully}
Christoph Freysoldt, J{\"o}rg Neugebauer, and Chris~G Van~de Walle.
\newblock Fully ab initio finite-size corrections for charged-defect supercell
  calculations.
\newblock {\em Phys. Rev. Lett.}, 102(1):016402, 2009.

\bibitem{Kaappa2018}
Sami Kaappa, Sami Malola, and Hannu H\"akkinen.
\newblock Point group symmetry analysis of the electronic structure of bare and
  protected metal nanocrystals.
\newblock {\em J. Phys. Chem. A}, 122(43):8576--8584, 2018.

\bibitem{Bertoldo2022}
Fabian Bertoldo, Sajid Ali, Simone Manti, and Kristian~S. Thygesen.
\newblock Quantum point defects in 2{D} materials - the {QPOD} database.
\newblock {\em npj Comput. Mater.}, 8(1):56, Apr 2022.

\bibitem{CORNWELL1997}
J.F. Cornwell.
\newblock {\em Group Theory in Physics}, volume~1 of {\em Techniques of
  Physics}.
\newblock Academic Press, San Diego, 1997.

\bibitem{popescu2012}
Voicu Popescu and Alex Zunger.
\newblock {Extracting E versus $\vec{k}$ effective band structure from
  supercell calculations on alloys and impurities}.
\newblock {\em Phys. Rev. B}, 85(8):085201, 2012.

\bibitem{qeh_model}
{Quantum Electrostatic Heterostructure Model}.
\newblock \url{https://qeh.readthedocs.io/en/latest/}, 2019.
\newblock [Online; accessed 15-May-2023].

\bibitem{andersen2015dielectric}
Kirsten Andersen, Simone Latini, and Kristian~S Thygesen.
\newblock Dielectric genome of van der {{Waals}} heterostructures.
\newblock {\em Nano Lett.}, 15(7):4616--4621, 2015.

\bibitem{qeh_tutorial}
{Calculating new building blocks}.
\newblock \url{https://qeh.readthedocs.io/en/latest/tutorials.html}, 2019.
\newblock [Online; accessed 15-May-2023].

\bibitem{held_simplified_2014}
Alexander Held and Michael Walter.
\newblock Simplified continuum solvent model with a smooth cavity based on
  volumetric data.
\newblock {\em The Journal of Chemical Physics}, 141(17):174108, 2014.

\bibitem{Alavi_JCP_2001}
A.~Y. Lozovoi, A~Alavi, Jorge Kohanoff, and R~M Lynden-Bell.
\newblock {Ab initio simulation of charged slabs at constant chemical
  potential}.
\newblock {\em The Journal of Chemical Physics}, 115(4):1661--1669, jul 2001.

\bibitem{Melander2019}
Marko~M Melander, Mikael~J Kuisma, Thorbj{\o}rn Erik~K{\o}ppen Christensen, and
  Karoliina Honkala.
\newblock {Grand-canonical approach to density functional theory of
  electrocatalytic systems: Thermodynamics of solid-liquid interfaces at
  constant ion and electrode potentials}.
\newblock {\em The Journal of Chemical Physics}, 150(4):041706, jan 2019.

\bibitem{Sakaushi2020}
Ken Sakaushi, Tomoaki Kumeda, Sharon Hammes-Schiffer, Marko~M Melander, and
  Osamu Sugino.
\newblock {Advances and challenges for experiment and theory for multi-electron
  multi-proton transfer at electrified solid–liquid interfaces}.
\newblock {\em Phys. Chem. Chem. Phys.}, 22(35):19401--19442, 2020.

\bibitem{Lindgren2022}
Per Lindgren, Georg Kastlunger, and Andrew~A. Peterson.
\newblock Electrochemistry from the atomic scale, in the electronically
  grand-canonical ensemble.
\newblock {\em The Journal of Chemical Physics}, 157(18):180902, nov 2022.

\bibitem{Sugino_PRB_2006}
M~Otani and O~Sugino.
\newblock {First-principles calculations of charged surfaces and interfaces: A
  plane-wave nonrepeated slab approach}.
\newblock {\em Phys. Rev. B}, 73(11):115407, mar 2006.

\bibitem{Vaspsol_2014}
Kiran Mathew, Ravishankar Sundararaman, Kendra Letchworth-Weaver, T~A Arias,
  and Richard~G Hennig.
\newblock {Implicit solvation model for density-functional study of nanocrystal
  surfaces and reaction pathways}.
\newblock {\em The Journal of Chemical Physics}, 140(8):84106, 2014.

\bibitem{Letchworth-Weaver2012}
Kendra Letchworth-Weaver and T.~A. Arias.
\newblock {Joint density functional theory of the electrode-electrolyte
  interface: Application to fixed electrode potentials, interfacial
  capacitances, and potentials of zero charge}.
\newblock {\em Phys. Rev. B}, 86(7):075140, aug 2012.

\bibitem{Hormann2019}
Nicolas~G. H{\"{o}}rmann, Oliviero Andreussi, and Nicola Marzari.
\newblock {Grand canonical simulations of electrochemical interfaces in
  implicit solvation models}.
\newblock {\em The Journal of Chemical Physics}, 150(4):041730, jan 2019.

\bibitem{Ringe2022implrev}
Stefan Ringe, Nicolas~G H{\"{o}}rmann, Harald Oberhofer, and Karsten Reuter.
\newblock {Implicit Solvation Methods for Catalysis at Electrified Interfaces}.
\newblock {\em Chem. Rev.}, 122(12):10777--10820, jun 2022.

\bibitem{Kastlunger2018}
Georg Kastlunger, Per Lindgren, and Andrew~A. Peterson.
\newblock Controlled-potential simulation of elementary electrochemical
  reactions: Proton discharge on metal surfaces.
\newblock {\em The Journal of Physical Chemistry C}, 122(24):12771--12781, jun
  2018.

\bibitem{Trasatti1986}
Sergio Trasatti.
\newblock The absolute electrode potential: an explanatory note
  (recommendations 1986).
\newblock {\em Journal of Electroanalytical Chemistry and Interfacial
  Electrochemistry}, 209(2):417--428, sep 1986.

\bibitem{10.1063/1.4976971}
Ravishankar Sundararaman and Kathleen Schwarz.
\newblock {Evaluating continuum solvation models for the electrode-electrolyte
  interface: Challenges and strategies for improvement}.
\newblock {\em The Journal of Chemical Physics}, 146(8), 02 2017.

\bibitem{10.1063/1.5047829}
Marko~M. Melander, Mikael~J. Kuisma, Thorbj{\o}rn Erik~K{\o}ppen Christensen,
  and Karoliina Honkala.
\newblock {Grand-canonical approach to density functional theory of
  electrocatalytic systems: Thermodynamics of solid-liquid interfaces at
  constant ion and electrode potentials}.
\newblock {\em The Journal of Chemical Physics}, 150(4), 11 2018.

\bibitem{PhysRevB.45.13709}
M.~Weinert and J.~W. Davenport.
\newblock Fractional occupations and density-functional energies and forces.
\newblock {\em Phys. Rev. B}, 45:13709--13712, Jun 1992.

\bibitem{Lindgren2020HER}
Per Lindgren, Georg Kastlunger, and Andrew~A Peterson.
\newblock {A Challenge to the G$\approx$0 Interpretation of Hydrogen
  Evolution}.
\newblock {\em ACS Catal.}, 10(1):121--128, jan 2020.

\bibitem{Kastlunger2022pH}
Georg Kastlunger, Lei Wang, Nitish Govindarajan, Hendrik~H Heenen, Stefan
  Ringe, Thomas Jaramillo, Christopher Hahn, and Karen Chan.
\newblock {Using pH Dependence to Understand Mechanisms in Electrochemical CO
  Reduction}.
\newblock {\em ACS Catal.}, 12(8):4344--4357, apr 2022.

\bibitem{Kastlunger2023selectivity}
Georg Kastlunger, Hendrik~H. Heenen, and Nitish Govindarajan.
\newblock {Combining First-Principles Kinetics and Experimental Data to
  Establish Guidelines for Product Selectivity in Electrochemical CO 2
  Reduction}.
\newblock {\em ACS Catal.}, 13(7):5062--5072, apr 2023.

\bibitem{melander_wu_honkala_2023}
Marko Melander, Tongwei Wu, and Karoliina Honkala.
\newblock {Constant inner potential DFT for modelling electrochemical systems
  under constant potential and bias}.
\newblock {\em ChemRxiv}, 2023.
\newblock DOI={10.26434/chemrxiv-2021-r621x-v3}.

\bibitem{PhysRevA.72.024502}
Qin Wu and Troy Van~Voorhis.
\newblock Direct optimization method to study constrained systems within
  density-functional theory.
\newblock {\em Phys. Rev. A}, 72:024502, Aug 2005.

\bibitem{10.1063/1.2360263}
Qin Wu and Troy Van~Voorhis.
\newblock {Extracting electron transfer coupling elements from constrained
  density functional theory}.
\newblock {\em The Journal of Chemical Physics}, 125(16), 10 2006.

\bibitem{doi:10.1021/cr200148b}
Benjamin Kaduk, Tim Kowalczyk, and Troy Van~Voorhis.
\newblock Constrained density functional theory.
\newblock {\em Chem. Rev.}, 112(1):321--370, 2012.

\bibitem{doi:10.1021/acs.jctc.6b00815}
Marko Melander, Elvar~{\"O}. J{\'o}nsson, Jens~J. Mortensen, Tejs Vegge, and
  Juan~Maria Garc{\'\i}a~Lastra.
\newblock {Implementation of Constrained DFT for Computing Charge Transfer
  Rates within the Projector Augmented Wave Method}.
\newblock {\em J. Chem. Theory Comput.}, 12(11):5367--5378, 2016.

\bibitem{doi:10.1021/acs.chemmater.7b04618}
Haesun Park, Nitin Kumar, Marko Melander, Tejs Vegge, Juan~Maria Garcia~Lastra,
  and Donald~J. Siegel.
\newblock {Adiabatic and Nonadiabatic Charge Transport in Li}--{S Batteries}.
\newblock {\em Chem. Mater.}, 30(3):915--928, 2018.

\bibitem{kumeda2022cations}
Tomoaki Kumeda, Laura Laverdure, Karoliina Honkala, Marko~M. Melander, and Ken
  Sakaushi.
\newblock {Cations determine the mechanism and selectivity of alkaline ORR on
  Pt(111)}, 2022.

\bibitem{Melander_2020}
Marko~M. Melander.
\newblock Grand canonical rate theory for electrochemical and electrocatalytic
  systems i: General formulation and proton-coupled electron transfer
  reactions.
\newblock {\em J. Electrochem. Soc.}, 167(11):116518, jul 2020.

\bibitem{LevyOFDFT}
Mel Levy, John~P. Perdew, and Viraht Sahni.
\newblock Exact differential equation for the density and ionization energy of
  a many-particle system.
\newblock {\em Phys. Rev. A}, 30:2745--2748, Nov 1984.

\bibitem{Lehtomaki2014}
Jouko Lehtomäki, Ilja Makkonen, Miguel~A. Caro, Ari Harju, and Olga
  Lopez-Acevedo.
\newblock {Orbital-free density functional theory implementation with the
  projector augmented-wave method}.
\newblock {\em The Journal of Chemical Physics}, 141(23), 12 2014.
\newblock 234102.

\bibitem{Lehtomaki2019}
Jouko Lehtomäki and Olga Lopez-Acevedo.
\newblock {Large-Z limit in atoms and solids from first principles}.
\newblock {\em The Journal of Chemical Physics}, 151(24), 12 2019.
\newblock 244101.

\bibitem{ivady2014pressure}
Viktor Iv{\'a}dy, Tam{\'a}s Simon, Jeronimo~R Maze, IA~Abrikosov, and Adam
  Gali.
\newblock Pressure and temperature dependence of the zero-field splitting in
  the ground state of nv centers in diamond: A first-principles study.
\newblock {\em Phys. Rev. B}, 90(23):235205, 2014.

\bibitem{rayson2008first}
MJ~Rayson and PR~Briddon.
\newblock First principles method for the calculation of zero-field splitting
  tensors in periodic systems.
\newblock {\em Phys. Rev. B}, 77(3):035119, 2008.

\bibitem{biktagirov2020spin}
Timur Biktagirov, Wolf~Gero Schmidt, and Uwe Gerstmann.
\newblock Spin decontamination for magnetic dipolar coupling calculations:
  Application to high-spin molecules and solid-state spin qubits.
\newblock {\em Physical Review Research}, 2(2):022024, 2020.

\bibitem{blochl2000first}
Peter~E Bl{\"o}chl.
\newblock First-principles calculations of defects in oxygen-deficient silica
  exposed to hydrogen.
\newblock {\em Phys. Rev. B}, 62(10):6158, 2000.

\end{thebibliography}
